\documentclass[12pt]{report}
\usepackage{amsmath, amssymb}
\usepackage[dvips]{graphicx}

\title{Gravitational Waves and Light Cosmic Strings}
\author{Matthew DePies\\
          University of Washington \\ 
          \\
           Advisor: Craig Hogan\\
           University of Chicago and Fermilab}
\date{2009}

\begin{document}

\maketitle

\setcounter{tocdepth}{1}  

%


%

%
%

\setcounter{page}{-1}
\abstract{
Gravitational wave signatures from cosmic strings are analyzed numerically.  Cosmic string networks form during phase transistions in the early universe and these networks of long cosmic strings break into loops that radiate energy in the form of gravitational waves until they decay.  The gravitational waves come in the form of harmonic modes from individual string loops, a ``confusion noise'' from galactic loops, and a stochastic background of gravitational waves from a network of loops.  In this study string loops of larger size $\alpha$ and lower string tensions $G\mu$ (where $\mu$ is the mass per unit length of the string) are investigated than in previous studies.  Several detectors are currently searching for gravitational waves and a space based satellite, the \textit{Laser Interferometer Space Antenna} (LISA), is in the final stages of pre-flight.  The results for large loop sizes ($\alpha=0.1$) put an upper limit of about $G\mu<10^{-9}$ and indicate that gravitational waves from string loops down to $G\mu \approx 10^{-20}$ could be detectabe by LISA.  The string tension is related to the energy scale of the phase transition and the Planck mass via $G\mu = \Lambda_s^2 / m_{pl}^2$, so the limits on $G\mu$ set the energy scale of any phase transition to $\Lambda_s < 10^{-4.5} m_{pl}$.  Our results indicate that loops may form a significant gravitational wave signal, even for string tensions too low to have larger cosmological effects.
} 
%
%
\tableofcontents
\listoffigures
\listoftables 
 
%

%
\newpage

  The author wishes to thank many who have helped during the long and arduous process that has led to this dissertation.  
  
In particular I would like to thank Oscar Vilches for overseeing my first fumbling foray into research and for his unwavering support ever since.  This dissertation would not have happened without his support.  Warren Buck has illuminated the postivie impact a physicist can make during a career, and his support has been invaluable.
  
Also, Craig Hogan has been a thoughtful and patient advisor, and is the inpsiration behind much of the work here.

The faculty at Humboldt State University led me into this wonderful subject and I would like to thank them all: Patrick Tam, Fred Cranston, Leung Chin, and Lester Clendenning.  In particular I would like to thank Richard Thompson for his support and time;  his enjoyment of physics and teaching were contagious.

My parents Michelle and Robert have always believed I would finish this work, and my siblings Nic and Camille have been supporters throughout.

Especially to Joel Koeth, fellow traveller from the battlefields of the soldier to the classrooms of the scholar.
  
  Supported by NASA grant NNX08AH33G at the University of Washington.   

%
%


\chapter{Introduction}

  \section{Cosmic Strings}
  
Quantum field theory predicts that the universe can transition between different vacuum states while cooling during expansion.  In these theories the vacuum expectation value of a field $\left\langle \phi\right\rangle$ can take on non-zero values when the ground state, at the minimum of the zero-temperature potential or ``true vacuum'',  breaks the symmetry of the underlying Lagrangian \cite{vil,hin}.  In the hot early universe, thermal effects lead to an effective potential with a temperature dependence,  such that the vacuum starts in a symmetric ``false'' vacuum and only later cools to its ground state  below a certain critical temperature~\cite{linde,vil,ba,va84,vil85}.  In general however the expansion of the universe occurs too quickly for the system to find its true ground state at all points in space~\cite{ki}. 

Depending upon the topology of the manifold of degenerate vacua, topologically stable defects form in this process, such as  domain walls, cosmic strings, and monopoles~\cite{vil,hin,va84,vil85,sak06}.  In particular, any field theory with a broken U(1) symmetry will have classical solutions extended in one dimension, and in cosmology these structures generically form a cosmic network of macroscopic, quasi-stable strings that steadily unravels but survives to the present day, losing energy primarily by gravitational radiation ~\cite{vi81,ho3,vi85,batt,turok,allen01}.  A dual superstring description for this physics is given in terms of one-dimensional branes, such as D-strings and F-strings~\cite{vi05,pol,dv,sas,jones1,pol06}.

Previous work has shown that strings form on scales larger than the horizon, and expand with the horizon with cosmic time~\cite{turok1984}.  These are referred to as ``infinite strings'', and through various mechanisms give rise to loops which form as a fraction of the horizon.  These loops are the primary potential source of gravitational radiation detectable from the earth.

Calculations are undertaken at much lower string masses than previously, with several motivations:
 
	\begin{enumerate}
\item
Recent advances in millisecond pulsar timing have reached new levels of precision and are providing better limits on low frequency backgrounds; the   calculations presented here provide a precise connection between the background limits and fundamental theories of strings and inflation~\cite{fir05,jenet,jones,tye06}.
   \item
The \emph{Laser Interferometer Space Antenna} (LISA) will provide much more sensitive limits over a broad band around millihertz frequencies.  Calculations have not previously been made for theories in this band of sensitivity and frequency, and are needed since the background spectrum depends significantly on string mass~\cite{lisa,armstrong,ci}.
   \item
Recent studies of string network behavior strongly suggest (though they have not yet proven definitively) a high rate of formation of stable string loops comparable to the size of the cosmic horizon.  This results in a higher net production of gravitational waves since the loops of a given size, forming earlier, have a higher space density. The more intense background means experiments are sensitive to lower string masses~\cite{ring,van2,martins}.
 \item
Extending the observational probes to the light string regime is an important constraint on field theories and superstring cosmology far below the Planck scale.  The current calculation provides a quantitative bridge between the parameters of the fundamental theory (especially, the string tension), and the properties of the observable background~\cite{ki04,sak06}.
\end{enumerate}

Cosmic strings' astrophysical properties are strongly dependent upon two parameters:  the dimensionless string tension $G\mu$ (in Planck units where $c=1$ and $G={m_{pl}}^{-2}$) or $G\mu/c^2$ in SI units, and the interchange probability p.  Our main conclusion is that the current pulsar data~\cite{jenet,det,sti} already place far tighter constraints on string tension than other arguments, such as microwave background anisotropy or gravitational lensing.  The millsecond pulsar limits are dependant upon the nature of the source, and are detailed in later chapters.  Recent observations from WMAP and SDSS \cite{wy} have put the value of $G\mu< 3.5\times10^{-7}$ due to lack of cosmic background anisotropy or structure formation consistent with heavier cosmic strings.  From \cite{wy} it is found that up to a maximum of 7\% (at the 68\% confidence level) of the  microwave anisotropy can be cosmic strings.  In other words, for strings light enough to be consistent with current pulsar limits, there is no observable effect other than their gravitational waves.    The limit is already a powerful constraint on superstring and field theory cosmologies.  In the future, LISA will improve this limit by many orders of magnitude.

The energy density of the cosmic strings is dependant upon the energy scale of the associated phase transition: for a GUT (Grand Unified Theory) scale phase transistion the energy is of the order $10^{16}$ GeV and the electroweak scale on the order of $10^{3}$ GeV.  In the case of cosmic strings, this energy determines the mass (energy) per unit length $\mu$ of the string, and the dimensionless string tensions $G \mu /c^2$ in SI units or $G \mu$ when $c=1$.  For this study lighter strings on the order of $10^{-15}< G \mu <  10^{-8}$ are used, as analysis of millisecond pulsars has ruled out higher mass strings~\cite{jenet}.

In general, lower limits on the string tension constrain theories farther below the Planck scale.  In field theories predicting string formation the string tension is related to the energy scale of the theory $\Lambda_s$ through the relation $G\mu\propto\Lambda_s^2/m_{pl}^2$~\cite{vil,ki,ba}.  Our limits put an upper limit on $\Lambda_s$ in Planck masses given by $\Lambda_s<10^{-4.5}$, already in the regime associated with Grand Unification; future sensitivity from LISA will reach $\Lambda_s\approx 10^{-8}$, a range often associated with a Peccei-Quinn scale, inflationary reheating, or supersymmetric B-L breaking scales~\cite{jean}.  In the dual superstring view, some current brane cosmologies predict that the string tension will lie in the range $10^{-6}<G\mu<10^{-11}$~\cite{tye05,fir05,tye06}; our limits are already in the predicted range and LISA's sensitivity will reach beyond their lower bound.

  \section{Equation of State and Thermodynamics}
  
  In this section a selection of the constituents of the universe and some of their thermal properties are analyzed~\cite{ba,carroll,dodson,hartle,liddle,much,narl,peacock,peebles,wein}.  These results are important in calculations of the rate of cosmic expansion which, in turn, affects the number of cosmic string loops formed and their size.  For a more thorough discussion of relativistic cosmology and general relativity see Appendices A and B and the aforementioned references.
  
We can define a region of space as a comoving volume denoted by $a(t)^3$ where $a(t)$ is the scale factor, which expands with the universe.  Of interest in this region are the equations of state of the constituents and how they evolve with time.  Of utmost importance to cosmic dynamics is the stress-energy tensor at different points in space.  More details on the stress-energy tensor are given in Appendices A and B, but it is a second rank tensor that describes the mass/energy content, energy flux, momentum density, and stess at a spacetime point.  

 As a useful and relevant example, a perfect fluid has a stress-energy tensor of the form,
\begin{equation}
    T_{\mu \nu}=(\rho + p) u_{\mu} u_{\nu} + p g_{\mu \nu},
 \end{equation}
where $u$ is the four velocity of the fluid.  In a local (Minkowski) frame the tensor is:
 \begin{equation}
       T_{00}=\rho \;\;\; \text{and} \;\;\; T_{ij}=p \delta_{ij},
 \end{equation}
where $\rho$ is the density and $p$ is the pressure of the cosmic constituent.  In this frame there is a simple definition of the stress-energy tensor:  the 00 component is the energy density and the diagonal components are the pressure.  Again, see Appendices A and B for more details.

  In general we can relate the equation of state to the energy density $\rho$ via,
\begin{equation}
p=w \rho.
\end{equation}
Conservation of the stress-energy tensor gives a relationship between the scale factor and density,
\begin{eqnarray}
  D_{\mu}T^{\mu\nu}  &=& 0,\\
  \partial_{\nu} T^{\mu\nu}+\Gamma^{\mu}_{\nu \lambda}T^{\lambda\nu} + \Gamma^{\nu}_{\nu \lambda}T^{\mu\lambda} &=& 0,
 \end{eqnarray}
where $\Gamma^{\beta}_{\alpha \lambda}$ are the Christoffel symbols (see Appendix A) and $D_{\mu}$ is the covariant derivate.   Using the $\mu=0$ component of the stress-energy tensor leads to,
 \begin{equation}
 \dot{\rho}+3(\rho+p)\frac{\dot{a}}{a}=0.
 \end{equation}
This equation gives the time dependance of the density of each constituent, and also describes the dependance of the densities on the scale factor (or volume).

For the constituents of the universe we have:
\begin{eqnarray}
  \text{Non-relativistic matter},\; w=0,\;              & &\rho\propto a^{-3},\\
  \text{Ultra-relativistic  matter},\; w=\frac{1}{3},\; & & \rho \propto a^{-4}\\
  \text{Vacuum energy (Dark energy)},\; w=-1,\;         & & \rho=\text{constant}.
 \end{eqnarray}
These relations indicate a universe composed of ``dark enegy'' eventually has the expansion dominated by it.  Both relativistic and non-relativistic matter density shrink as the universe expands, which corresponds to the observation that the very early universe was dominated by radiation, which was replaced by non-relativistic matter, finally to be overtaken today by vacuum energy.  A more complete description is given in Appendix B.

For reference we investigate the long ``infinte'' string equation of state~\cite{ca},
\begin{equation}
p_{\infty}=\frac{1}{3} \rho_{\infty} (2\left\langle v^2 \right\rangle -1),
\end{equation}
which varies with the rms speed of the long strings.  Assuming the infinite strings are ultra-relativistic we find,
\begin{equation}
p_{\infty}=\frac{1}{3} \rho_{\infty}.
\end{equation}
For the dynamics of the scale factor, the cosmic strings are a miniscule component, thus their equation of state is unimportant.   The upper limit on the density of strings is very small from observations of the cosmic microwave background looking for string related inhomogeneities using WMAP and SDSS~\cite{wy}.

     \section{Field Theory and Symmetry Breaking}
     
 The Lagrangian density $\mathcal{L}$ defines the spectrum of particles and interactions in a theory.  A number of requirements are imposed on any viable field theory to include:  the theory must be Lorentz invariant (relativistic), invariant under general gauge transformations, and result in a spectrum of particles and interactions that matches observation.  Said theory must also take into account local symmetries associated with the gauge group of the field.  For example, the electromagnetic interaction is symmetric under the $U(1)_{em}$ symmetry group, while the electroweak interaction is symmetric under the gauge group $SU(2)_{L} \times U(1)_Y$.  QCD is symmetric under $SU(3)_{QCD}$. 
 
For example, at some finite temperature the symmetry of the $SU(2)_{L} \times U(1)_Y$ group is broken to leave the $U(1)_{em}$ unbroken.   The resulting spectrum is the one in which we are familiar: spin-1 photons and spin-1/2 leptons.  This is predicted to occur at a temperature of about $10^3$ GeV, or $T \approx 10^{16}$K, using $k=8.617 \times 10^{-5}$ GeV/K.  Likewise, the group $SU(3)_{QCD} \times SU(2)_{L} \times U(1)_Y$ can be an unbroken group of another phase transition, e.g. the $SU(5)$ group.

 For more detail let us look at the electroweak model.  The Langrangian for the electroweak theory has a pure gauge and a Higgs boson part, given by:
  
 \begin{eqnarray}
 \mathcal{L}_{ew} =\bar{f}(\gamma^{\mu}D_{\mu}-m)f + \mathcal{L}_{gauge}+\mathcal{L}_{Higgs},
 \end{eqnarray}
 where   
 \begin{eqnarray}  
   \mathcal{L}_{gauge}&=&-\frac{1}{4}W^a_{\mu\nu} W^{a\:\mu\nu}-\frac{1}{4}B_{\mu\nu}B^{\mu\nu},\\
   \mathcal{L}_{Higgs}&=& (D_{\mu}\phi)^{\dagger} (D^{\mu}\phi)-V(\phi^{\dagger},\phi,T).
\end{eqnarray}
and
\begin{eqnarray}
     W^a_{\mu \nu} &=& \partial_{\mu} W^a_{\nu} - \partial_{\nu} W^a_{\mu} - g \epsilon^{abc} W^b_{\mu}W^c_{\mu},\\
     B_{\mu \nu}\  &=& \partial_{\mu}  B_{\nu} - \partial_{\nu}  B_{\mu}       
\end{eqnarray}

The $W^a_{\mu}$ are the three gauge fields associated with the generators of $SU(2)$ and $B_{\mu}$ is the guage field associated with $U(1)_Y$:
Here $D_{\mu}$ is the covariant derivative and is 
\begin{equation}
  D_{\mu}=\partial_{\mu}+\frac{ig}{2}\tau_a W_{\mu}^{a} +\frac {ig'}{2} B_{\mu}   
\end{equation} 
where the $\tau_a$ are the three generators of the group, represented by Pauli matricies for $SU(2)$.  $g$ and $g'$ are the gauge coupling constants for $SU(2)$ and $U(1)$.  

The potential $V(\phi, T)$ determines the behavior of the theory at varying temperatures, $T$.  At high $T$ the potential keeps the symmetry of the underlying theory.  When $T$ drops to a critical value the old symmetry is broken, and the spectrum of particles is changed.  For the electroweak theory, the massless exchange particles, the $W$ and $B$, which act like a photon, are gone and replace by the $Z$ and $W^\pm$ gauge bosons,which are massive.  The electromagnetic field $A_{\mu}$ is massless, from the superposition given by,
\begin{equation}
   A_{\mu}=\cos(\theta_W)B_{\mu}+\sin(\theta_W)W^3_{\mu},
 \end{equation}
 which remains massless after the break to $U(1)_{em}$ symmetry.  The $\theta_W$ is the Weinberg angle, which when coupled to the electromagnetic interaction with strength $e$ gives,
\begin{equation}
    e=g \sin(\theta_W)=g' \cos(\theta_W).
 \end{equation}
      
An elementary example of this symmetry breaking is the ``mexican hat'' potential.  At the critical temperature the minimum of the potential $V(\phi)$ takes on a set of degenerate values that break the symmetry of the theory.

Theories with global symmetry will tend to radiate Goldstone bosons, which typically reduce the lifetime of string loops to approximately 20 periods.  This reduces to possible gravitational wave signal, likely making it invisible.  Strings created by a locally symmetric gauge theory that radiate only gravitational waves are analyzed here, as most of the energy lost from the strings is in the form of gravitational waves.

     \section{Vacuum Expectation Value of Fields and Vacuum Energy}
     
Of great importance when dealing with topological defects is the vacuum expectation value (VEV) of the field.  The VEV is the  average value one expects of the field in the vacuum, denoted by $\left< \phi \right>$.  This can akin to the potential:  if the potential is 100 V throughout space that has no physical meaning.  But in the case of a homogeneous real scalar field, the VEV can lead to a non zero vacuum energy given by the stress-energy tensor,
\begin{eqnarray}
    T_{\mu \nu}=\frac{\partial \mathcal{L}}{\partial(\partial^{\mu}\phi)} \partial_{\nu} \phi  - g_{\mu \nu} \mathcal{L},
\end{eqnarray}  
which can be solved for the 00 term to get,
\begin{eqnarray}
   T_{00}=\rho=\frac{1}{2} [\dot{\phi}^2+  (\nabla\phi)^2]  +  V(\phi).
\end{eqnarray}

From above we find that even if $\left< \phi \right>$ is stationary, homogenous, and zero the potential $V(\left<\phi \right>)$ can result in a vacuum energy density.

During a phase transition the old VEV of zero changes to some non-zero value.  Given the finite travel time of information, for first order phase transitions bubbles of this new vacuum propagate at the speed of light.  Where these bubbles intersect can lead to cosmic turbulence as well as topological defects, to include cosmic strings.  For our strings, the zero VEV region is within the string and the new VEV without.

The regions of old vacuum (topological defects) are stable if the thermal fluctuations in the field are less than the differences in energy between regions, i.e. the old and new vacuums.  This is typically given by the Ginzburg temperature $T_G$ and is close to the critical temperature of the phase transition.  Below the Ginzburg temperature the defects ``freeze-out'' and are stable~\cite{vil}.

 \section{Cosmic String Motion}
    
Cosmic string motion can be characterized by paramaterization of the four space-time dimensional motion onto a two-dimensional surface called a worldsheet.  With the coordinate parameterization given by:
\begin{equation}
x^{\mu}=X^{\mu}(\sigma^a),\;\;a=0,1,
\end{equation}
where the $x^{\mu}$ are the four spacetime coordinates, $X^{\mu}$ are the function of $\sigma^a$, and the $\sigma^a$ are the coordinates on the worldsheet of the string.  In general $\sigma^0$ is considered timelike and $\sigma^1$ is considered spacelike.  On the worldsheet the spacetime distance is given by,
\begin{eqnarray}
ds^2 &=& g_{\mu\nu}dx^{\mu}dx^{\nu},\\
     &=& g_{\mu\nu}\frac{dx^{\mu}}{d\sigma^a} \frac{dx^{\nu}}{d\sigma^b} d\sigma^a d\sigma^b,\\
     &=& \gamma_{ab}d\sigma^a d\sigma^b,
\end{eqnarray}
where $\gamma_{ab}$ is the induced metric on the worldsheet.

The equations of motion for the string loops devised by Yoichiro Nambu were formulated by him assuming that that the action was a functional of the metric $g_{\mu\nu}$ and on functions $X_{\mu}$.  Also added were the assumptions:
\begin{itemize}
 \item{the action is invariant with respect to spacetime and gauge transformations,} 
 \item{the action contains only first order and lower derivatives of $X_{\mu}$.}
 \end{itemize}
Ths solution is dubbed the the Nambu-Goto action:
\begin{equation}
S=-\mu \int \sqrt{-\gamma} d^2 \sigma,
\end{equation}
where $\gamma$ is the determinant of $\gamma_{ab}$.  This is not the only relativistic action for cosmic strings, but is the only one that obeys the caveats above.  Including higher order derivatives corrects for the curvature of the string, but is only important in the neighborhood of cusps and kinks~\cite{an}

Minimizing the action leads to the following equations,
\begin{eqnarray}
\frac{\partial x^{\mu}}{\partial \sigma^a} ^{;a} + \Gamma^{\mu}_{\alpha \beta} \frac{\partial x^{\alpha}}{\partial \sigma^a} \frac{\partial x^{\beta}}{\partial \sigma^b}=0
\end{eqnarray}
where $\Gamma$ are the Christoffel symbols.

In flat spacetime with $g_{\mu\nu}=0$ and $\Gamma^{\mu}_{\alpha \beta}=0$ the string equations become,
\begin{equation}
   \frac{\partial    }{\partial \sigma^a} \left( \sqrt{-\gamma} \gamma^{ab} \frac{\partial x^{\mu}}{\partial \sigma^b} \right)=0.
 \end{equation}

Now we use the fact that the Nambu action is invariant under general coordinate reparameterizations, and we choose a specific gauge.  This does not disturb the generality of the solutions we find, since the action is invariant under these gauge transformations.  Here we choose:
\begin{eqnarray}
\gamma_{01} &=& 0,\\
\gamma_{00}+\gamma_{11} &=& 0,
\end{eqnarray}
this called the ``conformal'' gauge, because the worldsheet metric becomes conformally flat.

Our final choice is to set the coordinates by
\begin{eqnarray}
t=x^0=\sigma^0\\
\sigma \equiv \sigma^1,
\end{eqnarray}
so that we have fixed the time direction to match both our four dimensional spacetime and the worldsheet.  This is sometimes called the ``temporal'' gauge.  This leads to the equations of motion,
\begin{eqnarray}
\dot{\textbf{x}} \cdot \textbf{x}' &=& 0,\\
\dot{\textbf{x}}^2 + \textbf{x}'^2 &=& 1,\\
\ddot{\textbf{x}}-\textbf{x}''    &=& 0.
\end{eqnarray}
The first equation states the orthogonality of the motion in these coordinates, which leads to $\dot{\textbf{x}}$ being the physical velocity of the string.  The second equation is statment of causality as the total speed of the motion is c, which can also be interpreted as,
\begin{eqnarray}
   d\sigma &=& \frac{|dx|}{\sqrt{1-\dot{x}^2}} = dE/\mu, \\
   E       &=& \mu \int \frac{|dx|}{\sqrt{1-\dot{x}^2}} = \mu \int d\sigma.
\end{eqnarray} 
So the coordinates $\sigma$ measure the energy of the string.

  The final equation is a three dimensional wave equation.  This equation determines the behavior of the strings and has a general solution of the form:
 \begin{equation}
   \textbf{x}(\sigma,t)=\frac{1}{2}[\textbf{a}(\sigma-t)+\textbf{b}(\sigma-t)],
 \end{equation}
 where constraints from the equations of motion give,
 \begin{equation}
 \textbf{a}'=\textbf{b}'=1.
 \end{equation}

We can also write down the stress-energy tensor, momentum, and angular momentum in this choice of gauge:
\begin{eqnarray}
T^{\mu \nu} = \mu \int{d\sigma} [\dot{x}^{\mu} \dot{x}^{\nu}-x'^{\mu} x'^{\nu}] \delta^{(3)}(\textbf{x}-\textbf{x}(\sigma,t)),
\end{eqnarray}
which leads tot he energy given above, as well as,
\begin{eqnarray}
\textbf{p} &=& \mu \int d\sigma \; \dot{\textbf{x}}(\sigma,t),\\
\textbf{J} &=& \mu \int d\sigma \;  \textbf{x}(\sigma,t) \times \dot{\textbf{x}}(\sigma,t).
\end{eqnarray}

A cosmic string network is formed and behaves as a random walk.  As the long strings and string segments move and oscillate they often intersect one another.  Depending upon the characteristics of the strings a loop can be formed, which we discuss in the next section.

    \section{Cosmic String Loops}
    
\begin{figure}
   
	 \includegraphics[width=.85\textwidth]{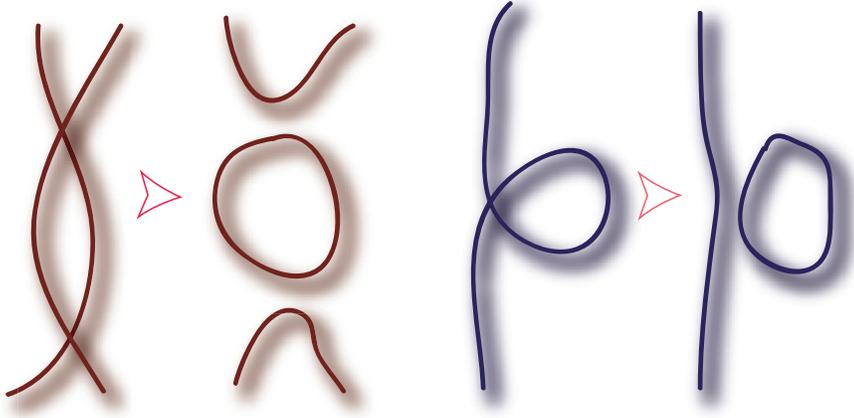}
  \caption{\label{figB} Loop formation from long strings.  Two possibilities for the creation of cosmic string loops from the intersection of long strings.  }
		
\end{figure}

Previous results have shown that the majority of energy radiated by a cosmic string network comes from loops which have detached from the infinite strings.  Loops are formed by interactions between the infinite strings, as shown in Fig.~\ref{figB}.  Of utmost importance is the interchange probability of the strings $p$.  If $p$=1 then each time strings intersect they exchange partners, and a loop is much more likely to form. 
    
\begin{figure}
   
	 \includegraphics[width=.85\textwidth]{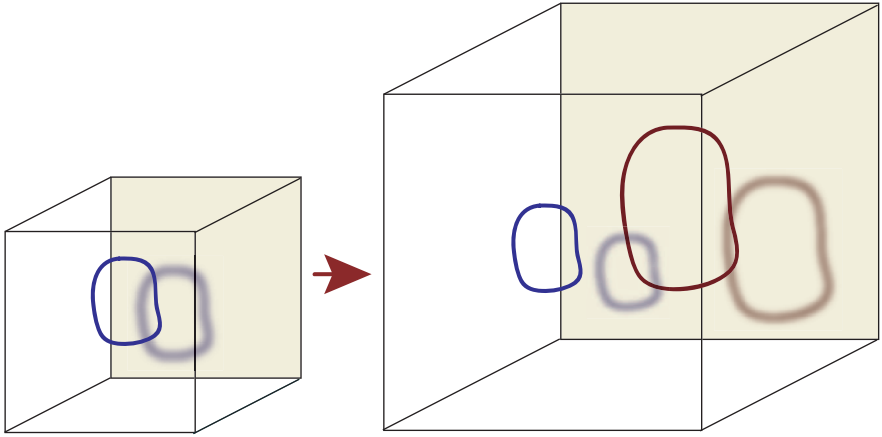}
  \caption{\label{figC} Scaling of cosmic string loops.  The figure on the left shows a newly formed loop, scaled to a certain fraction of the horizon.  When the horizon expands as on the right, the newly formed loop on the right scales with the expansion, while the old loop on the left continues to radiate gravitational waves and shrink.}
		
\end{figure}

Numerical simulations, analytic studies, and heuristic arguments indicate cosmic string loops form on the ``one-scale'' model; that is, their size scales with the horizon as in Fig.~\ref{figC}.  The ratio of the size of the loops to the horizon is given by $\alpha$, and ranges from the largest size of 0.1 down to the miniscule $10^{-6}$.   There is some debate about the stability of the larger string loops, and older numerical simulations predict that any large loops should decay to the smaller sizes.  More recent simulations indicate the possibility of large stable loops, which is part of the motivation for this study~\cite{ring,van2,martins}.

To analyze the strings in the universe we must make estimates of the loop formation rate, based on our models of string creation.  First we analyze the number of infinite strings created within the cosmic horizon.  By infinite we mean they extend beyond the horizon, and stretch as the horizon grows.  From this the number of loops formed per unit time can be analyzed.

To do this we define three parameters: $A$ is the number of infinite strings within the horizon, $B$ is the creation rate of loops within the horizon, and $\alpha$ is the size of loops formed relative to the horizon.  For the first two we have~\cite{ca},
\begin{eqnarray}
A=\rho_{\infty}\frac{L^2(t)}{\mu c^2},\\
B=\frac{dN}{dL} \frac{L^4(t)}{V(t)},\\
\end{eqnarray}
where $\rho_{\infty}$ is the energy density of infinite strings.  In principle the number density at any time can be calculated after these parameters have been chosen. 

For this study we set the number density of loops per log cosmic time interval in years of 0.1 ,
\begin{equation}
n(t_o,t_c)=\frac{N_t}{\alpha} \left( \frac{H(t_c)}{c} \right)^3 \left( \frac{a(t_c)}{a(t_o)} \right)^3,
\end{equation} 
where $t_c$ is the time of creation of the loops and $t_o$ is the time of observation.  In this equation our paramter $N_t$ has replaced $A$ and $B$ and incorporated $\rho_{\infty}$.  Results are discussed in more detail in the chapters describing the gravitational wave signal results.


\chapter{Introduction to Gravitational Waves}

Einstein's theory of general relativity is the foundation of modern cosmology, but at its inception it was a theory that seemed to be un-needed.  Today, though, it is found indespensable in explaining a large number of phenomena and in the proper functioning of such things as the \textit{Global Positioning System}.  One of the hallmarks of general relativity, and indeed of all metric theories of gravity, is the production of gravitational waves.  The motion of a gravitational mass alters the space-time nearby, and this alteration then propagates at the speed of light producing a gravitational wave.

In this chapter we discuss the basics of gravitational waves and introduce some of the equations that are used to describe them.  Of great import are the plane wave solutions, given that most of the sources are a distance from the detector.  For more details on general relativity see Appendix A and references cited there.  For more on gravitational waves in particular see~\cite{ci,magg}.

 \section{Einstein's Equations}

 Einstein's field equations are given by the relationships between the curvature of spacetime to the stress-energy tensor:
\begin{equation}
R_{\mu \nu}-\frac{1}{2}g_{\mu \nu}R=8\pi G T_{\mu \nu},
\end{equation}
where $R_{\mu \nu}$ is the Ricci tensor and $R$ is the Ricci scalar $g_{\mu \nu}R^{\mu \nu}$

For the weak field case we assume a small perturbation on our Minkowski metric so
\begin{equation}
g_{\mu \nu}=\eta_{\mu \nu}+h_{\mu \nu}
\end{equation}
where $h_{\mu\nu}<<1$.  The linearized Ricci tensor then becomes
\begin{equation}
R_{\mu\nu}=\partial_{\alpha} \Gamma^{\alpha}_{\mu\nu}+ \partial_{\nu} \Gamma^{\alpha}_{\mu\alpha},
\end{equation}
where the $ \Gamma^{\alpha}_{\mu\nu}$ are Christoffel symbols.

After inserting this into Einstein's equations we get:
\begin{equation}
-\partial_{\alpha} \partial^{\alpha} \bar{h}_{\mu \nu}-\eta_{\mu\nu} \partial^{\alpha} \partial^{\beta} \bar{h}_{\alpha \beta}+\partial_{\mu} \partial^{\alpha} \bar{h}_{\nu \alpha}=16\pi G T_{\mu\nu},
\end{equation}
where $\bar{h}_{\mu\nu}=h_{\mu \nu}-\frac{1}{2}\eta_{\mu \nu}h$ is the trace-reversed tensor.


   \section{Lorentz Gauge and Wave Equation}

From the above equations it isn't clear there are wave solutions, but if we choose the gauge
\begin{equation}
\partial^{\alpha}\bar{h}_{\alpha\beta}=0,
\end{equation}
which is known as the Lorentz gauge, we find the equation is now,
\begin{eqnarray}
\partial_{\alpha} \partial^{\alpha} \bar{h}_{\mu \nu}=-16\pi G T_{\mu \nu},\\
\label{eq:wav1}
(-\partial^2_t+\nabla^2)\bar{h}_{\mu \nu}=-16\pi G T_{\mu \nu}.
\end{eqnarray}

If we take the trace of both sides of the equation we find $\partial_{\alpha} \partial^{\alpha}h=16\pi G\;T$. If this is inserted into the above equation we get
\begin{equation}
\label{eq:wav2}
(-\partial_t^2 + \nabla^2) h_{\mu\nu}=-16\pi G \left( T_{\mu\nu}-\frac{1}{2}\eta_{\mu\nu}T \right),
\end{equation}
where $T=g_{\mu\nu}T^{\mu\nu}$.  Note that Eqs.~\ref{eq:wav1} and \ref{eq:wav2} are equivalent, although in practice Eq.~\ref{eq:wav2} is more useful when the stress-energy tensor is known.  This looks like a traditional wave equation, with the only difference being the tensor nature of the spacetime metric.  The general solution to Eq.~\ref{eq:wav2} is given by the retarded time solution,
\begin{equation}
h_{\mu\nu}(\textbf{x},t)=4 G \int{d^3 x' \frac {\bar{T}_{\mu\nu}(\textbf{x},t-\left|\textbf{x}-\textbf{x}' \right|)} {\left|\textbf{x}-\textbf{x}' \right|}}.
\end{equation}

Of course now we look at a region of space devoid of any matter, in which case the Einstein equations become,
\begin{equation}
(-\partial_t^2 + \nabla^2) h_{\mu\nu}=0.
\end{equation}
We should expect to get plane wave solutions from this.

   \section{Wave Solution and TT Gauge}
   
   \subsection{Plane Waves}
				
So solutions will be of the form
\begin{equation}
h_{\mu\nu}=A_{\mu\nu}e^{-ik^{\alpha}x_{\alpha}},
\end{equation}	
where $A_{\mu\nu}$ is a tensor describing the polarization of the plane waves. 

The \textit{Transervse Traceless} or TT gauge is used to greatly simplify the appearance of the wave solutions to Einstein's field equations.  The generality of the results is not damaged by this particular gauge choice, so it is to our benefit to make this change.

There are three main steps to creating the TT gauge: 
(a) from the Lorentz gauge we find that $A_{\mu\nu}k^{\nu}=0$ and $k_{\mu}k^{\mu}=0$.  
(b) We can also choose a Lorentz frame in which our velocity $\textbf{u}$ is orthogonal to $\textbf{A}$ so that $A_{\mu\nu}u^{\nu}=0$.  
(c) Finally we choose $tr(A)=A_{\;\mu}^{\mu}=0$.	

If we pick our particular frame to have the four velocity given by $u_{\mu}=(1,0,0,0)$,
\begin{displaymath}
h_{\mu\nu}= 
\left( \begin{array}{cccc}
             0&0         &0          &0 \\
             0&h_+       &h_{\times} &0 \\
             0&h_{\times}&-h_+       &0 \\
             0&0         &0          &0 \\
\end{array} \right) e^{-i(kz-wt)}.
\end{displaymath}
This can be rewritten in terms of polarization tensors $e_{\mu\nu}$,
\begin{equation}
h^{\mu\nu}=a\ e_{+}^{\mu\nu} e^{-i(kz-wt)} + b\ e_{\times}^{\mu\nu} e^{-i(kz-wt)}.
\end{equation}
where $a$ and $b$ are complex constants.
				
				\subsection{Spherical Waves}
				
For spherical waves we can use the same ansatz as before, but with $1/r$ dependence added:
\begin{equation}
h_{\mu\nu}=\frac{A_{\mu\nu}}{r}e^{-ik^{\alpha}x_{\alpha}}.
\end{equation}
In this case we make sure we have the proper sign in our exponential to ensure an outgoing wave.

			 \section{Particle Orbits from Gravitational Waves}
				
Now the important question: what happens to a particle as the wave passes?  Nothing!  Let's clarify that statement somewhat, lest the reader assume the author suggests there is no visible effect of a passing gravitational wave.  From the geodesic equation 
\begin{equation}
\frac{du^{\alpha}}{d\tau}=-\Gamma^{\alpha}_{\mu\nu} u^{\mu} u^{\nu},
\end{equation}
using the TT gauge we find $du^{\alpha}/d{\tau}=0$ and the particle remains at rest in its reference frame.  So in fact there is no observable effect on a single particle.  This is due to our choice of gauge, so a different approach must be taken.  We must look at two or more particles to observe an effect. 

More specifically, we want to look at the variation of the geodesics of two particles as a gravitational wave passes.  For two geodesics, one at $x$ and the other at $x+X$ their deviations can be measured by
\begin{equation}
\frac{D^2 X^{\alpha}}{D\tau^2}=-R^{\alpha}_{\; \beta \gamma \delta} u^{\beta} X^{\gamma} u^{\delta},
\end{equation}
where $X^{\alpha}$ is the four vector separation between the geodesics, and we use the definition:
\begin{equation} 
DX^{\alpha}/D\tau \equiv X^{\alpha}_{\; ;\gamma}= \frac{dX^{\alpha}}{d\tau}+ \Gamma^{\alpha}_{\mu\nu}X^{\mu} \frac{dx^{\nu}}{d\tau}.
\end{equation}
Only the $R^j_{\; 0k0}$ terms will contibute due to our TT gauge choice.

In reality we must look not at the TT gauge, but at an actuall laboratory frame.  This is denoted the proper lab frame and to first order in $h$ we find that $\tau=t$, so we can write, 
\begin{equation}
\frac{d^2 X^j}{dt^2}=-c^2 R^j_{\;0k0} X^k,
\end{equation}
which results in the differential equation,
\begin{equation}
\frac{d^2 X^j}{dt^2}=\frac{1}{2}\frac{\partial^2 h^{TT}_{jk}}{\partial t^2} X^k.
\end{equation}
This can be solved to give us
\begin{equation}
   \label{eq:hstrain}
\Delta X^j=\frac{1}{2} h^{TT}_{jk}X^k .
\end{equation}
Here the $X^j$ are the three vector components of the coordinate separation vector.  We see that the oscillations of spacetime are orthogonal to the direction of propogration of the wave.

\begin{figure}
   
	 \includegraphics[width=.85\textwidth]{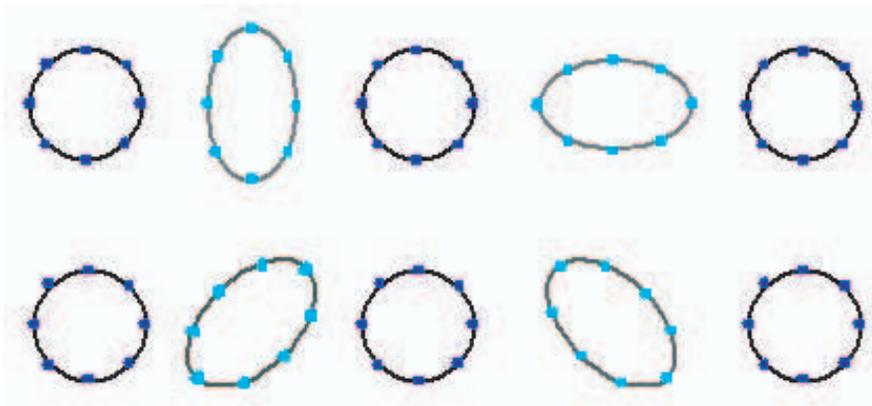}
  \caption{\label{fig33}Diagrams showing the motion of a ring of particles under the influence of a passing gravitational wave.  The top is plus polarization and the bottom is cross polarization.}
		
\end{figure}

In greater detail, the motion of the particles is determined by the polarization of the wave: ``cross'' and ``plus'' polarizations produce different relative motions.  Fig.~\ref{fig33} shows the motion of a group of test particles when a gravitational waves is incident upon them.  

        \section{Characteristic Strain}

In the transverse traceless gauge we can define the wave in terms of its Fourier components by,
\begin{eqnarray}
\label{hij}
h_{ij}(t,\textbf{x})&=&\int_{-\infty}^{\infty} df \int_{\Omega} d\hat{k} \int_0^{2\pi} d\psi \; \big( \tilde{h}_+(f,\hat{k},\psi) e_{ij}^+(\hat{k},\psi)+{}\nonumber \\
    &&{}+ \tilde{h}_{\times}(f,\hat{k},\psi) e_{ij}^{\times}(\hat{k},\psi) \big) e^{-i2\pi f(t-\hat{k}\cdot \textbf{x})},
\end{eqnarray}
where, $\tilde{h}_A(-f,\hat{k},\psi) = \tilde{h}^{*}_A(f,\hat{k},\psi)$.  Also $\hat{k}$ is a unit vector pointed in the direction of propagation and $\psi$ is the angle of polarization of the wave.  The two polarization tensors have the orthogonality relation $e_{ij}^A(\hat{k})e^{ij}_{A'}(\hat{k})=2\delta^{AA'}$, no sum over $A$.  Now we define the \textit{spectral density} $S_h(f)$ from the ensemble average of the fourier amplitudes,
\begin{eqnarray}
\left\langle \tilde{h}^*_A(f,\hat{k},\psi) \tilde{h}_{A'}(f',\hat{k}',\psi') \right\rangle=\delta_{AA'} \delta(f-f')\nonumber\\ 
\times \frac{\delta^2(\hat{k},\hat{k}')}{4\pi} \frac{\delta(\psi-\psi')}{2\pi} \frac{1}{2} S_h(f),
\end{eqnarray}
where $\delta^2(\hat{k},\hat{k}')=\delta(\phi-\phi') \delta(\cos (\theta)-\cos(\theta'))$. 

Now we write down the average of the real waves and the result is,
\begin{eqnarray}
\left\langle   h_{ij}(t,\textbf{x})h^{ij}(t,\textbf{x})\right\rangle &=& 2\int_{-\infty}^{\infty} df S_h(f),\nonumber\\
     &=& 4 \int_{f=0}^{f=\infty} d(\ln f)\:f \:S_h(f),
\end{eqnarray}
where we have used a one-sided function for $S_h$.

The characteristic strain is defined by,
\begin{equation}
 \label{eq:strain}
 \left\langle   h_{ij}(t,\textbf{x})h^{ij}(t,\textbf{x}) \right\rangle=2 \int_{f=0}^{f=\infty} d(\ln f)\:h_c^2(f).
\end{equation}
Therefore, 
\begin{eqnarray}
h_c^2(f)=2 f \:S_h(f),\\
\end{eqnarray}
which is an important result used in the analysis of spectra from a gravitational wave source.

In the normalization used in the LISA sensitivity plots~\cite{shane1} we get the relation,
\begin{equation}
S_h(f)=\left| \tilde{h}(f) \right|^2.
\end{equation}
      
There are numerous definitions of strain in the multitude of publications on the subject of gravitational waves.  The author wishes to maintain the notation as close as possible to the literature, and to be consistent throughout this work.  To that end, the above definitions are adhered to in all that follows.
    
We have derived the basic formulas for gravitational waves from linearized general relativity.  Of note is the use of the transervse, traceless (TT) gauge, in which the waves are transerve and are superpositions of two polarizations, $+$ and $\times$.  A single particle feels no acceleration from the passing wave as its coordinate system moves with the waves, but several particles will observe relative accelerations between them.


  \chapter{Power Radiated in Gravitational Waves}
  
  The universe is rich in sources of gravitational waves; from black hole binaries to cosmological phase transitions, the spectrum should be quite full.  The keys to detection are the power radiated by the source, its distance from the observer, and the nature of the signal.  In this chapter we overview the primary equations of the power radiated by a source, and discuss a few of the sources.  For a more detailed survey readers should consult~\cite{ci,magg,strau}.  
  
  The linearized solutions to Einstein's equations are expanded in the retarded time for a low velocity source, and terms are collected into multipole moments.  These moments are analyzed for their far field contributions, which are then time averaged to give the power in gravitational waves radiated by the source.    

     \section{Retarded Solution to Field Equations}  
  
  Let us first start with the equation for the retarded solution to Einstein's linearized field equations,     
\begin{equation}
h_{\mu\nu}(\textbf{x},t)=4 G \int{d^3 x' \frac {\bar{T}_{\mu\nu}(\textbf{x}',t-\left|\textbf{x}-\textbf{x}' \right|)} {\left|\textbf{x}-\textbf{x}' \right|}}.
\end{equation} 

If we make the approximation for $x>>x'$ then we get
\begin{equation}
h_{\mu\nu}(\textbf{x},t)=\frac{4 G}{r} \int d^3 x' \bar{T}_{\mu\nu}(\textbf{x}',t-r+\frac{\textbf{x} \cdot \textbf{x}'}{r}). 
\end{equation} 
This approximation is valid for a source with any velocity.  

To facilitate the expansion we switch to the TT gauge and find the Fourier components of $h$:
\begin{equation}
h^{TT}_{jk}(\textbf{x},t)=\frac{4 G}{r} \int d^3 x' \int \frac{d^4 k}{(2\pi)^4} \tilde{T}_{jk}(\omega,\textbf{k})e^{-i\omega(t-r+\frac{\textbf{x} \cdot \textbf{x}'}{r})} e^{i\textbf{k} \cdot \textbf{x}'}.
\end{equation}
Next we expand the exponential.  

One can equivalently expand the stress energy tensor around the retarded time $t_r=t-r$,
\begin{eqnarray}
{T}_{kl}(\textbf{x}',t-r+\frac{\textbf{x} \cdot \textbf{x}'}{r}) & \approx & {T}_{kl}(t-r,\textbf{x'})\\ \nonumber
                                                                   & &      + x'^i n^i \partial_{t_r} T_{kl} + \frac{x'^i n^i x'^j n^j}{2} \partial^2_{t_r} T_{kl}+\ldots
\end{eqnarray}
We then collect the terms around the powers of the coordinates into the multiple moments.

   \section{Mass Multipole Moments}
   
  Gravitational radiation in General Relativity is only quadrupolar and higher, so dipole or monopoles terms do not create gravitational waves.  This is due to the photon being spin-2 and massless, thus the only spin states of $j=2$ are allowed.  This also reveals itself in the first order solutions of the field equations, although it is more general than that.  In this first section we write down some of the moments from the gravitational field, given by:
\begin{eqnarray}
M&=&\int d^3 x T^{00} \;\;\text{monopole},\\
M^j &=& \int d^3 x T^{00} x^j \;\; \text{dipole},\\
M^{jk} &=& \int d^3 x T^{00} x^j x^k \;\; \text{quadrupole}.
\end{eqnarray}

From the equation of conservation of mass/energy in Minkowski space for linearized gravity we get,
\begin{equation}
\partial_{\beta}T^{\alpha\beta}=0,
\end{equation}
leads us to the additional restrictions:
\begin{eqnarray}
\dot{M}&=&0 \\
\dot{M}^{j}&=&P^j=\int d^3 x T^{0j}\;\;\text{momentum} \\
\ddot{M}^j &=& \dot{P}^j=0 \\
\dot{M}^{jk}&=&2\int d^3 x T^{0j}x^k \\
\ddot{M}^{jk}&=&2\int d^3 x T^{jk}.
\end{eqnarray}
 Of importance are the higher order derivates of the quadrupolar terms, as they lead to gravitational waves.

  \section{Solutions for $h_{\mu\nu}$}

If we expand the solution of $h_{\mu\nu}$ in terms of the retarded time we find that 
\begin{eqnarray}
h_{00}&=& \frac{2G}{r}\left( M-n_j P^j \right), \\
h_{0j}&=& \frac{4G}{r} P_j, \\
h_{jk}&=& \frac{2G}{r}\left( -\eta_{jk}( M-n_j P^j) + f_{jklm}\ddot{Q}^{lm} \right).
\end{eqnarray}
where the quadrupole tensor $Q_{jk}$ is given by,
\begin{eqnarray}
Q_{jk}&=& M_{jk}-\frac{1}{3} \eta_{jk}M^n_n,   \\
Q_{jk}    &=& \int d^3x T_{00} \left( x_j x_k -\frac{1}{3} \eta_{jk} r^2 \right),
\end{eqnarray}
and the projection tensor is given by,
\begin{eqnarray}
f_{jklm}&=&\Pi_{jl} \Pi_{km}-\frac{1}{2}\Pi_{jk}\Pi_{lm},\\
\Pi_{jk}&=&\eta_{jk}-n_j n_k,\\
n_j&=&\frac{x_j}{r}.
\end{eqnarray}

For our transverse-traceless (TT) gauge we take $\textbf{n}=(0,0,1)$, and our equations simplify to,
\begin{equation}
h_{jk}^{TT}=\frac{2G}{r}\: \ddot{Q}_{jk} .
\end{equation}
Nice.  Here we have ignored the monopole and dipole terms, as they don't contribute to the gravitational radiation.

     \section{Power Radiated}
     
Now for the power radiated by a source: we must deal with both the energy from the source and energy in the waves.  First we note that the $\textit{effective}$ stress energy tensor is a sum of both the gravitational wave contribution and the source,
\begin{eqnarray}
 \tau_{\mu\nu} &=& ( T_{\mu\nu}+t^{LL}_{\mu\nu}), \\
               &=& \frac{c^4}{8 \pi G}\left( R_{\mu\nu}-\frac{1}{2}g_{\mu\nu}R \right),
  \end{eqnarray} 
 where $g$ is the determinant of the spacetime metric.  If the second equation is solved using the Christoffel symbols and assuming a flat background metric we find the gravitational wave portion, $t^{LL}_{\mu\nu}$, (known as the Landau-Lifshitz $\textit{pseudo}$ stress energy tensor) is given by,
\begin{equation}
t^{LL}_{\mu\nu}=\frac{c^4}{32\pi G}\:  \partial_{\mu} h_{\alpha\beta}\partial_{\nu}h^{\alpha\beta},
\end{equation}   
Averaging this gives the effective tensor, also known as the $\textit{Isaacson}$ tensor: 
\begin{eqnarray}
T^{gw}_{\mu\nu}&=&\left< t^{LL}_{\mu\nu} \right> ,\\
 								&=&\frac{c^4}{32\pi G}\:  \left< \partial_{\mu} h_{\alpha\beta}\partial_{\nu}h^{\alpha\beta} \right>.
\end{eqnarray}

Conservation of the stress-energy tensor is given by the covariant derivative,
\begin{eqnarray}
D_{\mu} T^{\mu\nu}_{GW}=0,
\end{eqnarray}
which in a flat background is
\begin{equation}
\partial_0 T_{gw}^{00} + \partial_j T_{gw}^{j0} = 0
\end{equation}
when integrated over a volume gives,
\begin{equation}
\int d^3 x \left( \partial_0 T_{gw}^{00} + \partial_j T_{gw}^{j0} \right) = 0.
\end{equation}
The 00 component is the energy density, so we find,
\begin{equation}
L=\int d^3 x \partial_j T_{gw}^{j0}
\end{equation}
when evaluated on the surface gives,
\begin{equation}
L=\int d^2 x T_{gw}^{0k}n_k.
\end{equation}
This is a statement of Green's theorem.  

From this we find the flux is given by,
\begin{equation}
F_{gw}=c T_{gw}^{00},
\end{equation}
where the speed of the wave is $c$.

The average gravitational wave power radiated from a source is given by,
\begin{equation}
L_{GW}=\int d^2 x \ T^{gw}_{0k}n^k.
\end{equation}
The final result is given by,
\begin{equation}
L_{gw} = \frac{G}{5c^5} \left< \dddot{Q}_{jk} \dddot{Q}^{jk} \right>,
\end{equation}
which was first derived by Einstein a long time ago.  In this case $T_{00}$ has units of mass per unit volume, rather than energy per unit volume.   Note c has been reinserted.  

     \subsection{Plane Waves}
     
Assuming the source very far from the detector, we can make a plane wave approximation.  For plane waves in the TT gauge the energy flux can be calculated to be,
\begin{eqnarray}
F_{gw} &=& \frac{\pi c^3}{4G}f^2 h_{rms}^2,\\
 h_{rms} &=& \sqrt{\frac{1}{2}\left\langle h_+^2 + h_{\times}^2 \right\rangle}.
\end{eqnarray}
From this $h$, which is $h_{rms}$, can be determined if the luminosity and the distance to the source are known,
\begin{equation}
	h = \frac{\sqrt{L_{gw} G}}{\pi r f}.
\end{equation}
Again, this is valid if the source of the waves is distant enough to approximate a plane wave solution.

		\section{Astrophysical Sources and the Stochastic Background}

					\subsection{Binary Systems}

Let us consider a binary system of equal mass $m$ in circular orbit of radius $R$, with the masses separated by a distance of 2R.  The moments of inertia can be easily calculated and the time derivatives found using the equation above, which leaves us with:
\begin{eqnarray}
L_{GW}&=&\frac{128 G}{5 c^5} m^2 R^4 \omega^6,\\
   &=&\frac{32 G^{7/3}}{4^{1/3} 5 c^5}\ (m \omega)^{\frac{10}{3}},
\end{eqnarray}
for the last equality we have used $\omega^2 R=G m/(2R)^2$. If we seperate our factors of $G$ and $c$ we get,
\begin{eqnarray}
L_{GW}&=&\frac{32 c^5}{4^{1/3}5 G}\ \left( \frac{Gm \omega}{c^3} \right)^{10/3},\\
      &=& 3.021\times10^{26} W \left( \frac{m}{M_{\odot}} \frac{1 hr}{T} \right)^{10/3},
\end{eqnarray}
where $T$ is the period in hours.

			\subsection{Stochastic Background}
			
There are a number of sources of a stationary, isotropic gravitational wave background, to include:
\begin{itemize}
	\item Unresolved whited dwarf and neutron star binaries.
	\item Supernovae.
	\item Quantum fluctuations from post-inflation reheating.
	\item First order cosmological phase transitions and subsequent turbulence from bubble interactions.
	\item Topological defects: cosmic strings.
\end{itemize}

The universe is expected to be rich in a number of these sources, with the possibility of cosmic strings as well.  Cosmic string loops are a very discernible source due the very flat spectrum in frequency space.  The one-scale model ensures that the radiation from past epochs is added in such a way that the spectrum is very flat.  Current loops have the effect of creating a peak at approximately the inverse Hubble time of decay. 

     \subsubsection{\textbf{$\Omega_{gw}$}}

 The stochastic gravitational wave background can be characterized by the dimensionless quantity,
\begin{equation}
\Omega_{gw}(f)=\frac{1}{\rho_c}\frac{d\rho_{gw}}{d\ln f},
\end{equation}
where $\rho_c=3H_o^2/(8\pi G)$.
      
Now we relate $h_c$ to $\Omega_{gw}$.  The equation for the gravitational wave energy density from the waves is,
\begin{equation}
\rho_{gw}=\frac{1}{32\pi G}\left\langle \dot{h}_{ij}\dot{h}^{ij} \right\rangle.
\end{equation}
From eqn.~\ref{eq:strain} we find,
\begin{equation}
\rho_{gw}=\frac{1}{16\pi G} \int_{f=0}^{f=\infty} d(\ln f)\:(2\pi f)^2\:h_c^2(f).
\end{equation}
If we take the derivative with respect to $\ln f$,
\begin{equation}
\frac{d\rho_{gw}}{d\ln f}=\frac{1}{16\pi G}\:(2\pi f)^2\:h_c^2(f).
\end{equation}
Finally we divide by the critical density,
\begin{eqnarray}
\frac{1}{\rho_c} \frac{d\rho_{gw}(f)}{d\ln f}&=&\frac{1}{16\pi G \rho_c}\:(2\pi f)^2\:h_c^2(f),\\
\Omega_{gw}(f)  &=&\frac{2 \pi}{3 H_o^2}\:f^2\:h_c^2(f),\\
								&=&\frac{4 \pi}{3 H_o^2}\:f^3\:S_h(f).
\end{eqnarray}

       \section{Millisecond Pulsars}
       
An ingenious method for detecting a stochstic background of gravitational waves is the use of millisecond pulsars~\cite{jenet}.  The frequency of millisecond pulsars is measured for extended periods of time, and all known effects on the frequency are removed.  The resulting residual frequencies are analyzed for evidence of a stochastic background.  

The low frequencies of this method, on the order of inverse 20 years, are the only method of detecting the stochastic background predicted in many models.  This range is below that of LISA, and puts tangible limits on the background of cosmic strings.

		\section{Alternate Theories of Gravity}
		
All viable metric theories of gravity have gravitational waves~\cite{will}.  For GR the photon is a massless spin-2 particle and thus there are only the plus and cross polarizations.  Also, the lowest order contribution is from the quadrupole moment while other theories of gravity, to include the scalar-tensor theory, include a dipole moment~\cite{will}.  Thus LISA could potentially be a test of GR for a binary mass in-spiralling system.  

Other effects include a difference between the speeds of light and gravitational waves, and up to six possible polarizations aside from the aforementioned two.   

What follows is a brief summary of dipole radiation taken from Will,1993~\cite{will}.    

If we define the dipole moment of the self-gravitational binding energy of two bodies as,
\begin{equation}
\textbf{D}=\sum_{a} \Omega_a \textbf{x}_a,
\end{equation}
where $\Omega_a$ is the self gravitational binding energy of object a given by,
\begin{equation}
\Omega_a=\int_{a} \frac{\rho_a(\textbf{x}) \rho_a(\textbf{x}')}{\left|\textbf{x}-\textbf{x}'\right|} d^3x\: d^3x'.
\end{equation}
So at least representatively one can write,
\begin{equation}
L_{GW}^D=\frac{1}{3}\kappa_D \left\langle \ddot{\textbf{D}} \cdot \ddot{\textbf{D}} \right\rangle,
\end{equation}
where $\kappa_D$ is a constant that depends upon the theory.

\emph{Scalar-tensor theory}:  As an example we look at a binary system with Brans-Dicke (BD) scalar-tensor theory.  For the luminosity we have the dipole contribution,
\begin{equation}
L_{gw}^{D}=\frac{1}{3} \kappa_D \Omega^2,
\end{equation}
where $\Omega=\Omega_1/m_1-\Omega_2/m_2$ is the difference in the  self-gravitational binding energy of two bodies.  For BD theory $\kappa_D=2/(2+\omega)$, which is very small given the current solar system bound with $\omega>500$.

Even though this value is quite small given the current limits, it is still discernible as an additional source of gravitational wave energy.


 \chapter{Gravitational Wave Detectors}

        \section{Gravitational Waves on the Detector} 
        
A point which the reader should take note of is the extremely small amplitude of the signal produced by gravitational waves.  One way to understand this is via the Einstein equation $G_{\mu\nu}=16\pi G T_{\mu\nu}$ which leads to Eqs.~\ref{eq:hstrain} and~\ref{eq:wav1}.  Eq.~\ref{eq:hstrain} indicates $h \approx \Delta x/x$ is a strain, and is often referred to as the ``dimensionless strain of space'' or the ``time-integrated shear of space''~\cite{hawking}.  Eq.~\ref{eq:wav1} shows that $T^{\mu\nu}$ is comparable to the stress.  If we use the analog of the stress-strain relationship of a solid we find the Young's modulus is given by $c^4/(16 \pi G)$; this is a huge number!  In other words, spacetime is very ``stiff'' and strongly resists the stress produced by a passing gravitational wave.

Thus the signal on any detector is going to be minuscule, even for a very luninous source.  In order to extract this signal the sensitivity of the device must, apparently, be very high.  This sensitivity then leads to problems with background noise, especially on the earth.  A typical strain is $h \approx 10^{-21}$, which for a detector with arms of length 10 m is a sensitivity on the order of a nucleus.  For LISA with arms of length $10^8$ m the length of the detectable oscillations is the size of an atom.  Getting a signal from the noise at these small values is indeed no small feat. 

In general there are two different types of detectors: resonant-mass detectors and interferometers.  Resonant-mass detectors were first pioneered by Joseph Weber in the 1960's.  Today the detectors are four orders of magnitutde more sensitive than Weber's original bars, but even this sensitivity only allows the detection of very powerful emitters within the galaxy and local galactic neighborhood.  Given the relative rarity of such events, it is unlikely they are to be detected.  On the positive side, resonant-mass detectors can be fabricated on small scales when compared to the interferometers.

Interferometers were first theorized for use in 1962~\cite{magg} but their complexity and cost prevented implementation for the next thirty years.  The current detectors (LIGO, VIRGO, GEO600, and TAMA) are all taking data now, but have had no detections.  These are all ground based detectors and must deal with noise from the earth, especially at frequencies from 1-10 Hz.  This means that ground based detectors are limited to frequencies above 10 Hz.  The first space based detector, LISA, is planned to launch early in the next decade.  Being free from the earth's seismic noise, this detector is planned to have a detection band in the millisecond range.

For details on the calculations of strain from single and multiple detectors see Appendix C.
 

    \section{Ground Based Detectors}
  
  Ground based detectors typically come in two varieties: resonant bars and laser interferometers.  Their detection bands are ultimately limited by seismic noise from the earth, so they typically cannot go below 10 Hz.
  
 First we discuss the resonant-mass detectors.  These are of less interest in our study as their frequency range is of the order 700-900 Hz~\cite{ci}, which corresponds to very small loops.  This corresponds to loop sizes for which our simplified model is not necessarily accurate.
 
 The laser interfermometers have a much larger sensitivity range, but are still limited by seismic events.  This puts their range fairly high.

    \section{Laser Interferometer Space Antenna}  
    
 The \textit{Laser Interferometer Space Antenna} LISA, is in the final stages of pre-flight and is scheduled to launch in the near future.  It is a technological marvel, using three satellites in an equilateral triangle with sides of $5 \times 10^6$ km .  It is to be in orbit around the sun at the earth's radius $20^o$ behind the earth, or a distnce of $5 \times 10^7$ km .  The triangle of satellites are dipped an angle of $60^o$ from the ecliptic~\cite{ci,shane1,shane2,lisa,tinto}.
 
 The length of the arms correspond to a peak sensitivity of about 10 mHz, see for example ~\ref{fig51}.  The frequencies below about 0.1 mHz are determined by the drift of the spacecraft, which every attempt is made to minimize.  Our plots show theoretical values of the sensitivity for those low frequencies, although it is by no means certain they will be attained.
 
 The range of frequences in LISA's detection band is quite rich.  A number of astrophysical sources to include emission by white dwarf binaries and the inspiral of black holes and stars with supermassive black holes can be found here.  Also included are the most luminous events in the universe: the inspiral of two supermassive black holes.
 
Some of the signals are so rich that they form a signal that washes out all other signals.  This ``confusion noise'' is expected in the higher frequency range from white dwarf binaries in the galaxy.  Several studies have been done on this~\cite{nel,farm} and their results are indicated on our plots of sensitivity.


\chapter{Gravitational Radiation from Individual Cosmic Strings}

Perhaps the most exciting aspect of gravitational waves from cosmic strings is that the radiation from the loops is a series of harmonic modes.  This allows even small signals to be extracted from the background.  Also, for the light strings, the quality factor is very large which ensures the frequencies of the harmonic modes stay constant during detection.  Given that the maximum time frame for detection is on the order of three years, we can safely assume the signal is stationary in time.

There has been much investigation into the radiation from cusps, kinks, and at reconnection points~\cite{jack}.  These sources are typically directed  bursts of gravitational waves and are not harmonic in nature.  They also tend to be at much higher frequencies than the fundamental modes of the larger string loops.  For these reasons they are not included in this study.

In this chapter we first re-analyze the equations of motion for a cosmic string loop and determine that the motion is indeed a series of harmonic modes.  Subsequently the distribution of the loops is estimated using the distribution of dark matter as a model.  It is assumed that the cosmic string loops follow the dark matter as they are bound gravitationally to the galaxy.  Finally the power and spectra of an individual loop are estimated using numerical calculations.

    \section{Linearized Gravity}
 Let us first start with a review of the equation for the retarded solution to Einstein's linearized field equations,     
\begin{equation}
h_{\mu\nu}(\textbf{x},t)=4 G \int{d^3 x' \frac {\bar{T}_{\mu\nu}(\textbf{x}',t-\left|\textbf{x}-\textbf{x}' \right|)} {\left|\textbf{x}-\textbf{x}' \right|}}.
\end{equation} 
For a cosmic string we get,
\begin{equation}
T^{\mu\nu}= \frac{\mu}{ \sqrt{-g}} \int d\sigma d\tau \sqrt{-\gamma} \gamma^{AB} \partial_{A} X^{\mu} \partial_{B} X^{\nu} \delta^{(4)}(x-X)),
\end{equation}
where $X^{\mu}(\sigma,\tau)=x^{\mu}=(\tau,\textbf{r}(\tau,\sigma))$, and $\gamma_{AB}=g_{\mu\nu} \partial_{A} X^{\mu} \partial_{B} X^{\nu}$.  Substituting values and assuming a locally Minkowski spaceime we get:
\begin{equation}
\label{eq:Tstring}
T_{\mu\nu}= \mu \int d\sigma ( N_{\mu}V_{\nu}-N_{\mu}N_{\nu}) \delta^{(3)}(\textbf{x}'-\textbf{r}(t',\sigma ')),
\end{equation}
where $V^{\mu}=(1,\textbf{r}_{\tau})$, $N^{\mu}=(0,\textbf{r}_{\sigma})$, and $t'=\tau$.

If we make the approximation for $x>>x'$ then we get
\begin{equation}
h_{\mu\nu}(\textbf{x},t)=\frac{4 G}{r} \int d^3 x' \bar{T}_{\mu\nu}(\textbf{x}',t-r+\frac{\textbf{x} \cdot \textbf{x}'}{r}). 
\end{equation} 
We then use Eq.~\ref{eq:Tstring} to find our $h_{\mu\nu}$ at a given direction and distance from the source. 

In principle this equation can be used to find the field far from a cosmic string loop.  In practice, the waves from a cosmic string loop are harmonic so we can simply use plane wave solutions and sum over several wavelengths to find the luminosity.  This is detailed in later sections of this chapter.

 \section{Periodicity of Cosmic String Loop Motion}
 
 We reiterate previous results from Chapter 1 and then add a final calculation which involves cosmic string loops.  This last part is important as it characterizes the loop motion as a series of harmonic modes.
    
Cosmic string motion can be characterized by paramaterizing the motion on a two-dimensional surface called a worldsheet.  With the coordinate parameterization given by:
\begin{equation}
x^{\mu}=x^{\mu}(\sigma^a),\;\;a=0,1,
\end{equation}
where the $x^{\mu}$ are the four spacetime coordinates and the $\sigma^a$ are the coordinates on the worldsheet of the string.  In general $\sigma^0$ is considered timelike and $\sigma^1$ is considered spacelike.  On the worldsheet the spacetime distance is given by,
\begin{eqnarray}
ds^2 &=& g_{\mu\nu}dx^{\mu}dx^{\nu},\\
     &=& g_{\mu\nu}\frac{dx^{\mu}}{d\sigma^a} \frac{dx^{\nu}}{d\sigma^b} d\sigma^a d\sigma^b,\\
     &=& \gamma_{ab}d\sigma^a d\sigma^b,
\end{eqnarray}
where $\gamma_{ab}$ is the induced metric on the worldsheet.

The equations of motion for the string loops can be found via the Nambu-Goto action:
\begin{equation}
S=-\mu \int \sqrt{-\gamma} d^2 \sigma,
\end{equation}
where $\gamma$ is the determinant of $\gamma_{ab}$.  Minimizing the action leads to the following equations,
\begin{eqnarray}
\frac{\partial x^{\mu}}{\partial \sigma^a} ^{;a} + \Gamma^{\mu}_{\alpha \beta} \frac{\partial x^{\alpha}}{\partial \sigma^a} \frac{\partial x^{\beta}}{\partial \sigma^b}=0
\end{eqnarray}
where $\Gamma$ are the Christoffel symbols.

In flat spacetime with $g_{\mu\nu}=\eta_{\mu\nu}$ and $\Gamma^{\mu}_{\alpha \beta}=0$ the string equations become,
\begin{equation}
   \frac{\partial    }{\partial \sigma^a} \left( \sqrt{-\gamma} \gamma^{ab} \frac{\partial x^{\mu}}{\partial \sigma^b} \right)=0.
 \end{equation}

Now we use the fact that the Nambu action is invariant under general coordinate reparameterizations, and we choose a specific gauge.  This does not disturb the generality of the solutions we find, since the action is invariant under these gauge transformations.  Here we choose:
\begin{eqnarray}
\gamma_{01} &=& 0,\\
\gamma_{00}+\gamma_{11} &=& 0,
\end{eqnarray}
this called the ``conformal'' gauge, because the worldsheet metric becomes conformally flat.

Our final choice is to set the coordinates by
\begin{eqnarray}
t=x^0=\sigma^0\\
\sigma \equiv \sigma^1,
\end{eqnarray}
so that we have fixed the time direction to match both our four dimensional spacetime and the worldsheet.  This is sometimes called the ``temporal'' gauge.  This leads to the equations of motion,
\begin{eqnarray}
\dot{\textbf{x}} \cdot \textbf{x}' &=& 0,\\
\dot{\textbf{x}}^2 + \textbf{x}'^2 &=& 1,\\
\ddot{\textbf{x}}-\textbf{x}''     &=& 0.
\end{eqnarray}
The first equation states the orthogonality of the motion in these coordinates, which leads to $\dot{\textbf{x}}$ being the physical velocity of the string.  The second equation is statment of causality as the total speed of the motion is c, which can also be interpreted as,
\begin{eqnarray}
   d\sigma &=& \frac{|dx|}{\sqrt{1-\dot{x}^2}} = dE/\mu, \\
   E       &=& \mu \int \frac{|dx|}{\sqrt{1-\dot{x}^2}} = \mu \int d\sigma.
\end{eqnarray} 
So the coordinates $\sigma$ measure the energy of the string.

  The final equation is a three dimensional wave equation.  This equation determines the behavior of the strings and has a general solution of the form:
 \begin{equation}
   \textbf{x}(\sigma,t)=\frac{1}{2}[\textbf{a}(\sigma-t)+\textbf{b}(\sigma-t)],
 \end{equation}
 where constraints from the equations of motion give,
 \begin{equation}
 \textbf{a}'=\textbf{b}'=1.
 \end{equation}

We can also write down the stress-energy tensor, momentum, and angular momentum in this choice of gauge:
\begin{eqnarray}
T^{\mu \nu} = \mu \int{d\sigma} [\dot{x}^{\mu} \dot{x}^{\nu}-x'^{\mu} x'^{\nu}] \delta^{(3)}(\textbf{x}-\textbf{x}(\sigma,t)),
\end{eqnarray}
which leads tot he energy given above, as well as,
\begin{eqnarray}
\textbf{p} &=& \mu \int d\sigma \; \dot{\textbf{x}}(\sigma,t),\\
\textbf{J} &=& \mu \int d\sigma \;  \textbf{x}(\sigma,t) \times \dot{\textbf{x}}(\sigma,t).
\end{eqnarray}
It is found for loops that the large angular momentum tends to reduce rotation rates dramatically.

      \subsection{Loop Motion}

For a loop of length $L$ we have the identity, $\textbf{x}(\sigma+L,t)=\textbf{x}(\sigma,t)$.  Thus the functions from our general solution also have the same periodicity:
\begin{equation}
\textbf{a}(\sigma \pm t +L)=\textbf{a}(\sigma \pm t),
\end{equation}
and similarly for $\textbf{b}(\sigma,t)$.  From above, the smallest value of $t$ that is periodic is $t=L/2$ since,
\begin{equation}
\textbf{a}([\sigma+L/2] \pm [t+L/2])=\textbf{a}(\sigma \pm t).
\end{equation}
Hence the fundamental frequency is given by
\begin{equation}
f=\frac{1}{T}=\frac{2c}{L},
\end{equation}
where $c$ has been reinserted for clarity.

The general solution may be represented in terms of its Fourier components as,
\begin{eqnarray}
\textbf{x}(\sigma,t) &=& \textbf{x}_0 +\frac{\textbf{p}}{\mu} t \nonumber\\
  &+& \sum_{n=1}^{\infty} \left(\textbf{A}_n^{\pm} e^{i2 \pi n [\sigma_{\pm}]/L}+\textbf{B}_n^{\pm} e^{-i2 \pi n [\sigma_{\pm}]/L} \right),
\end{eqnarray}
where $\sigma_{\pm}=\sigma \pm t$.  So in general, one can write the string motion in terms of notmal modes which are harmonic in time.  This fact is important and dramatically increases the detectability of the normal mode oscillations of the string loops.


  \section{Distrubtion of Loops}
     
The average number density of loops of size $L$ is $n(L)$, computed numerically.  Thus the total number density of loops is given by,
\begin{equation}
\frac{N(L)}{V}=\int n(L) dL
\end{equation}
and the average distance between loops is $(N/V)^{-1/3}$.

The loops are assumed to be clustered around galaxies within the dark matter halo.  From the average distance between galaxies the total number density, the number of loops within the galactic halo is estimated.   

The distribution detected from the Earth is a function of both the length of loops and their distance from the earth.  The number density in the galaxy will be matched to the dark matter halo, given by the NFW density distribution $\rho_{NFW}(r)$~\cite{NFW95}.  The number density is then a function of the length of loops and distance from the center of the galaxy $n(L,r)$.

      \subsection{Calculations of Loop Density}

We start by figuring out the size of the loops at the fundamental frequency:
\begin{equation}
L=2c/f_1
\end{equation}
We then find the time in the past when loops formed and have decayed via graviatational waves to the size now given by $2c/f_1$.  Using the one-scale model loops form at a given time with size $L \approx \alpha c H^{-1}(t)$ and start decaying at a rate $\dot{L}=-\gamma G \mu$.  So loops created at a time $t_c$ will be of the size $L(t_c,t)$ at the time $t$ given by the equation,
\begin{equation}
L(t_c,t)=\alpha c H^{-1}(t_c)- \gamma G \mu (t-t_c)/c,
\end{equation}
From the present time $t_o$ we insert the value of $L$ to get:
\begin{eqnarray}
2c/f_1 &=& \alpha c H^{-1}(t_c)- \gamma G \mu (t-t_c)/c.
\end{eqnarray}

The number density as a function of the time of observation, $t_o$, and time of creation, $t_c$ within a $\Delta t$ in log scale of 0.1, is given by:
\begin{equation}
n(t_o, t_c)=\frac{ N_t }{\alpha} \left( \frac{H(t_c)}{c}\right)^3 \left( \frac{a(t_c)}{a(t_o)}\right)^3,
\end{equation}
Using the time $t_c$ we solve numerically to find the number density and taking $n^{-1/3}$ we find the number within the galaxy.

If we assume the average distance between galaxies is 5 Mpc, the number of loops within the galaxy for $G\mu=10^{-12}$ and a fundamental frequency of 1 mHz is approximately $10^4$.  This number is distributed within the halo of each galaxy, which leads to an overall increase within the halo and a large increase in the density near the center of a galaxy.

        \subsection{Number Density in the Milky Way}
        
The density of galactic string loops is matched to the dark matter distribution in the Milky Way using the NFW density $\rho_{NFW}(r)$:
\begin{eqnarray}
\rho_{NFW}(r) &=& \frac{\rho_s}{x(1+x)^2},\\
x &=& \frac{r}{r_s},
\end{eqnarray}
where $\rho_o$ and $r_s$ are determined by observation.  Representative values are given by~\cite{klypin02} using the favored model with $r_s$=21.5 kpc.    From above we see that for $r<<r_s$, $\rho \propto r^{-1}$ and for $r>>r_s$, $\rho \propto r^{-3}$.  

The NFW mass density can be matched to the number density of the same form:
\begin{equation}
n(r,L)=\frac{n_s(L)}{x(1+x)^2}
\end{equation}
where again $x=r/r_s$.  The number density constant $n_s$ is matched to this curve by integrating over the halo volume up to a truncation value of $r_t$ and normalizing to the total number of string loops $N$:
\begin{eqnarray}
N(L)   &=& \int_0^{r_t} n(r,L) d^3 x,\\
n_s(L) &=& \frac{N(L)\;(r_t+r_s)}{4 \pi r_s^3 \left[ (r_t+r_s) \text{ln}\left(\frac{r_t+r_s}{r_s}\right)-r_t \right]}.
\end{eqnarray}
In general $r_t$ is of the order 100 kpc.  The value of $n_s(L)$ must be determined for each length (or equivalently fundamental frequency) of loop, in log bins of 0.1.

             \subsubsection{Sample Calculation of Number Density Distribution}
      
A reprentative value from the previous example of strings of tension $G \mu=10^{-12}$, $N=10^4$, $r_t/r_s = 10$, and $r_s=21.5$ kpc, at the earth's distance form the galactic center, $r=8$ kpc, we find $n \approx 9.4 \times 10^{-2}$ kpc$^{-3}$.  Inverting and taking the cube root we find $n^{-1/3} \approx 2.2$ kpc.  Thus on average the distace to the nearest loop is appoximately 2.2 kpc.  

    \subsection{Distribution of Loops Near the Earth} 

The signal at the earth will be a function of both distance from the earth $r'$ and size $L$.  The number of loops $dN$ of length $L$ a distance $r'$ from the Earth and $R$ from the center of the milky way, within an interval $dr'$ and a solid angle $d\Omega'$ is given by:
\begin{equation}
dN(L)=n(L, \textbf{R}+\textbf{r$'$}) d^3r',
\end{equation}
where
\begin{eqnarray}
n(L,\textbf{R}+\textbf{r$'$}) = n(L, r', \theta', \phi') = \frac{n_s(L)}{x'(1+x')^2},
\end{eqnarray}
and
\begin{eqnarray}
   x' = \frac{\sqrt{r'^2+R^2-2R r' \sin\theta' \cos\phi'}}{r_s}.
\end{eqnarray}
$R$ is the distance of the Earth from the galactic center and $r_s$ is defined above.  Here $\theta'$ and $\phi'$ define the angle measured from the earth.  Here $\phi'=0$ is the direction of the galactic center and $\theta'=\pi/2$ defines the plane of the galactic disk.  The signal in the direction of the galactic center has an increased luminosity over other directions, especially directly away from the center. 

The number of particles per unit solid angle from the earth is defined by,
\begin{equation}
\frac{dN}{d\Omega'}(L, \theta', \phi')=\int_0^R dr' \frac{n_s r'^2}{x'(1+x')^2}.
\end{equation}
This is useful for determining the direction of highest gravitational wave luminosity.

    \subsection{Results for Number Density and Distributions}
    
 The number density of loops in the Milky Way is strongly influenced by the size and mass density of the cosmic string loops.  Tables~\ref{table51}--\ref{table53} give results for the number of loops in the Milky Way for various parameterizations of the string loops.  The results clearly show the large predominance of loops for very light and large strings.
    
    \begin{table}[H] 
 \caption{\label{table51}Number of Loops in the Milky Way for the given fundamental frequencies and $\alpha$=0.1}
 
 \begin{tabular}{|c|| c| c| c| c|}
  \hline
            
  $G\mu$   &  $f_1$(Hz)$10^{-4}  $          &       $ 10^{-3}  $     &    $ 10^{-2}$     &     $ 10^{-1}  $                    \\   
\hline
 $10^{-11}$  &     338            &         338          &      338        &       338                       \\

$10^{-12}$   & 1.05 $\times10^{4}$            &          1.05 $\times10^{4}$     &    1.05 $\times10^{4}$       &    1.05 $\times10^{4}$                \\
 
$10^{-13}$   & 3.29 $\times10^{5}$& 3.33 $\times10^{5}$  & 3.34 $\times10^{5}$   &3.34 $\times10^{5}$       \\
 
$10^{-14}$   & 9.22 $\times10^{6}$& 1.04 $\times10^{7}$  & 1.05 $\times10^{7}$   &1.05 $\times10^{7}$       \\

$10^{-15}$  & 1.25 $\times10^{8}$& 2.92 $\times10^{8}$  & 3.28 $\times10^{8}$   &3.32 $\times10^{8}$        \\

$10^{-16}$   & 3.21 $\times10^{8}$& 3.94 $\times10^{9}$  & 9.22 $\times10^{9}$   &1.04 $\times10^{10}$        \\
\hline
 \end{tabular}
 
 \end{table}

  \begin{table}[H] 
 \caption{\label{table52}Number of Loops in the Milky Way for the given fundamental frequencies, varying $\alpha$ and G$\mu=10^{-12}$}
 
 \begin{tabular}{|c||c| c| c| c|}
  
  \hline
  
  $\alpha$        & $f_1$(Hz)  $ 10^{-4}  $      &       $ 10^{-3}  $   &    $ 10^{-2}$       &     $ 10^{-1}  $               \\   
\hline
 $10^{-1}$  & 1.05 $\times10^{4}$             &    1.05 $\times10^{4}$       &     1.05 $\times10^{4}$   &    1.05 $\times10^{4}$            \\

$10^{-2}$   & 3400              &          3400      &     3400             &        3400               \\
 
$10^{-3}$   & 1100              &           1100       &      1100           &         1100                \\
 
$10^{-4}$   & 385               &          386        &     386               &       386               \\

$10^{-5}$  &  160               &          161       &     161                &      161                  \\
 
$10^{-6}$   &  93                &          93       &       93                &       93                 \\
  \hline
 \end{tabular}
 
 \end{table}

  \begin{table}
 \caption{\label{table53}Number of Loops in the Milky Way for the given fundamental frequencies, varying $\alpha$ and G$\mu=10^{-16}$}
 
\begin{tabular}{|c||c |c |c |c|}
  
  \hline
  $\alpha$  & $f_1$(Hz) $ 10^{-4}  $      &       $ 10^{-3}  $   &    $ 10^{-2}$       &     $ 10^{-1}  $  \\   
\hline
 $10^{-1}$  & 3.21 $\times10^{8}$& 3.94 $\times10^{9}$  & 9.22 $\times10^{9}$   &1.04 $\times10^{10}$      \\

$10^{-2}$   & 1.02 $\times10^{8}$& 1.33 $\times10^{9}$  & 3.30 $\times10^{9}$   &3.76 $\times10^{9}$       \\
 
$10^{-3}$   & 3.21 $\times10^{7}$& 3.97 $\times10^{8}$  & 9.32 $\times10^{8}$   &1.05 $\times10^{9}$      \\
 
$10^{-4}$   & 1.02 $\times10^{7}$& 1.25 $\times10^{8}$  & 2.92 $\times10^{8}$   &3.29 $\times10^{8}$     \\

$10^{-5}$   & 3.26 $\times10^{6}$& 3.97 $\times10^{7}$  & 9.27 $\times10^{7}$   &1.04 $\times10^{8}$      \\
 
$10^{-6}$   & 1.07 $\times10^{6}$& 1.27 $\times10^{7}$  & 2.96 $\times10^{7}$   &3.33 $\times10^{7}$      \\
  \hline
 \end{tabular}
 
 \end{table}

  \section{Power from each loop}

A loop will radiate power in each mode $n$ modeled by,
\begin{equation}
\dot{E}_n(r') \propto \frac{\dot{E}_n}{r'^{2}},
\end{equation}
where $\dot{E}_n$ is power radiated per mode from a loop.  The total power is 
\begin{equation}
\dot{E}=\gamma G \mu^2=\sum P_n G \mu^2,
\end{equation}
where $\gamma$ is found to be between 50 and 100, and the $P_n$ are power coefficents of the individual modes.  Note that the power radiated is independent of the loop size.  To first order we ignore directionality of the emitted gravitational radiation.

Each loop has a spectrum given by each of the power coefficients, $P_n$, where the sum $\sum{P_n}=\gamma$ is fixed by the fundamental properties of the strings.  This is an average over a wide array of loops: it is expected individual loops vary somewhat from this.  Regardless is expected that the most power is in the fundamental mode and the lowest modes, while the higher modes have significantly reduced power.

    \subsection{Quality Factor}
For a given mode the quality factor is given by,
\begin{eqnarray}
Q_n &=& \frac{2 \pi E}{\Delta E_n},\\
 &=& \frac{f_n}{\Delta f}.
\end{eqnarray}   
Upon inserting values for a given string loop, we find
\begin{eqnarray}
Q_n &=& \frac{2 \pi f_n E}{\dot{E}_n},\\
&=&2 \pi \frac{2 n}{L}    \frac{ \mu L}{P_n \gamma G \mu^2},
\end{eqnarray}
which simplifies to the result:
\begin{equation}
Q_n=\frac{4 \pi n}{P_n G \mu}.
\end{equation}
For light strings with $G\mu=10^{-12}$ we find,
\begin{equation}
Q_n \approx 10^{12} \frac{n}{P_n}.
\end{equation}
For the fundamental mode $P_n$ is of order unity at most, so $Q$ is large, and increases with higher $n$.  Thus the full width at half maximum should be small at each mode, and justifies the use of delta functions in other applications.

Given the extremely large value of $Q_n$ it can be assumed that the signal from each mode in the loop is monochromatic for the duration of observation with regard to decay, with is at most three years.  Potential frequency shifts due to gravitational lensing, gravitaional doppler effect, and relative motion have not been analyzed~\cite{dubath07}.


  \section{Strain Produced by Gravitational Radiation}

The relative weakness of gravitational waves from cosmic strings allows the use of linearized gravity.  The stress energy tensor can then be given by the Isaacson tensor:
\begin{equation}
T_{\alpha \beta}^{gw}=\frac{1}{32 \pi G} \left\langle h^{TT}_{ij,\alpha} h^{ij,TT}_{,\beta}\right\rangle,
\end{equation}
and the total luninosity in gravitational waves is given by
\begin{equation}
L_{gw}=\int T_{0k}^{gw} n^k d^2 x.
\end{equation}

Assuming plane wave solutions, we find for the energy flux:
\begin{equation}
F_{gw}=\frac{\pi c^3}{4 G} f_n^2 h_n^2.
\end{equation}

Inserting the radiated power $L_{gw}=\dot{E_n}$ into the equation above and solving for $h_n$ we find,
\begin{eqnarray}
h_n &=& \frac{\sqrt{\dot{E}_n G}}{\pi r f_n},\\
h_n &=& \frac{\sqrt{P_n} G \mu}{\pi r f_n}.
\end{eqnarray}

Inserting $c$ we find :
\begin{eqnarray}
h_n &=& \sqrt{P_n} \frac{c}{\pi} \frac{G \mu}{c^2} \frac{1}{r f_n},\\
h_n &=& 3.095 \times 10^{-12} \sqrt{P_n} \; [G \mu (c=1)] \times \nonumber \\
     & & \;    \left(\frac{1 \text{Hz}}{f_n}\right)  \left( \frac{1 \text{kpc}}{r} \right).
\end{eqnarray} 
For $G \mu=10^{-12}$, $f$=1 mHz, and $r$=1 kpc, we find $h_1 \approx 10^{-21}$, within the detection limits of LISA.

   
   \section{Spectrum of a Single Loop}
   
Using previous results for the strain produced by a gravitational wave, we can add the harmonic time dependence to find the strain at each mode as follows:
 \begin{eqnarray}
h_n(t)= \frac{2}{f_n}\sqrt{\frac{G F_n^{gw}}{\pi c^3}}e^{-i 2 \pi f_n t}
\end{eqnarray}  
where,
\begin{equation}
F_n^{gw}=\frac{\dot{E}_n}{4 \pi r^2}=\frac{P_n c^5}{4 \pi G r^2} \left(\frac{G \mu}{c^2}\right)^2.
\end{equation}
Thus we have,
\begin{equation}
h_n(t)=\frac{c\sqrt{P_n}}{\pi f_n} \frac{G \mu}{c^2} \frac{1}{r}\,e^{-i 2 \pi f_n t}
\end{equation}
Again, the $P_n$ are the ``power coefficients'' and has been estimated to go asymptotically as $n^{-4/3}$.  From previous we recall that the sum of our coefficients is given by $\sum P_n=\gamma$.  In practice most of the power is in the fundamental mode, so the the power from lower modes drops off, allowing truncation of the sum at a reasonable value of $n$.   As an example, we set
\begin{equation}
P_n= P_1 n^{-4/3},
\end{equation}
and $P_1=18$.  At $n=20$ we find $\sum P_n \approx 45$, giving us a reasonable estimate.

For a single loop the total strain will be a sum of all the modes,
\begin{equation}
h(t)=\sum_{n=1}^{\infty} h_n(t),
\end{equation}
This sum is typically taken to $n=10$ to facilitate computational efficiency.  It is assumed most of the power is in or near the fundamental mode, with power dropping off as $n^{-4/3}$ as mentioned previously.  We expect to lose little by ignoring the higher modes given their small contribution.  Other research has been done on bursts and cusps which are effectively part of this high frequency tail.

      \subsection{Estimate of Strain compared to Stochastic Background}

For comparison we investigate the strain produced by the stochastic background.  Of use is the value $\Omega_{gw}(f)$ which is given by:
\begin{eqnarray}
\Omega_{gw}(f)&=&  \frac{1}{\rho_c}\frac{d\rho_{gw}}{d \text{ln}(f)},\\
              &=&  \frac{2 \pi}{3 H_o^2} f^3 h_f^2,
  \end {eqnarray} 
where $\rho_c$ is the critical density of the universe, $H_o$ is the current Hubble constant, and $h_f$ is essentially the Fourier transform of the gravitational wave signal~\cite{shane1}.

From above we get,
\begin{eqnarray}
h_f &=& \sqrt{\frac{3 \Omega_{gw}}{2 \pi}} \frac{H_o}{f^{3/2}}\\
    &=& 2.239 \times 10^{-18} \sqrt{\Omega_{gw} h^2} \frac{1 \text{Hz}^{3/2}}{f^{3/2}},
\end{eqnarray}
where $h=H_o / 100 \approx 74$.

From previous numerical calculations we find $\Omega h^2 = 10^{-9.5}$ for $f$=1 mHz, $\alpha=0.1$, $G \mu = 10^{-12}$, and $\gamma=50$, gives $h_f \approx 10^{-19}$.  Thus the background appears to overwhelm the loops farther than 1 kpc from the earth.  For distances from the earth to a single loop less than about 100 pc, the gravitational wave signal from the individual loops is greater than the stochastic background.  The periodic signal from the individual loops makes it likely that even strains less than the background can be picked up by filtering out the periodic signal from the background.

   \subsection{Fourier Transform}
   
The frequency spectrum from a gravitational wave on a detector is found by taking the Fourier transform of the detected time signal.  In practice this is done over a finite time, thus introducing resolution issues even though, in principle, the spectrum of a harmonic signal should be delta functions.  For the $n^{\text{th}}$ component of the strain we find,
\begin{eqnarray}
\tilde{h}_n(f) &=&\frac{1}{\sqrt{T}} \int_{-T/2}^{T/2} dt e^{i 2 \pi f t} h_n(t),\\
               &=& \frac{c\sqrt{P_n}}{\sqrt{T} \pi f_n} \frac{G \mu}{c^2} \frac{1}{r}\; \int_{-T/2}^{T/2} dt e^{i 2 \pi (f-f_n) t},
 \end{eqnarray}
which can be readily integrated to yield, up to a phase,
 \begin{eqnarray}
 \tilde{h}_n(f)&=&\frac{c\sqrt{P_n}}{\sqrt{T} \pi f_n} \frac{G \mu}{c^2} \frac{1}{r}\; \left[\frac{ \sin[\pi (f-f_n)T]}{\pi (f-f_n)}\right],\\
               &=&\frac{c\sqrt{P_n}}{\pi^2 \sqrt{T}} \left(\frac{G \mu}{c^2}\right) \frac{1}{r}\; \frac{ \sin[\pi (f-f_n)T]}{f_n (f-f_n)}
 \end{eqnarray}
Note the introduction of the factor of the square root of the integration time $T$.  This is done to match our results to the sensitivity curves of LISA~\cite{shane1,shane2}.  This leads to the the result,
\begin{equation}
S_h(f)=\left|\tilde{h}(f) \right|^2,
\end{equation}
where $S_h(f)$ is the spectral density, whose square root is plotted on the sensitivity curves for LISA.
 
 Thus the spectrum from a loop is the sum of the Fourier components for each mode, given by,
 \begin{eqnarray}
 \tilde{h}(f)&=& \sum_{n=1}^{\infty} \tilde{h}_n(f),\\
              &=& \frac{c}{\sqrt{T} \pi^2} \left(\frac{G \mu}{c^2}\right) \frac{1}{r}\; \sum_{n=1}^{\infty}\frac{\sqrt{P_n} \sin[\pi (f-f_n)T]}{f_n (f-f_n)},\\
   &=& \frac{c\sqrt{P_1}}{\sqrt{T} f_1 \pi^2} \left(\frac{G \mu}{c^2}\right) \frac{1}{r}\; \sum_{n=1}^{\infty}\frac{ \sin[\pi (f-n f_1)T]}{n^{5/3} (f-n f_1)}.
 \end{eqnarray}
 
 Given that gravitational wave detectors sample the data as a discrete time series, we use a discrete-time Fourier transform, as discussed in the next section.
 
 \subsection{Signal to Noise}

A gravitational wave of Fourier transform $\tilde{h}(f)$ incident on a dector has a signal to noise ratio of,
\begin{equation}
(SNR)^2=4\int_0^{\infty} df \frac{|\tilde{h}(f)|^2}{S_n(f)}.
\end{equation}
This is an upper limit assuming optimal filtering.
  
            \subsection{Calculation of Single Loop Spectrum using Discrete Fourier Transform}

Since the signal detected by LISA will be as a discrete set of points, it is more appropriate to take a discrete Fourier transform of a sample signal.  In general LISA has a sampling rate of $f_{samp}=10$Hz, which leads to large numbers of data points for integration times of one to three years.  This large number of data points also leads to a very clear signal due to the harmonic nature of the gravitaional waves produced by the cosmic strings.

Given the signal $h(j)$ where $j$ is an integer and the $j^{th}$ data point corresponding to $h(t)=h(j \Delta T)$, we define the discrete-time Fourier transform as,
\begin{equation}
\label{eq:dft}
\tilde{h}(k)=\frac{1}{\sqrt{T}}\sum_{j=1}^N h(j)e^{-i 2 \pi \frac{(j-1)(k-1)}{N}}.
\end{equation} 
$N$ is the total number of data points and $k$ is the integer value in Fourier frequency space.  Note again the introduction of the square root of the integration time $T$ as a normalization.

The gravitational wave signal from the cosmic strings has the form,
\begin{eqnarray}
h(j)=\sum_{n=1}^{\infty} \frac{c\sqrt{P_n}}{ \pi f_n} \frac{G \mu}{c^2} \frac{1}{r} \sin(2 \pi j \omega_n/N),
\end{eqnarray}
where $\omega_n$ is not the physical frequency, but the discrete time frequency, which is different than the physical frequency given by $f_n=2nc/L$, and $P_n$ is the magnitude of the $n^{th}$ power mode.  Using $P_n=P_1 n^{-4/3}$ we find,
\begin{eqnarray}
h(j)=\frac{c \sqrt{P_1} }{ f_1} \frac{G \mu}{c^2} \frac{1}{r}  \sum_{n=1}^{\infty} \frac{\sin(2 \pi j  \omega_n/N)}{n^{5/3}},
\end{eqnarray}
which is then inserted into~\ref{eq:dft}.


  \section{ Results}

In this study an estimate of the distances to the loops nearest the solar system are calculatd for each frequency and string tension, and from this we calculate the signal of the loop from our location, the solar system.

     \subsection{General Results}
     
Cosmic string loops have a distinctive spectrum of harmonic modes labelled by $n$, with a high frequency tail that goes as $n^{-4/3}$.  

We use the one year results for the LISA sensitivity plots and most of our simulations are run for one year, aside from the higher frequency, $f_1\geq10^{-2}$ Hz, loops.  In these cases we use one month sampling time.

Investigations of the parameter space of the loops indicates the heaviest detectable loops individually are of tension $G\mu>10^{-10}$ for $\alpha=0.1$, due to their reduced numbers in the Milky Way, see Table~\ref{table1}.

\begin{figure}
   
	 \includegraphics[width=.98\textwidth]{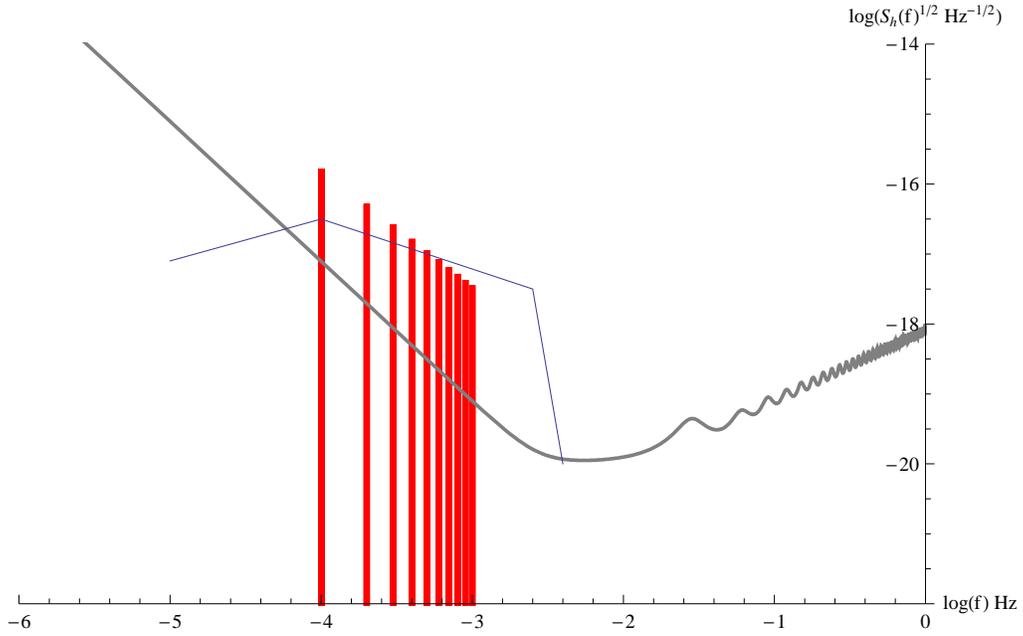}
  \caption{\label{fig51}Plot of the cosmic string loop spectrum for $G\mu=10^{-11}$, $\alpha$=0.1, at a distance of $r=7$ kpc.  Sampling time is $T=1$ year, the sampling rate is 0.1 Hz, $P_1=18$, and the fundamental frequemcy is $f_1=10^{-4}$ Hz.  The first 10 modes are shown.  This signal stands above the confusion noise from galactic binaries which is shown on the graph.  The LISA sensitivity curve is given by the thick gray line.}
		\end{figure}

For smaller loops the detectable range decreases due to a smaller number in the galaxy.  Shown in Fig.~\ref{fig51} is the largest likely signal from an individual loop at the frequency $10^{-4}$.  This signal stands above the noise from binaries.

For the large loops, $\alpha=0.1$, it is found that the smallest loops likely detectable are of tension $G\mu=10^{-16}$.   Fig.~\ref{fig53} gives an example within the binary confusion noise, but the loops that have decayed are detectable at higher frequencies.

   \subsection{Varying Loop Size}
   
\begin{figure}
	 \includegraphics{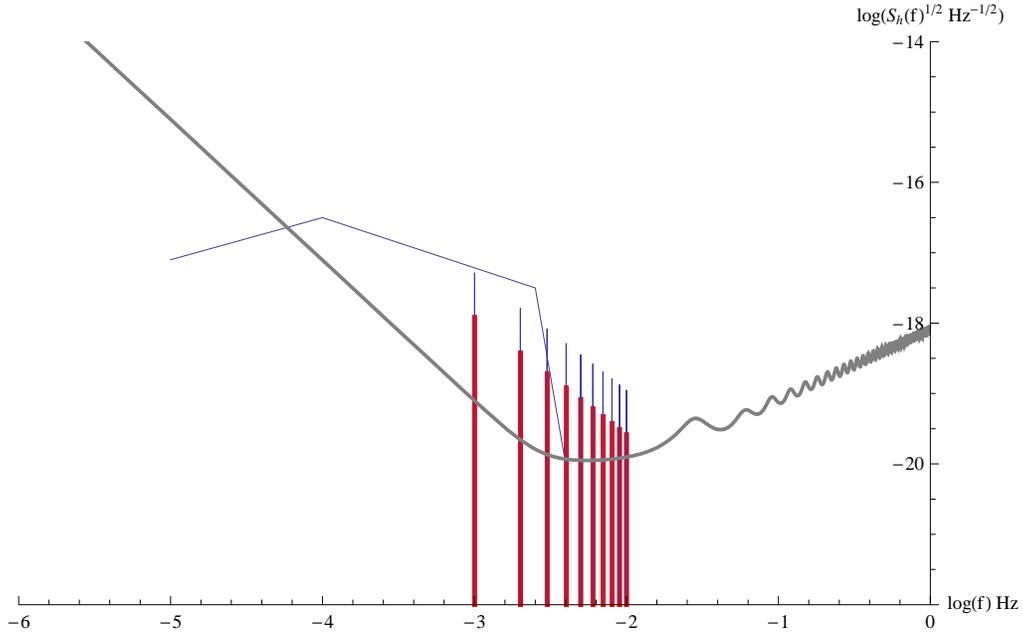}
  \caption{\label{fig52}Plot of two cosmic string loop spectra with $G\mu=10^{-12}$: the ``thin'' spectrum $\alpha$=0.1 at a distance $r=2.2$ kpc from the solar system, and the ``thick'' spectrum $\alpha=10^{-5}$ with $r$=8.9 kpc.  Sampling time is $T=1$ year, the sampling rate is 0.1 Hz, and $P_1=18$.  The fundamental frequemcy is $f_1=1$ mHz and the first 10 modes are shown as is the confusion noise from galactic binaries.}
		\end{figure}

The size of the loops formed as a fraction of the horizon, $\alpha$, affects the number of loops currently radiating.  For large $\alpha$ the number of loops remaining is large, thus at a given frequency the distance to the nearest loop will be shorter.  This is indicated in Fig.~\ref{fig52}.  Note the amplitude varies as $r^{-1}$, which shows only a small difference on our log plots.  

The total number of loops is significantly larger for large $\alpha$.  This indicates that the total signal from all loops in the galaxy at a particular frequency is significantly enhanced for large loops campared to that for the individual loops.

  Shown in Fig.~\ref{fig52} is a plot of the two frequency spectra with varying $\alpha$: one is 0.1 the other is $10^{-5}$.  The sampling time is $T=1$ year, sampling rate is 0.1 Hz, $G\mu=10^{-12}$, $P_1=18$, $r=2.2$kpc, and the fundamental frequency $f_1=1$ mHz.  In this case the string tensions differe by four orders of magnitude, while the distances differ by a factor of approximately 20.  The ratio of approximately three orders of magnitude matches the overall difference between the signals from the loops.

  \subsection{Varying String Tension}

\begin{figure}
	 \includegraphics[width=\textwidth]{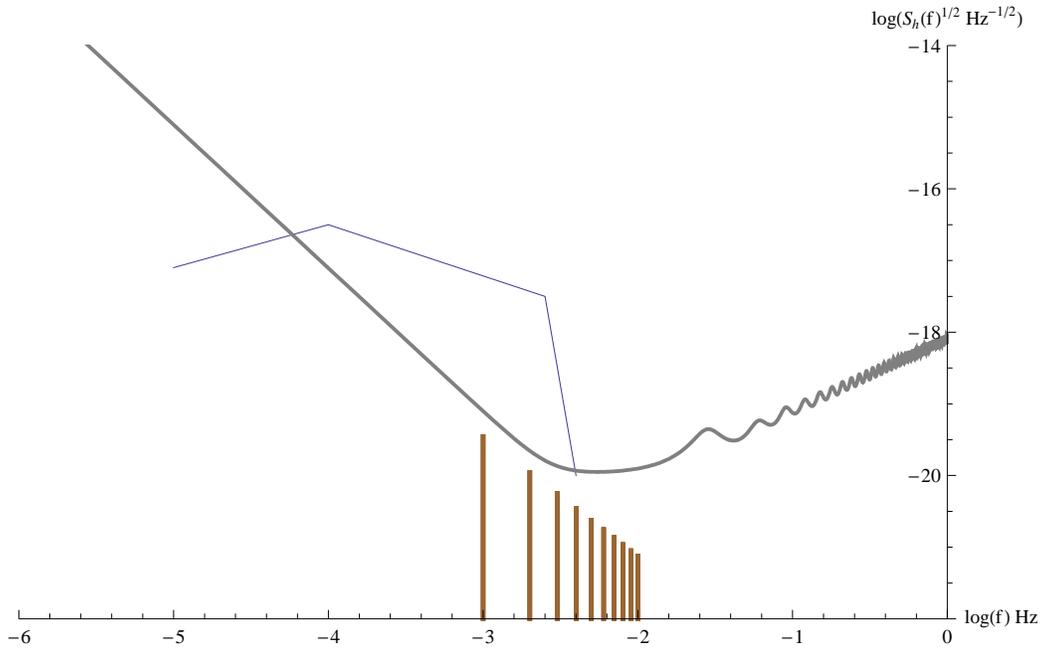}
  \caption{\label{fig53}Plot of the Fourier spectrum of cosmic string loops with tension $G\mu=10^{-16}$ and a distance of $r=0.066$ kpc.  Sampling time is $T=1$ year, and the sampling rate is 0.1 Hz, $\alpha=0.1$, and $P_1=18$.  The fundamental frequemcy is $f_1=1$ mHz and the first 10 modes are shown.  Note the heavier loops ($G\mu=10^{-12}$) have a much larger signal, in spite of their greater distance from the solar system.  The confusion noise from galactic binaries is also shown.  At this fundamental frequency, the loop is not detectable, but shifter to higher frequencies it is.}
		\end{figure}

The increase in the radiated power of the heavier string loops results in greatest variation in signals potentially detectable.  Fig.~\ref{fig53} shows the difference in signal via a string of tension $G\mu=10^{-16}$.   The heavier string loops are, on average, farther away but their larger output makes up for increased distance.

This is, on average, a typical result.  The upper limit occurs at approximately $G\mu>10^{-10}$, where the number of loops per log frequency bin becomes of order one (and smaller).  At these large string tension we expect the probability of there being a loop within the galaxy to vary from somewhat likely to very unlikely.

 \section{Discussion}  

These results dramatically increase the detectable regime of cosmic string loops.  Due to the constraints by recent observations of the cosmic microwave background and millisecond pulsars, light strings appear to be the most likely candidates remaining for cosmic string phenomena.  These light strings have a very low stochastic background signal, which is reduced further if the loops fracture into tiny sizes.  

These results indicate that very light cosmic strings are detectable due to the number density of loops and subsequent clustering around dark matter, as well as the harmonic nature of the cosmic string gravitational wave emissions.  The signal of an individual loop is a series of harmonic modes with a very large quality factor.  Thus the frequency spectrum is static over the several years of observation, further increasing detectability of the signal.

This study has taken a somewhat simplified approach to the distribution of the loops, simply matching them to the dark matter halo of the Milky Way.  No attempt was made to take into account redshift or any large scale anisotropies in cosmic string production.  Neither have we taken into account directionality in the gravitational radiation from the string.  There has been much work in the literature on this subject for particular string configurations~\cite{dubath07}.

The following chapter gives the results of numerical calculations of the spectrum of all loops within the Milky Way.


\chapter{Multiple Loops in the Milky Way}

  \section{Multiple Loops}
  
The previous chapter went into great detail about the spectrum from one loop, so this chapter is dedicated to the actual signal the sum of many loops in the galaxy produce.  Previous studies of binary systems are relevant as they provide a model for detection~\cite{nel,farm}.  The total of all the loop spectra produces a background noise that obscures other signals, while a few of the closer loops stand out above this noise.  The signal is strongly enhanced by the harmonic nature of the gravitational waves from the loops.

Using the number of loops in the galaxy we found earlier, we integrate over the whole sky to sum all of the loops.  Of significance is the relatively large numbers loops present if the progenitor strings are light.  This could produce a fairly significant signal, especially luminous in the direction of the galactic center.

The plots of all of the spectra in this work show an average signal akin to the confusion noise from the white dwarf binaries, as they obscure a fairly significant range of frequencies.  It is expected that the noise from cosmic strings is similar, with variations in its magnitude and frequency based on the string parameters.  The larger the string tension the greater the potential signal, limited by the the diminished numbers in the galaxy.  At $G \mu > 10^{-11}$ it is found that the number of loops in the galaxy drops significantly (see Table~\ref{table51}).

              \section{Power and Strain from Loops}

We reiterate here that the flux of gravitational wave energy from a loop in each mode $n$ is modeled by,
\begin{equation}
F^{gw}_n(r') \propto \frac{\dot{E}_n}{r'^{2}},
\end{equation}
where $\dot{E}_n$ is power radiated per mode from a loop.  The total power is 
\begin{equation}
\dot{E}=\gamma G \mu^2=\sum P_n G \mu^2,
\end{equation}
where $\gamma$ is found to be between 50 and 100, and the $P_n$ are power coefficents of the individual modes.  Each loop has a spectrum given by each of the power coefficients, $P_n$, where the sum $\sum{P_n}=\gamma$ is fixed by the fundamental properties of the strings.  This is an average over a wide array of loops: it is expected individual loops vary somewhat from this.  Regardless, it is expected that the most power is in the fundamental mode and the lowest modes, while the higher modes have significantly less power.

The relative weakness of gravitational waves from cosmic strings allows the use of linearized gravity.  The stress energy tensor can then be given by the Isaacson tensor:
\begin{equation}
T_{\alpha \beta}^{gw}=\frac{c^4}{32 \pi G} \left\langle h^{TT}_{ij,\alpha} h^{ij,TT}_{,\beta}\right\rangle,
\end{equation}
and the total luninosity and the flux in gravitational waves are given by
\begin{eqnarray}
L_{gw}=\int T_{0k}^{gw} n^k d^2x,\\
F_{gw}=c\;T_{00},
\end{eqnarray}
which are found from the conservation of the stress-energy tensor $D_{\mu}T_{gw}^{\mu\nu}=0$ in a flat (Minkowski) background.

Assuming plane wave solutions in the TT gauge, we find for the energy flux:
\begin{equation}
F_{gw}=\frac{\pi c^3}{4 G} f_n^2 h_n^2,
\end{equation}
where $h$ is the root-mean-square of the strain.

Solving for the rms strain we get,
 \begin{eqnarray}
h_n= \frac{2}{f_n}\sqrt{\frac{G F_n^{gw}}{\pi c^3}}
\end{eqnarray}  
where each loop has a flux given by,
\begin{eqnarray}
F_n^{gw}&=&\frac{\dot{E}_n}{4 \pi r^2},\\
        &=&\frac{P_n c^5}{4 \pi G r^2} \left(\frac{G \mu}{c^2}\right)^2.
\end{eqnarray}
Thus from the total flux we can find the rms strain at the detector.

    \section{Total Galactic Flux}

We then sum over all of the loops in the galaxy, grouping the results in bins of $\Delta f \approx 1/T$ where $T$ is the total integration time.  Typically this is between 1 and 3 years, for these calculations we use 1 year.  This value is indicated on the LISA sensitivity curves, which shows the minimum $h_{rms}$ that is detectable for the given integration time.

	\subsection{GW Signal from Galactic Loops}

Another key ingredient is the gravitational wave signal from loops within the galactic halo.  These loops are not resolvable on an individual basis, but do contribute to the background.  The isotropic and stationary background from both current and evaporated loops has been calculated previously.  To this background we add the signal from loops within the dark matter halo of the milky way, which is not isotropic from the solar system.  The results are given in a later chapter.

So for the $m^{\text{th}}$ mode of the flux at a frequency $f_m$ we sum all the contributions in the galaxy to get the spectrum:
\begin{eqnarray}
F_{gw}(f_m)^{net}= \int F_m^{gw}(r')\; n(\textbf{x}',L)\; d^3\textbf{x}' 
\end{eqnarray}
where $n(\textbf{x},L)$ is the number density of loops and the primes denote coordinates centered on the solar system.  Both $L$ and $f_m$ are related by $L=2c/f_1$.  We repeat here for clarity,
\begin{eqnarray}
n(L, r', \theta', \phi') &=& \frac{n_s(L)}{x'(1+x')^2},\\
x'                       &=& \frac{\sqrt{r'^2+R^2-2R r' \sin\theta' \cos\phi'}}{r_s}.
\end{eqnarray}
The number density constant $n_s$ is matched to this curve by integrating over the halo volume up to a truncation value of $r_t$ (about 150 kpc) and normalizing to the total number of string loops $N$:
\begin{eqnarray}
N(L)   &=& \int_0^{r_t} n(r,L) d^3 x,\\
n_s(L) &=& \frac{N(L)\;(r_t+r_s)}{4 \pi r_s^3 \left[ (r_t+r_s) \text{ln}\left(\frac{r_t+r_s}{r_s}\right)-r_t \right]}.
\end{eqnarray}
In general $r_t$ is of the order 100 kpc.  The value of $n_s(L)$ must be determined for each length (or equivalently fundamental frequency) of loop.

One can also calculate the expected gravitational wave energy flux from a region of the sky:
\begin{eqnarray}
    \frac{ dF_m^{gw} }{d\Omega'}(\Omega') &=& F_m^{gw}\; n(L, \textbf{r$'$}) r'^2.
\end{eqnarray}

\begin{figure}
   
	 \includegraphics[width=.98\textwidth]{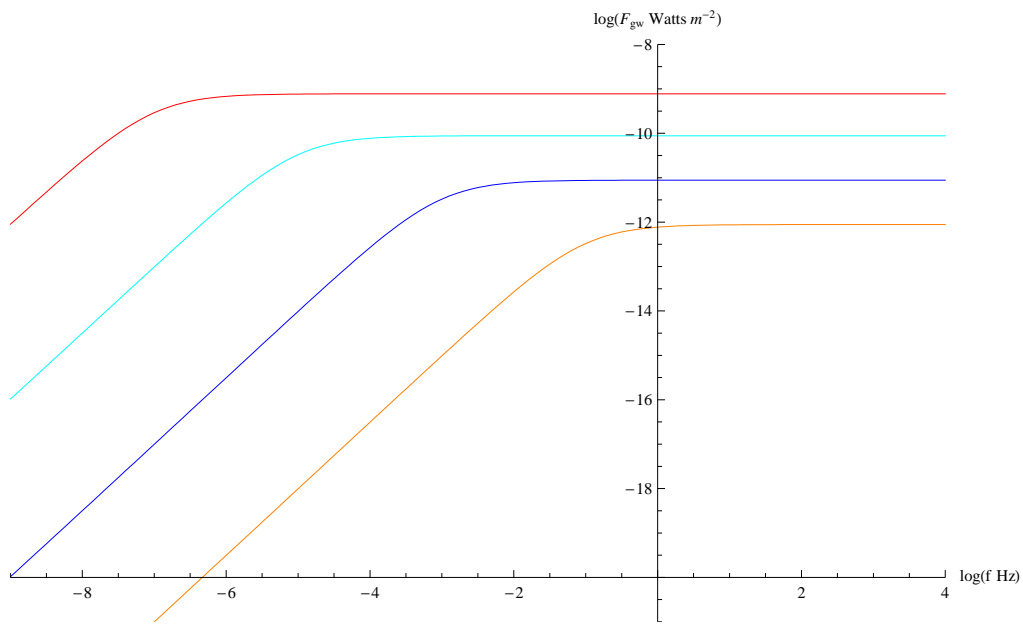}
  \caption{\label{fig61}  Plot of the cosmic string loop gravitational wave flux at the Earth for large loops $\alpha=0.1$.  The uppermost curve is string tension $G\mu=10^{-12}$ down to $10^{-18}$ in increments of $10^2$.}
		\end{figure}

\begin{figure}
   
	 \includegraphics[width=.98\textwidth]{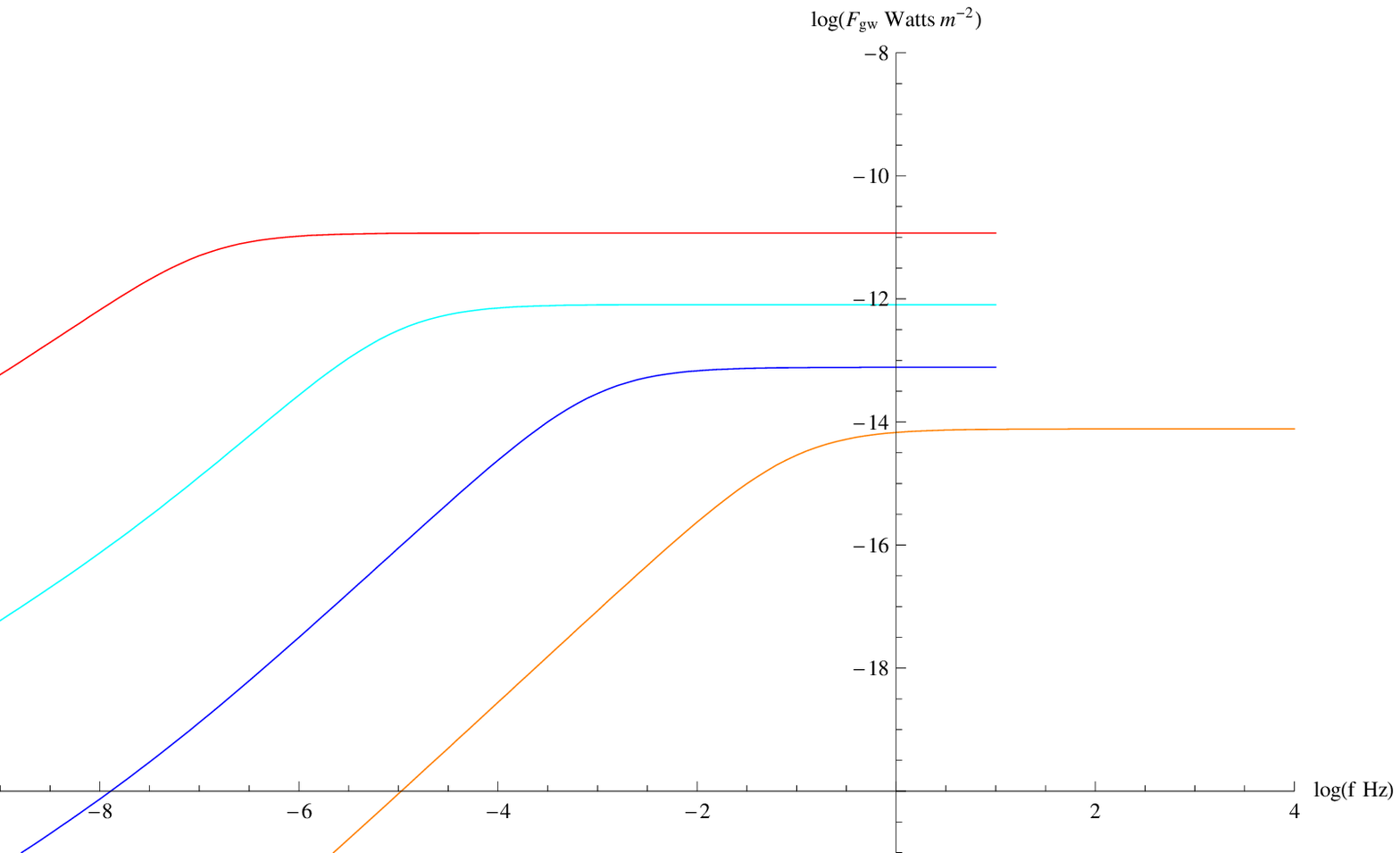}
  \caption{\label{fig62}  Plot of the cosmic string loop gravitational wave flux at the Earth for small loops $\alpha=10^{-5}$.  The uppermost curve is string tension $G\mu=10^{-12}$ down to $10^{-18}$ in increments of $10^2$.}
		\end{figure}

Results of numerical calculations of the flux are given in Figs.~\ref{fig61} and~\ref{fig62}.

   \section{Strain Spectra for Light Loops}
   
 From the total flux at a given frequency we find the strain,
 \begin{eqnarray}
h(f_n)= \frac{2}{f_n}\sqrt{\frac{G F_{gw}(f_n)^{net}}{\pi c^3}},
\end{eqnarray}  
which should be interpreted as the rms strain at a frequency $f_n$.

\begin{figure}
   
	 \includegraphics[width=.98\textwidth]{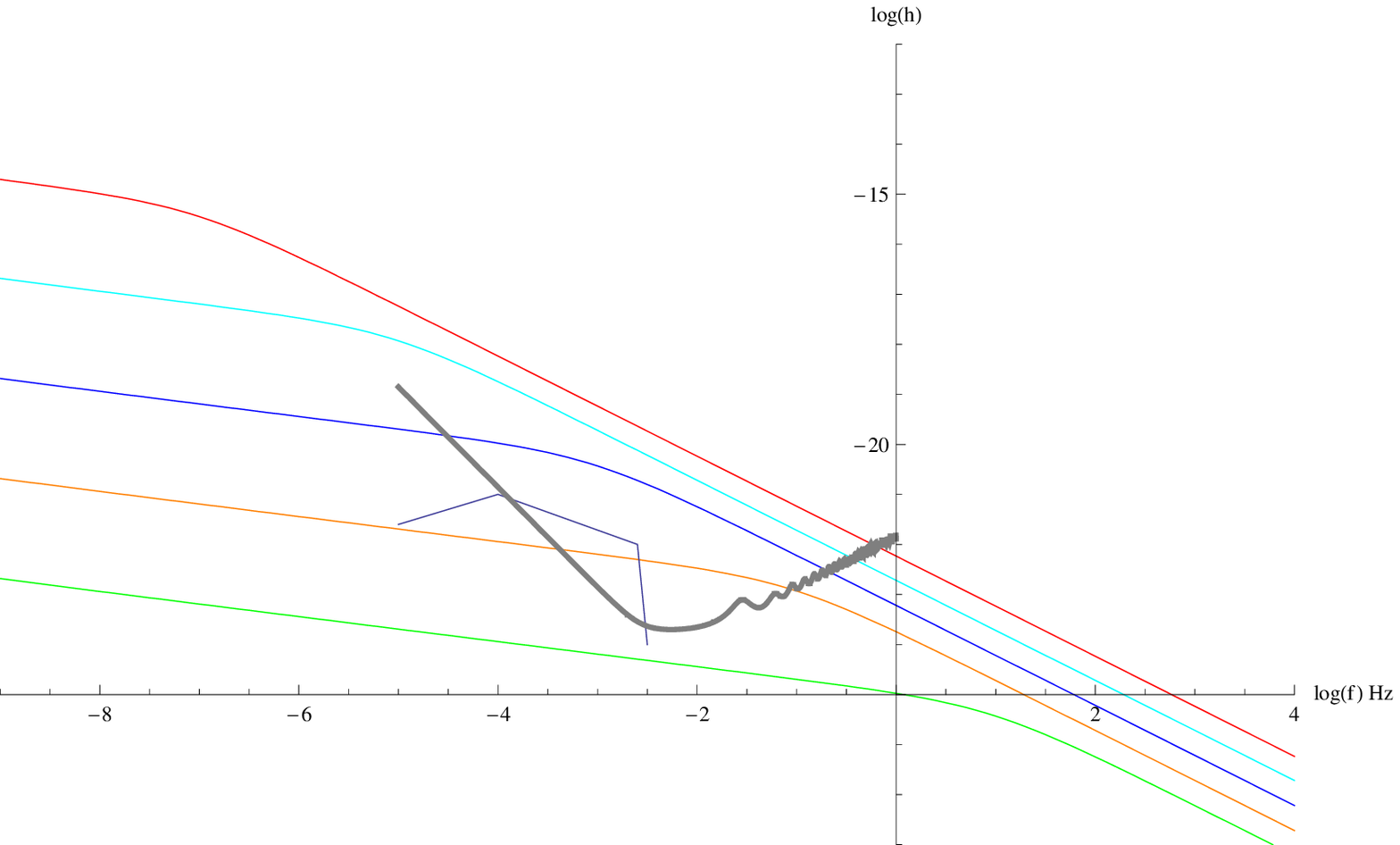}
  \caption{\label{fig63} Plot of the cosmic string loop strain spectrum for large loops $\alpha=0.1$ in the galaxy.  The top curve is of string tension $G\mu=10^{-12}$ and the bottom curve is $G\mu=10^{-20}$, in increments of $10^2$.  Also included are the LISA sensitivity curve with an integration time of 1 year, and the galactic white dwarf noise.}
		\end{figure}

\begin{figure}
   
	 \includegraphics[width=.98\textwidth]{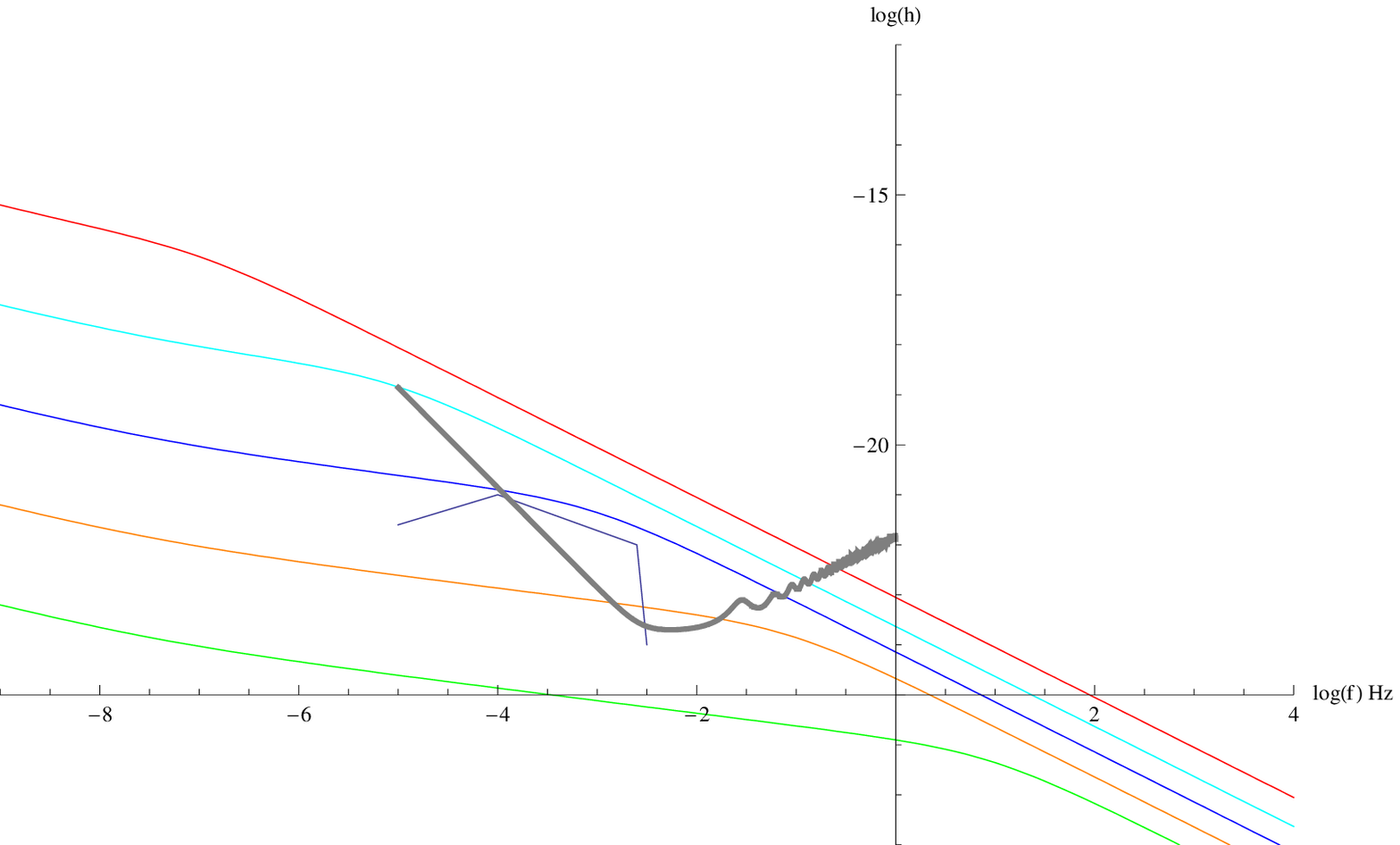}
  \caption{\label{fig64} Plot of the cosmic string loop strain spectrum for small loops $\alpha=10^{-5}$ in the galaxy.  The top curve is of string tension $G\mu=10^{-12}$ and the bottom curve is $G\mu=10^{-20}$, in increments of $10^2$.  Also included are the LISA sensitivity curve with an integration time of 1 year, and the galactic white dwarf noise.}
		\end{figure}

Results of numerical calculations are plotted in Figs.~\ref{fig63} and ~\ref{fig64}.

 \begin{figure}
   
	 \includegraphics[width=.98\textwidth]{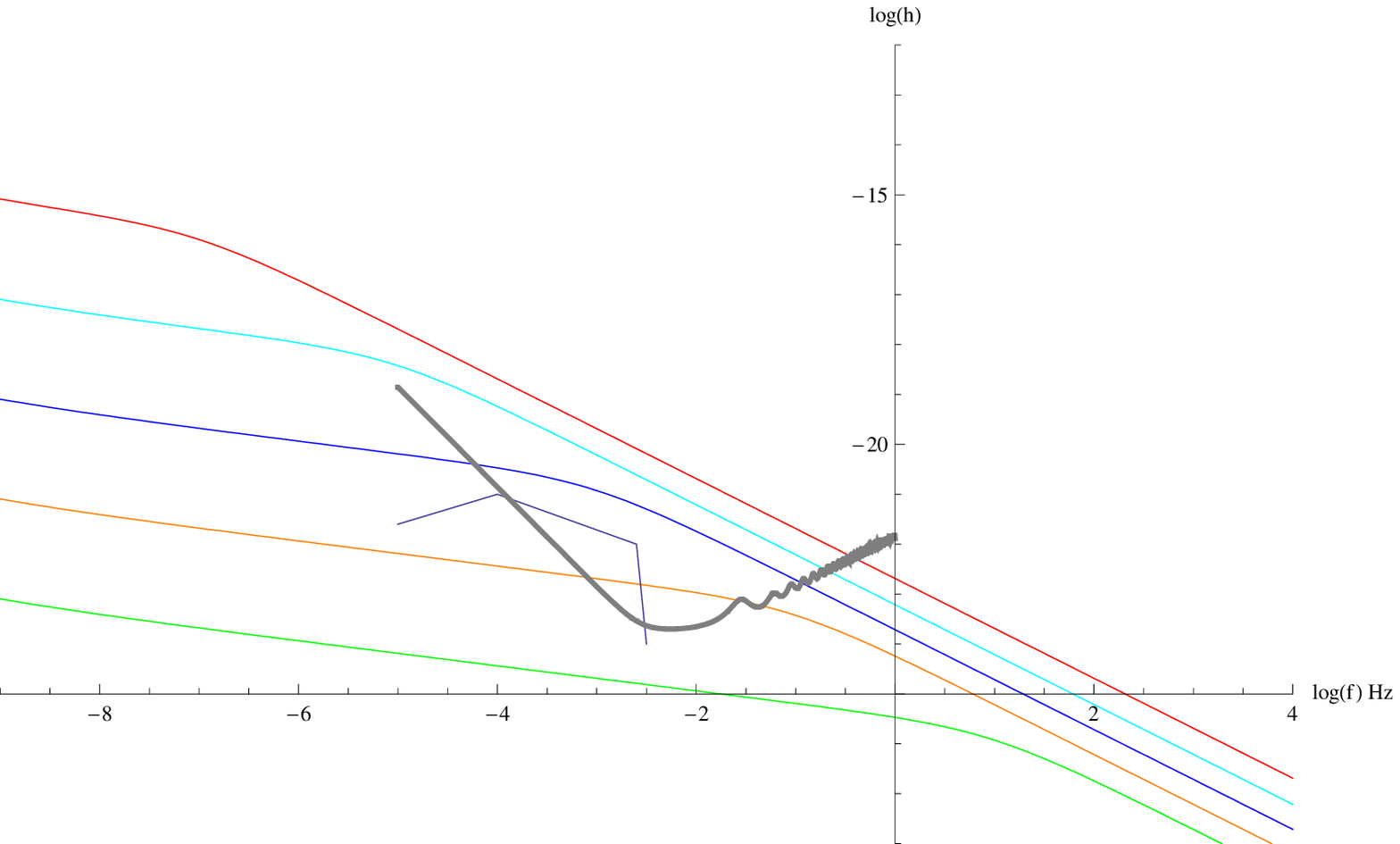}
  \caption{\label{fig65} Plot of the cosmic string loop strain spectrum for small and large loops in the galaxy.  It is assumed 90\% of the loops shatter into small loops of $\alpha=10^{-5}$.  The top curve is of string tension $G\mu=10^{-12}$ and the bottom curve is $G\mu=10^{-20}$, in increments of $10^2$.  Also included are the LISA sensitivity curve with an integration time of 1 year, and the galactic white dwarf noise.}
		\end{figure}

    \section{Heavier Loops}

The average spectra calculated above do not show individual string signals that may stand out.  For the light loops the number of loops is quite large so the average is an accurate description of the likely signal.  For heavier loops Table~\ref{table51} shows the sharp decrease in number of loops at a particular frequency.  This reduction requires an analysis of each loop and its potential signal, while using an average is less accurate.

For loops of tension $G\mu>10^{-10}$ the number of loops in the galaxy is likely of order $10^4$ overall, and though their distribution is roughly that of the dark matter the small numbers require more care in analysis.  

In each frequency bin there are few loops, so its distribution is random with probability given by the NFW distribution.  Thus the signal may be very low--perhaps not detectable--if the loop is far away, or very large if the loops is close to the earth.

     \section{Discussion}
     
     The detectable strain of light loops is strongly enhanced by the periodic nature of the signal and the clustering of string loops within the galactic halo.  This effect is greater for large loops, but even small loops show a detectable signal for LISA.  
     
     For large loops, detectable string tensions are as low as $G\mu=10^{-19}$, three orders of magnitude lower than previous estimates~\cite{depies}.  Even if the loops shatter into very small loops, the detectable signal is still appreciably larger with string tensions down to $G\mu=10^{-18}$.


\chapter{Gravitational Wave Background from Cosmic Strings}

 \section{Introduction}

 In this chapter we calculate the spectrum of  background gravitational radiation from cosmic string loops.  The results of these calculations confirm the estimates made in an earlier exploratory analysis, that also gives simple scaling laws, additional context and background~\cite{ho}.  Here we add a realistic concordance cosmology and detailed numerical integration of various effects.  We estimate that the approximations used to derive the current results are reliable to better than about fifty percent, and are therefore useful for comparison of fundamental theory with real data over a wide range of string parameters. 

Cosmic strings radiate the bulk of their gravitational radiation from loops~\cite{an,vil,vi85,turok,vi81}.  For a large network of loops we can ignore directionality for the bulk of their radiation which then forms a stochastic background~\cite{allen96}.  Each loop contributes power to the background:
\begin{equation}
\frac{dE}{dt}=\gamma G \mu^2 c,
\end{equation}
where $\gamma$ is generally given to be on the order of 50 to 100 \cite{ho,vil,burden85,ga87,allen94}.  

Gravitational radiation from strings is characterized for our purposes by the dimensionless luminosity of gravitational waves $\gamma$, which depends on the excitation spectrum of typical radiating modes on string loops.  Loop formation is characterized by $\alpha$, the typical size of newly formed loops as a fraction of the Hubble scale $H(t)^{-1}$.  Estimates based on numerical simulations have in the past  suggested very small values, leading to the hypothesis that $\alpha \sim G \mu$ or smaller~\cite{sie2,bennett88,allen90}; those ideas suggest that all but a fraction $\sim G \mu$ or smaller of loops shatter into tiny pieces.  Recent simulations designed to avoid numerical artifacts suggest a radically different picture, that in fact a significant fraction of loops land on stable non-self-intersecting orbits at a fraction $\sim0.1$ of the horizon scale~\cite{van1,van2,ring,martins}.  Our study is oriented towards this view, which leads to a larger density of loops and a more intense background for a given string mass.

For units of the gravitational energy in a stochastic background a convenient measure we adopt is the conventional dimensionless quantity given by
\begin{equation}
\label{eqs:omega}
\Omega_{gw}(f)=\frac{1}{\rho_c}\frac{d\rho_{gw}}{d\ln f},
\end{equation}
where $\rho_{gw}$ is the energy density in the gravitational waves and $\rho_c$ is the critical density.  $\Omega_{gw}(f)$ gives the energy density of gravitational waves per log of the frequency in units of the critical density.  $\Omega_{gw}$ is proportional to the mean square of the strain on spacetime from a stochastic background of gravitational waves.  

There has been much interest in cusps and kinks as sources of gravitational radiation from cosmic strings \cite{da00,da01,da05,sie}.  In this paper an estimate of  the radiation power from cusps and kinks  is included as higher frequency behavior in the loop gravitational wave spectrum,  but the beaming and bursts of gravitational waves is not discussed explicitly.  As estimated in \cite{ho}, in the regime of very light strings, observable bursts are expected to be rare, and harder to detect than the stochastic background.

Finally we would like to state some additional uncertainties associated with the model of strings described below:

\begin{itemize}

\item{ There is the possibility of multiple scales in loop formation: some string theories suggest multiple
stable intersection points and loops connected by these intersections. This may strongly affect the
accuracy of the one scale model, and indeed some of these theories also have significant other modes of decay that  invalidate the  
entire framework used here~\cite{jackson}.}
\item{ The reconnection probability p can be less than one. This may affect loop formation and the gravitational wave background. In general it is thought that this effect always increases the predicted background although it is not agreed by how much~\cite{dv}.}
\item{ There may be transient behavior of cosmic strings due to the formation process of cosmic quenching.   The model below  
assumes that transient effects have long since settled out by the time our model begins; this seems likely at LISA frequencies which  
correspond to initial loop sizes much larger than the horizon at the end of inflation~\cite{jones1}.}
\end{itemize}
Further numerical string network studies will be needed to resolve these issues.

           \section{Model of String Loop Populations}

  \subsection{Loop Formation}

The behavior of strings on cosmological scales has been thoroughly discussed in the literature~\cite{vil,vil85,hin,ho3,van1,van2,turok1984,allen90,bennett88}.  By the Kibble mechanism strings are created as a random walk with small coherence length, and they evolve following the Nambu-Goto action.  The strings form in large threadlike networks \cite{van1}, and on scales longer than the Hubble length $cH(t)^{-1}$ the network is frozen and stretches with expansion.  The strings move with speed c, and they interact to form loops with a probability $p$ by different mechanisms: one involves two strings, the other a single string forming a loop.  These loops break off from the infinite string network and become the dominant source of gravitational waves.  They oscillate with a fundamental frequency given by $f=2c/L$, where L is the length of the loop.  In general the loops form a discrete set of frequencies with higher modes given by  $f_n=2nc/L$.

The exchange probability $p$ measures the likelihood that two crossing string segments interact and form new connections to one another.  If $p$=0 then the string segments simply pass through each other; but if $p$=1 then they can exchange ``partners'', possibly forming a loop.  The value of $p$ depends upon the model~\cite{dv,hash,eto}; it is close to unity in models most commonly discussed but in principle is an independent parameter (and in a broad class of models, the only one) that can modulate the amplitude of the spectrum for a given string tension.  In this work it is taken to be $p$=1 and is parameterized as part of the number of loops formed at a given time~\cite{ho,vi81,ho3}.  The number of loops is normalized to the previous results of R.R. Caldwell and B. Allen~\cite{ca}, which is described in detail later in this section. 

  \subsection{Loop Radiation}

For the dynamics of a string the Nambu-Goto action is used:
\begin{equation}
S=-\mu \int \sqrt{-g^{(2)}} d\tau d\sigma,
\end{equation}
where $\tau$ and $\sigma$ are parameterized coordinates on the world sheet of the loop, and $g^{(2)}$ is the determinant of the induced metric on the worldsheet.  The energy-momentum tensor of the string can be given by \cite{vil},
\begin{equation}
T^{\mu\nu}(t,\textbf{x})=\mu \int d\sigma \left(\dot{x}^{\mu}\dot{x}^{\nu}-\frac{dx^{\mu}}{d\sigma} \frac{dx^{\nu}}{d\sigma} \right) \delta^{(3)}(\textbf{x}-\textbf{x}(t,\sigma)),
\end{equation}
where $\tau$ is taken along the time direction.  The radiated power from a loop at a particular frequency, $f_n$, into a solid angle $\Omega$ is,
\begin{equation}
\frac{d\dot{E_n}}{d\Omega}=4\pi G f_n^2 \left(T_{\mu\nu}^*(f_n,\textbf{k}) T^{\mu\nu}(f_n,\textbf{k})-\frac{1}{2} \left|T_{\nu}^{\nu}(f_n,\textbf{k})\right|^2 \right).
\end{equation}
Here $T_{\mu\nu}(f_n,\textbf{k})$ is the Fourier transform of $T_{\mu\nu}(t,\textbf{x})$ \cite{vil}.  Dimensionally the power is given by $G\mu^2$~\cite{turok} with numerical coefficients determined by the frequency mode and the luminosity of the loop.

In general for the power output of a loop at a given mode we can write $\dot{E_n}=P_n G \mu^2 c$~\cite{vil,ca}, where directionality is ignored by effectively inserting the loop into a large network of loops and the $P_n$ are dimensionless power coefficients.  Note that c has been reinserted.  The power coefficients are generally given the form $P_n\propto n^{-4/3}$ \cite{vil} which will create an $f^{-1/3}$ high frequency tail to the power spectrum of a loop.  The total power from a loop is a sum of all the modes:
\begin{subequations}
\begin{eqnarray}
\dot{E}&=&\sum_{n=1}^{\infty}P_n G \mu^2 c,\\
       &=&\gamma G \mu^2 c.
\end{eqnarray}
\end{subequations}
Thus $\gamma$ measures the output of energy and is labeled the ``dimensionless gravitational wave luminosity'' of a loop.  For most calculations we use the value $\gamma=50$~\cite{allen94}.

For the majority of these calculations we assume only the fundamental mode contributes, effectively a delta function for $P(f)$, and $f=2c/L$ is the only emission frequency of each loop.  This assumption is relatively good.  To check this, some calculations include higher frequency contributions and show that the effect on the background is small overall, with almost no difference at high frequencies.  This is discussed in more detail in Results.  The high frequency additions also represent the contribution from kinks and cusps, which add to the high frequency region with an $f^{-1/3}$ tail from each loop.  This high frequency dependence is the same as that for the high frequency modes, so their contributions should be similar, particularly for light strings.  Light strings contribute at higher frequencies because the loops decay more slowly and more small loops are able to contribute at their fundamental frequency longer throughout cosmic history.

The populations of loops contributing to the background fall naturally into two categories~\cite{ho}: the high redshift ``H'' population of long-decayed loops, and the present-day ``P'' population, where the redshift is of the order unity or less.  The spectrum from the H population, which have long decayed and whose radiation is  highly redshifted, is nearly independent of frequency for an early universe with a $p=\rho/3$ equation of state.  The P loops have a spectrum of sizes up to the Hubble length;  they create a peak in the spectrum  corresponding approximately to the fundamental mode of the  loops whose decay time is the Hubble time today, and a high frequency tail  mainly from higher modes of those loops.  For all of the observable loops (that is, less than tens of light years across), formation occurred during radiation era (that is, when the universe was less than $~10^5$ years old).  As in Hogan \cite{ho} we use scaling to normalize to the high frequency results from R.R. Caldwell and B. Allen \cite{ca}:  $\Omega_{gw}=10^{-8} (G\mu/10^{-9})^{1/2} p^{-1} (\gamma \alpha/5)^{1/2}$, where $p$=1.  At the high frequency limit of our curves this relation is used to normalize the parameterization for the number of loops created at a given time.

  \subsection{Loop Lengths}

Loop sizes are approximated by the ``one-scale'' model in which the length of created loops scale with the size of the Hubble radius.   The infinite strings create loops with a size that scales with the expansion, creating a distribution with a characteristic size $\alpha/H$.  

The expression for the average length of newly formed loops is $\left<L(t)\right>=\alpha c H(t)^{-1}$, where $\alpha$ measures the length of the loop as a fraction of the Hubble radius.  It is assumed that the newly formed, stable orbiting loops form a range of lengths at a given time, and this range peaks near $\alpha$.  For the numerical calculations, and to facilitate comparison with previous work, we use a delta function for the length.  Thus only one length of loop is created at a given time and this is the average length defined above.  Because of the averaging introduced by the expansion, this introduces only a small error unless the distribution of sizes is large, on the order of several orders of magnitude or more~\cite{ca}, which would no longer be a one scale model.  As a check, models of loop formation with several loop sizes are also analyzed, and it is found that the larger loops tend to dominate the background over the smaller.

The loops start to decay as soon as they are created at time $t_c$, and are described by the equation:
\begin{equation}
\label{eqs:length}
L(t_c,t)=\alpha c H(t_c)^{-1}-\gamma G \mu \frac{t-t_c}{c}.
\end{equation}
(This expression  is not valid as $L\rightarrow0$ but is an adequate approximation.  As the length approaches zero Eqn.~\ref{eqs:length} becomes less accurate, but in this limit the loops are small and contribute only to the high frequency red noise region of the gravitational wave background.  This region is very insensitive to the exact nature of the radiation so Eqn.~\ref{eqs:length} is accurate within the tolerances of the calculations.)  

Loops that form at a time $t_c$ decay and eventually disappear at a time $t_d$.  This occurs when $L(t_c,t_d)=0$ and from Eqn.~\ref{eqs:length}:
\begin{subequations}
\begin{eqnarray}
0&=&\alpha c H(t_c)^{-1}-\gamma G \mu \frac{t_d-t_c}{c},\\
t_d&=&t_c+\frac{\alpha c^2 H(t_c)^{-1}}{\gamma G \mu}.
\end{eqnarray}
\end{subequations}
At some time $t=t_d$ there are no loops created before time $t_c$, so they no longer contribute to the background of gravitational waves.  This is discussed in more detail below. 

  \subsection{Loop Number Density}

For the number density of strings we parameterize the number created within a Hubble volume at a given time by $N_t / \alpha$.  The newly formed loops redshift with the expansion of the horizon by the inverse cube of the scale factor $a(t)^{-3}$:
\begin{eqnarray}
n(t,t')= \frac{N_t}{\alpha} \left(\frac{H(t)}{c}\right)^3 \left(\frac{a(t)}{a(t')}\right)^3.
\end{eqnarray}
$n(t,t')$ is the number density at time $t$ as seen by an observer at time $t'$, and the redshift has been incorporated into the function.  In the Appendix there is a description of the calculation of the scale factor.  Analyzing the high frequency background equation given by C.J. Hogan \cite{ho} from R.R. Caldwell and B. Allen \cite{ca} the curves are normalized by setting $N_t=8.111$.  

At an observation time $t'$ the total number density of loops is found by summing over previous times $t$ excluding loops that have disappeared,
\begin{equation}
\label{eqs:nsum}
n_s(t')=\sum_{t=t_e}^{t'} n(t,t'),
\end{equation}
where $t_e$ is the time before which all the loops have decayed when making an observation at time $t'$.  The earliest time in our sum, $t_e$, is found by solving Eqn.~\ref{eqs:length}:
\begin{eqnarray}
t'=t_e+\frac{\alpha c^2 H(t_e)^{-1}}{\gamma G \mu},
\end{eqnarray}
at a given $t'$.  Thus $t_e(t')$ tells us that at some time $t'$, all of the loops formed before $t_e(t')$ have decayed and are not part of the sum.  For large loops and light strings $t_e$ trends toward earlier in the history of the universe for a given $t'$, allowing more loops to contribute longer over cosmic history.  For the final calculations, which are described in detail in the next section, $t'$ and the number density are summed over the age of the universe.

As the loops decay the energy lost becomes gravitational radiation, which persists in the universe and redshifts with expansion.  This gravitational wave energy is detectable as a stochastic background from the large number of emitting loops.

               \section{Gravitational Radiation from a Network of Loops}

  \subsection{Frequency of Radiation}

The frequency of radiation is determined by the length of the loop, and has normal modes of oscillation denoted by $n$.  The frequency of gravitational radiation from a single loop is given by,
\begin{eqnarray}
f_n(t_c,t)=\frac{2n c}{L(t_c,t)},
\end{eqnarray}
where $t_c$ is the creation time of the loop and $t$ is the time of observation.  There is also redshifting of the frequency with expansion which must be accounted for as this strongly affects the shape of the background curves.  The strings radiate most strongly at the fundamental frequency so $n=1$ for most of the calculations.  This also facilitates computation times and introduces only small errors in the amplitude of the background, which is shown in Results.  Ultimately the higher modes do not greatly alter the results at the resolution of the computations.

  \subsection{Gravitational Wave Energy}

A network of loops at a given time $t'$ radiates gravitational wave energy per volume at the rate:
\begin{equation}
\label{eqs:rhogw}
\frac{d\rho_{gw}(t')}{dt}=\gamma G \mu^2 c\;n_s(t').
\end{equation}
Recall that $n_s(t')$ is the number density $n(t,t')$ summed over previous times $t$ at a time of observation $t'$; given by Eqn.~\ref{eqs:nsum}.

To find the total gravitational wave energy density at the present time the energy density of gravitational waves must be found at previous times $t'$ and then summed over cosmic history.  Since the energy density is given by Eqn.~\ref{eqs:rhogw} the total density is found by summing over $n_s(t')$ for all times $t'$ while redshifting with the scale factor, $a^{-4}$.  Thus the total gravitational energy per volume radiated by a network of string loops at the current age of the universe is given by:
\begin{subequations}
\label{eqs:densitytime}
\begin{eqnarray}
\rho_{gw}(t_{univ})=\gamma G \mu^2 c \int^{t_{univ}}_{t_0} \frac{a(t')^4}{a(t_{univ})^4} n_s(t') dt',\\
=\gamma G \mu^2 c \int^{t_{univ}}_{t_0} \frac{a(t')^4}{a(t_{univ})^4} \left(\sum^{t'}_{t=t_e} n(t,t')\right)dt'.
\end{eqnarray}
\end{subequations}
Here $t_{univ}$ is the current age of the universe and $t_0$ is the earliest time of loop formation.  The internal sum is over $t$ and the overall integral is over $t'$.  In practice we can make $t_0$ small to observe higher frequency, H loops, earlier during the radiation era.    
 
Eqs.~\ref{eqs:densitytime} do not have explicit frequency dependence, which is needed to calculate the background gravitational wave energy density spectrum $\Omega_{gw}(f)$ determined by Eqn.~\ref{eqs:omega}.  This requires that the gravitational energy density be found as a function of the frequency: the number density as a function of time is converted to a function of frequency at the present time.  This is discussed in detail in Appendix D.

For most calculations it is assumed each length of loop radiates at only one frequency, $f=2c/L$.  The number density, as a function of frequency and time $t'$, is redshifted and summed from the initial formation time of loops $t_0$ to the present age of the universe:
\begin{equation}
\rho_{gw}(f)=\gamma G \mu^2 c \int_{t_0}^{t_{univ}} \frac{a(t')^4}{a(t_{univ})^4} n(f\frac{a(t')}{a(t_{univ})},t') dt'.
\end{equation}
Note that the frequency is also redshifted.

We have checked the effects of including higher mode radiation from each loop.  The time rate of change of the gravitational energy density of loops at time $t'$ is found by summing over all frequencies $f_j$ and weighting with the power coefficients $P_j$: 
\begin{equation}
\frac{d\rho_{gw}(f',t')}{dt'}=\sum_{j=1}^{\infty} P_j G\mu^2 c\: n(f_j',t').
\end{equation}
This is then integrated over cosmic time and redshifted to give the current energy density in gravitational waves:
\begin{eqnarray}
  \rho_{gw}(f)&=&G\mu^2 c \int_{t_0}^{t_{univ}}dt' \frac{a(t')^4}{a(t_{univ})^4}
                 \nonumber\\
                 & &\times\left\{\sum_{j=1}^{\infty} P_j\:n(f_j\frac{a(t')}{a(t_{univ})},t')\right\} .
\end{eqnarray}
The power coefficients $P_j$ are functions of the mode and have the form $P_j\propto j^{-4/3}$~\cite{vil,ca}.  This includes the behavior of cusps and kinks, which also contribute an $f^{-1/3}$ tail to the power spectrum.  Depending upon the percentage of power in the fundamental mode, the sum to infinity is not needed.  As an example, if 50\% of the power is in the fundamental mode only the first six modes are allowed.  In general it is found that the higher modes do not significantly effect the values of the background radiation.  Details are given in Results.

From the gravitational radiation density we can find the density spectrum readily by taking the derivative with respect to the log of the frequency and dividing by the critical density:
\begin{subequations}
\begin{eqnarray}
\Omega_{gw}(f)&=&\frac{f}{\rho_c}\frac{d\rho_{gw}(f)}{df},\\
&=&\frac{8 \pi G f}{3 H_o^2} \frac{d\rho_{gw}(f)}{df}.
\end{eqnarray}
\end{subequations}
A convenient measure is to take $\Omega_{gw} h^2$ to eliminate uncertainties in $H_o=H(t_{univ})=100 h\ km\ s^{-1} Mpc^{-1}$ from the final results.  (Note here $h$ is the value of the Hubble parameter, not the strain on spacetime.)

               \section{Results for the Stochastic Background}

  \subsection{General Results}
    
\begin{figure*}
\includegraphics[width=.95\textwidth]{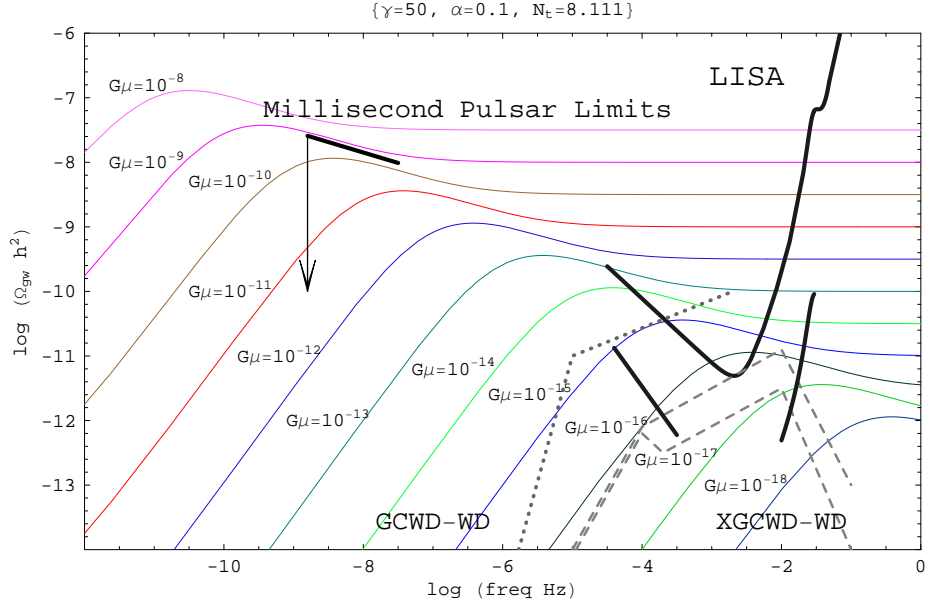}
 \caption{\label{fig1} The gravitational wave energy density per log frequency from cosmic strings is plotted as a function of frequency for various values of $G\mu$, with $\alpha=0.1$ and $\gamma=50$.  Note that current millisecond pulsar limits have excluded string tensions $G\mu>10^{-9}$ and LISA is sensitive to string tensions $\sim10^{-16}$ using the broadband Sagnac technique, shown by the bars just below the main LISA sensitivity curve.  The dotted line is the Galactic white dwarf binary background (GCWD-WD) from A.J. Farmer and E.S. Phinney, and G. Nelemans, et al.,~\cite{farm,nel}.  The dashed lines are the optimistic (top) and pessimistic (bottom) plots for the extra-Galactic white dwarf binary backgrounds (XGCWD-WD)~\cite{farm}.  Note the GCWD-WD eliminates the low frequency Sagnac improvements.  With the binary confusion limits included, the limit on detectability of $G\mu$ is estimated to be $>10^{-16}$.}
\end{figure*}

All of the computations show the expected spectrum for loops:  old and decayed ``H'' loops leave a flat spectrum at high frequencies and matter era ``P'' loops contribute to a broad peak, with smooth transition region connecting them that includes comparable contributions from both.  The frequency of the peak is strongly dependent upon the string tension, with lighter strings leading to a higher frequency peak.  This is a result of the lighter loops decaying at a lower rate, so smaller loops survive for a Hubble time.  This behavior is exhibited in Fig.~\ref{fig1}.  

Fig. 7.2 shows the peaks of power output per unit volume per log frequency as a function of frequency at different times in cosmic history.  The most recent times contribute to the broad peak, while later times contribute to the flat red noise region.  Although the power output per volume is larger earlier, the summation time is foreshortened so the contributions overlap to form the lower amplitude flat region in the background.

\begin{figure}
  \includegraphics[width=.95\textwidth]{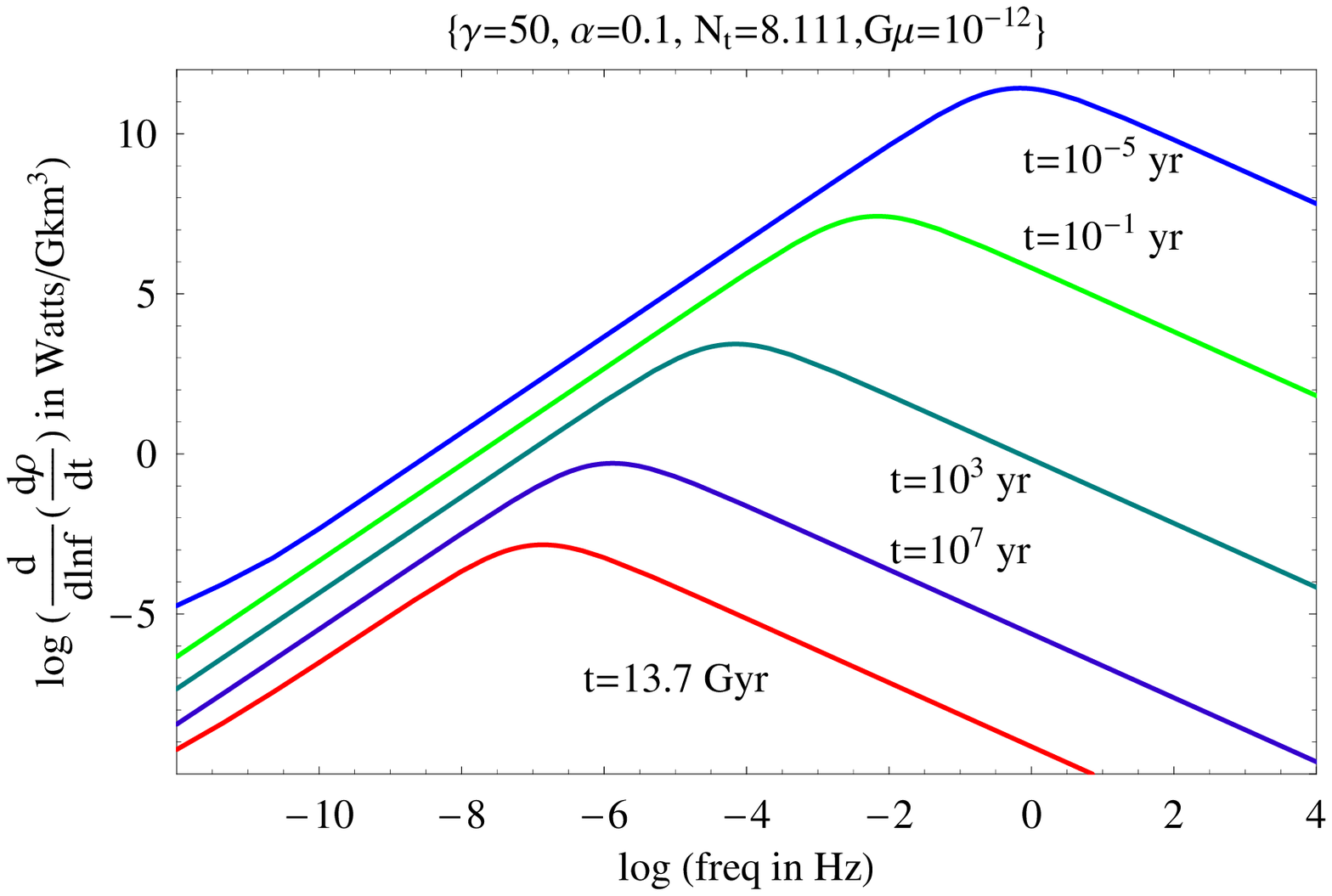}
  \caption{\label{fig2}The power per unit volume of gravitational wave energy per log frequency from cosmic strings is measured at different times in cosmic history as a function of frequency for $G \mu=10^{-12}$, $\alpha=0.1$, and $\gamma=50$.  Note that early in the universe the values were much larger, but the time radiating was short.  The low frequency region is from current loops and the high frequency region is from loops that have decayed. This plot shows the origin of the peak in the spectrum at each epoch; the peaks from earlier epochs have smeared together to give the flat high frequency tail observed today.}
\end{figure}

The differences in these calculations from previous publications can be accounted for by the use of larger loops and lighter strings as predicted in~\cite{ho}.  The change in cosmology, i.e. the inclusion of a cosmological constant, is not a cause of significant variation.  Since the cosmological constant becomes dominant in recent times the increased redshift and change in production rate of loops does not strongly affect the background.

   \subsection{Varying Lengths}

The effect of changing the size of the loops is shown in Fig. 3.  A larger $\alpha$ leads to an overall increase in the amplitude of $\Omega_{gw}$ but a decrease in the amplitude of the peak relative to the flat portion of the spectrum. As expected there is little if no change in the frequency of the peak.  In general the red noise portion of the spectrum scales with $\alpha^{1/2}$ and this is seen in the high frequency limits of the curves in Fig. 3 \cite{ho}.  An increase in $\alpha$ leads to longer strings and an increase in the time that a given string can radiate.  This leads to an overall increase in amplitude without an effect on the frequency dependence of the spectrum peak.

\begin{figure}
 \includegraphics[width=.95\textwidth]{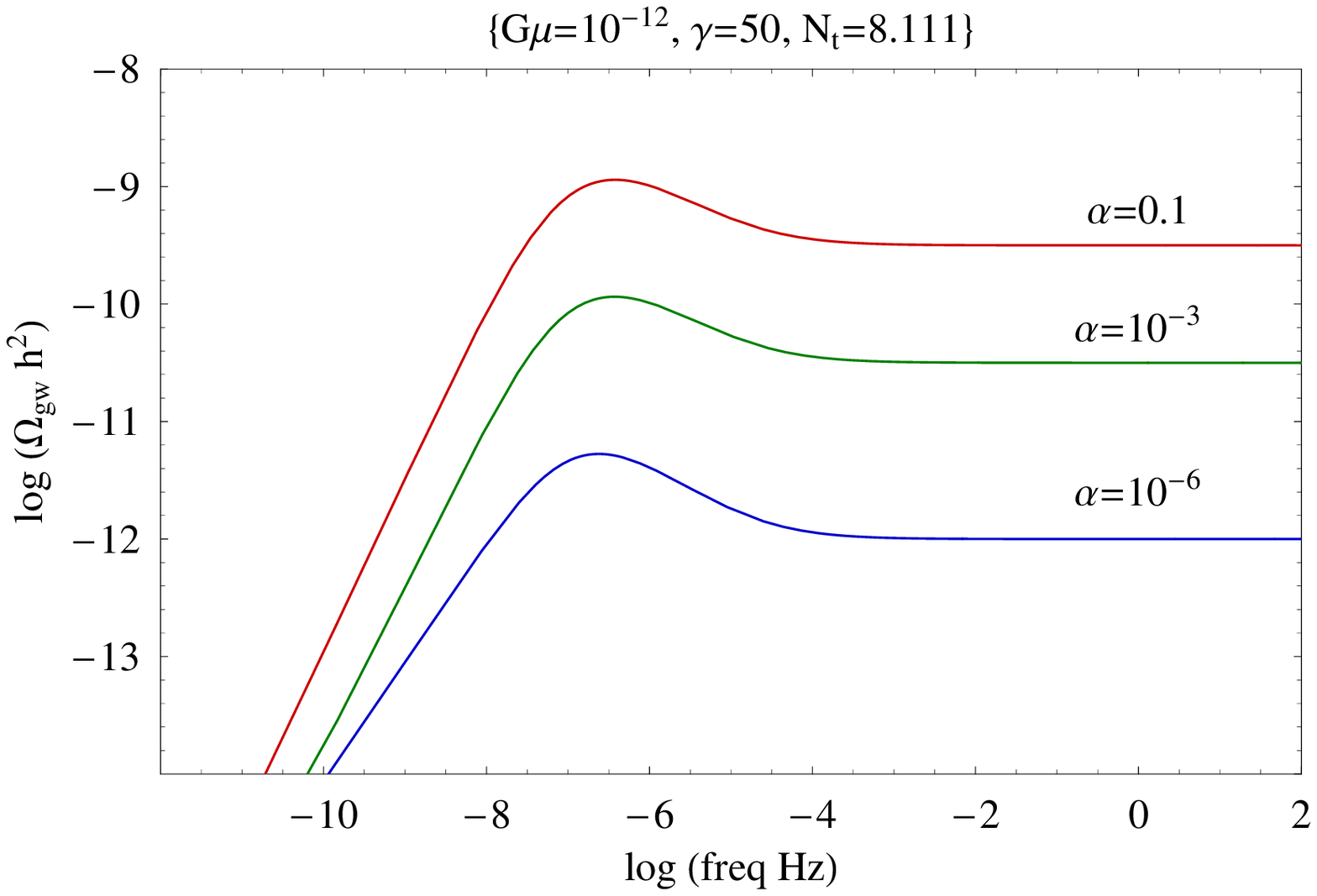}
 \caption{\label{fig3} The gravitational wave energy density per log frequency from cosmic strings is shown with varying $\alpha$ for $G\mu=10^{-12}$ and $\gamma=50$.  Here $\alpha$ is given values of 0.1, $10^{-3}$, and $10^{-6}$ from top to bottom.   Note the overall decrease in magnitude, but a slight increase in the relative height of the peak as $\alpha$ decreases; while the frequency remains unchanged.  The density spectrum scales as $\alpha^{1/2}$, and larger loops dominate the spectrum over small loops.}
\end{figure}

\begin{figure}
 \includegraphics[width=.95\textwidth]{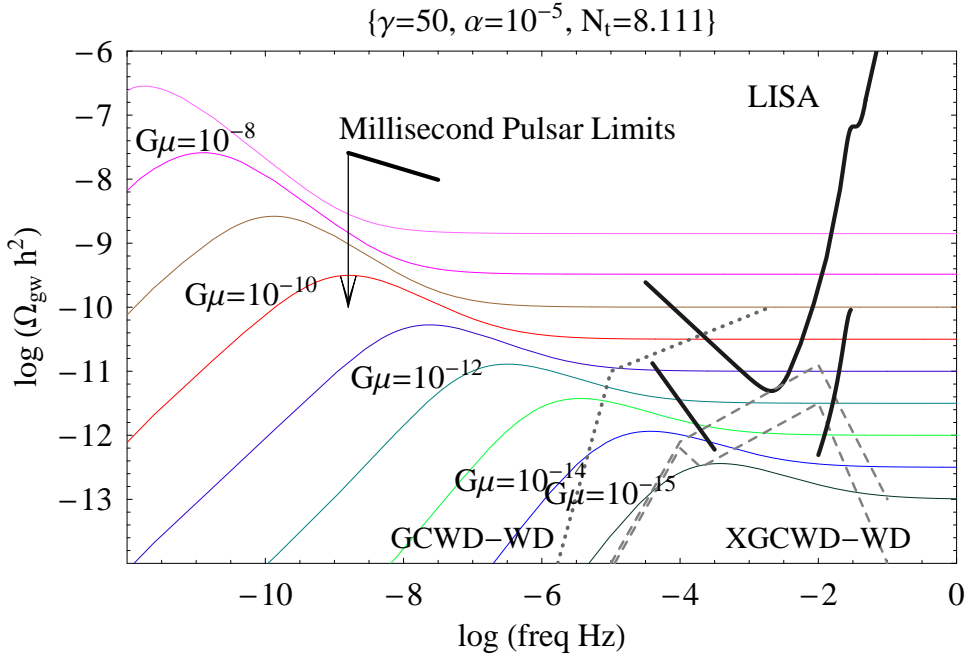}
 \caption{\label{fig4}The gravitational wave energy density per log frequency from cosmic strings for $\alpha=10^{-5}$ and $\gamma=50$.  Smaller $\alpha$ reduces the background for equivalent string tensions, with a LISA sensitivity limit of $G\mu>10^{-14}$ if the Sagnac sensitivity limits are used.  The bars below the main LISA sensitivity curve are the Sagnac improvements to the sensitivity, while the dotted line is the Galactic white dwarf binary background (GCWD-WD) from A.J. Farmer and E.S. Phinney, and G. Nelemans, et al.,~\cite{farm,nel}.  The dashed lines are the optimistic (top) and pessimistic (bottom) plots for the extra-Galactic white dwarf binary backgrounds (XGCWD-WD)~\cite{farm}.  Note the GCWD-WD eliminates the low frequency Sagnac improvements, and increases the minimum detectable $G\mu$ to $>10^{-12}$.  $G\mu$ ranges from the top curve with a value $10^{-7}$ to the bottom curve with $10^{-15}$.} 
 \end{figure}

In Fig.~\ref{fig4} a smaller $\alpha$ is plotted and shows the reduction in the background.  The general shape of the background remains unchanged, except for some small differences in the heavier strings.  For heavier strings the amplitude of the peak tends to increase relative to the red noise portion of the background spectrum, and the frequency of the peak shifts less with $G\mu$.  These differences become more pronounced for heavy and very small strings; and at this limit our results are found to match very closely those computed in~\cite{ca}. 

   \subsection{Luminosity}
\begin{figure}
 \includegraphics[width=.95\textwidth]{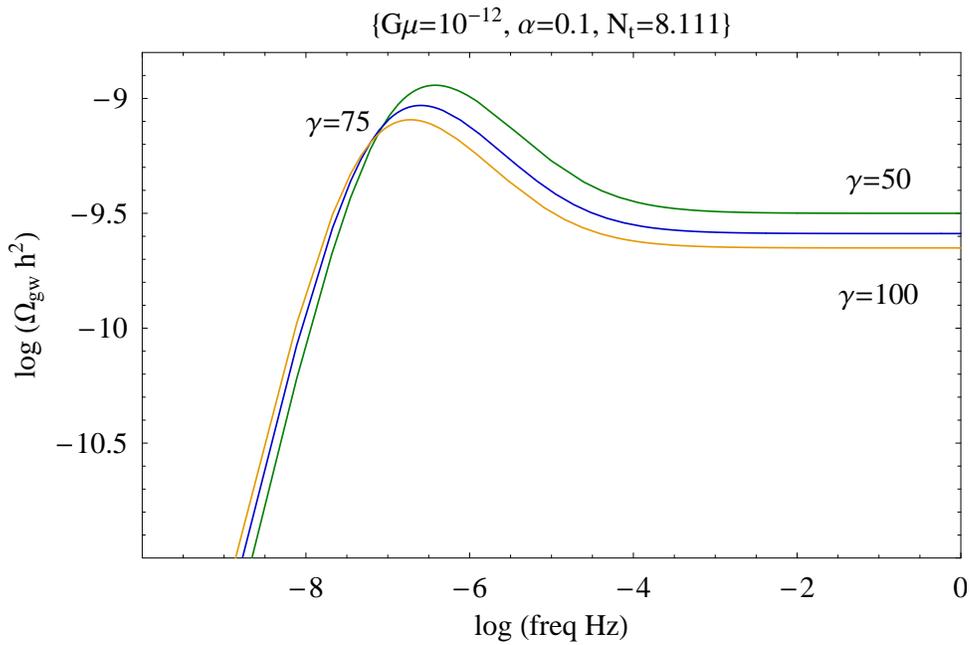}
  \caption{\label{fig5}The energy density in gravitational waves per log frequency from cosmic strings with varying luminosity $\gamma$ for $G\mu=10^{-12}$ and $\alpha=0.1$.  Here the $\gamma$'s are given as follows at high frequencies: 50 for top curve, 75 for the middle curve, and 100 for the bottom curve.  The peak frequency, peak amplitude, and amplitude of flat spectrum all vary with the dimensionless gravitational wave luminosity $\gamma$.}
 \end{figure}

A residual factor of the order of unity arises from uncertainty in the typical radiation losses from loops when all modes are included averaged over an entire population.  Changing the dimensionless luminosity $\gamma$ affects the curves as shown in Fig.~\ref{fig5}.  Larger $\gamma$ decreases the amplitude of the red noise region of the spectrum, but increases the lower frequency region.  Smaller $\gamma$ increases the amplitude of the peak and the flat portion of spectrum, while leading to a decrease in the high frequency region.  This occurs because the loops decay more quickly for larger $\gamma$ and have less time to contribute over the long time periods of cosmology.  In the matter dominated era the higher luminosity allows the loops to contribute more to the high frequency region, although it does not increase the peak amplitude.  Higher luminosity also decreases the frequency of the overall peak by contributing more power in current loops.  The power and change in length depend on the luminosity by $\dot{E}\propto\gamma$ and $\dot{L}\propto-\gamma$.  Note that the gravitational wave power depends on the square of the mass density but the first power of the luminosity, thus different behavior is expected varying $\gamma$ or $\mu$.  An analytic description of the scaling of the amplitude and frequency of $\Omega_{gw}$ with  $\gamma$ is given in~\cite{ho}. 

  \subsection{Multiple Lengths}

\begin{figure}
    \includegraphics[width=.95\textwidth]{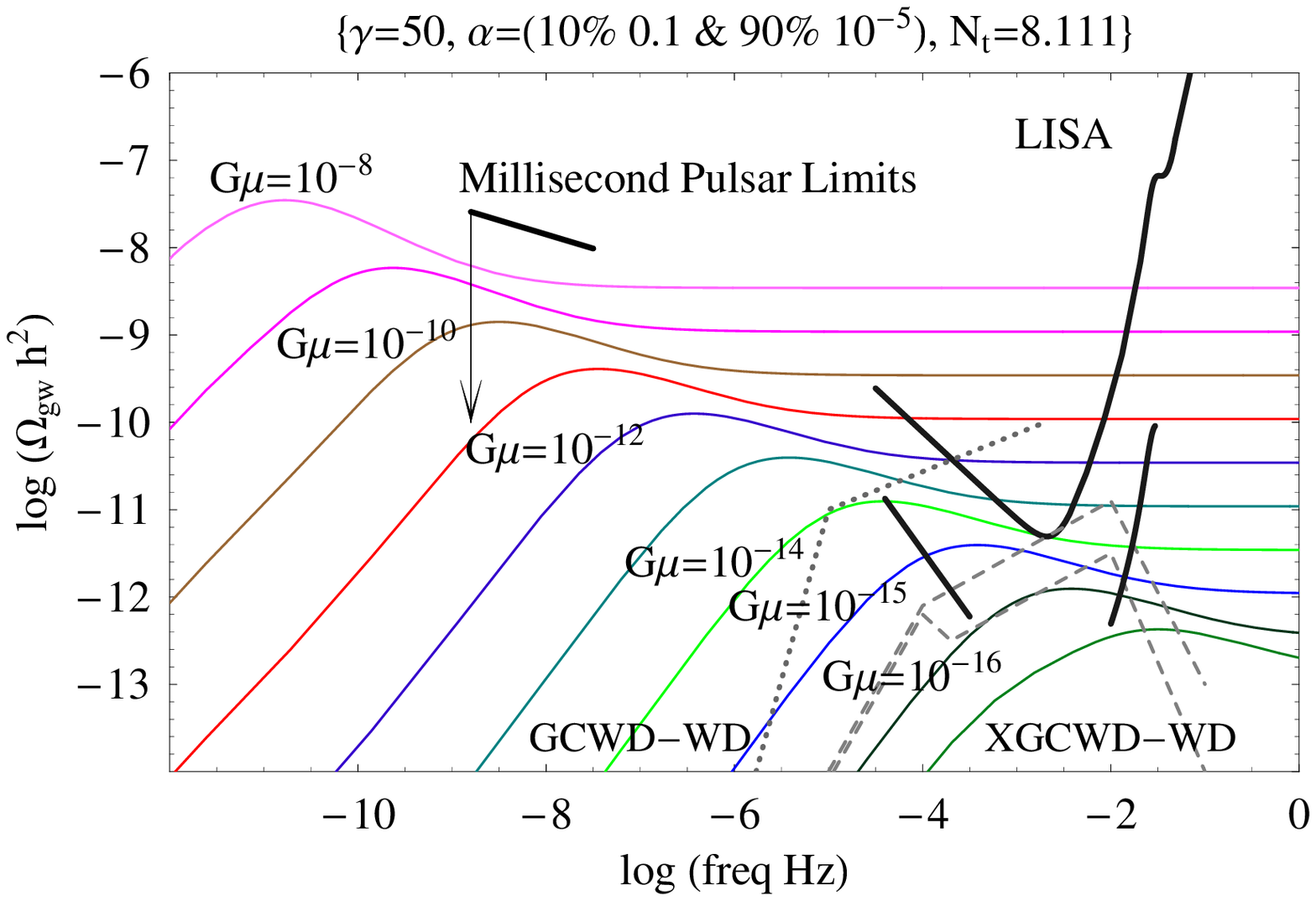}
       \caption{\label{fig6} The gravitational wave energy density per log frequency from cosmic strings for $\gamma$=50, $G\mu=10^{-12}$, and a combination of two lengths ($\alpha$'s).  90\% of the loops shatter into loops of size $\alpha=10^{-5}$ and 10\% remain at the larger size $\alpha=0.1$.  The overall decrease is an order of magnitude compared to just large alpha alone.  From bottom to top the values of $G\mu$ run from $10^{-17}$ to $10^{-8}$. }
    \end{figure}

Some models indicate several characteristic scales for the loop populations, with some combination of several scales.  This implies multiple distributions of lengths, which can be described as a combination of different $\alpha$'s.  Figures~\ref{fig1}, \ref{fig3}, and \ref{fig4} indicate the eventual dominance of the larger loops over the smaller in the final spectrum.  Even though, in this model, the number of loops created goes as $\alpha^{-1}$, their contribution to the background is still much smaller and scales as $\alpha^{1/2}$.  Thus an admixture of loop scales tends to be dominated by the largest stable radiating loops, which is shown in Fig.~\ref{fig6}.

In Fig.~\ref{fig6} we assume 90\% of the loops shatter into smaller sizes $\alpha=10^{-5}$ while 10\% are larger with $\alpha=0.1$.  A ``two scale'' model is used and two relatively narrow distributions are assumed, as in the one scale model.  It is imagined that the inital loops are large and splinter into smaller loops, with the percentage that shatter depedent upon the model.  Overall the amplitude decreases by an order of magnitude below a pure $\alpha=0.1$ amplitude, but it is an order of magnitude above the $\alpha=10^{-5}$ spectra.  Thus the large loops, even though they form a small fraction, contribute more to the background.  Because of this weighting a one-scale model is likely to be suitable if appropriately normalized.  Note that a string simulation requires a dynamic range of scale greater than $\alpha$ to make a good estimate of $\alpha$ and to choose the right normalization.

    \subsection{High Frequency Modes, Kinks, and Cusps}

\begin{figure}
    \includegraphics[width=.95\textwidth]{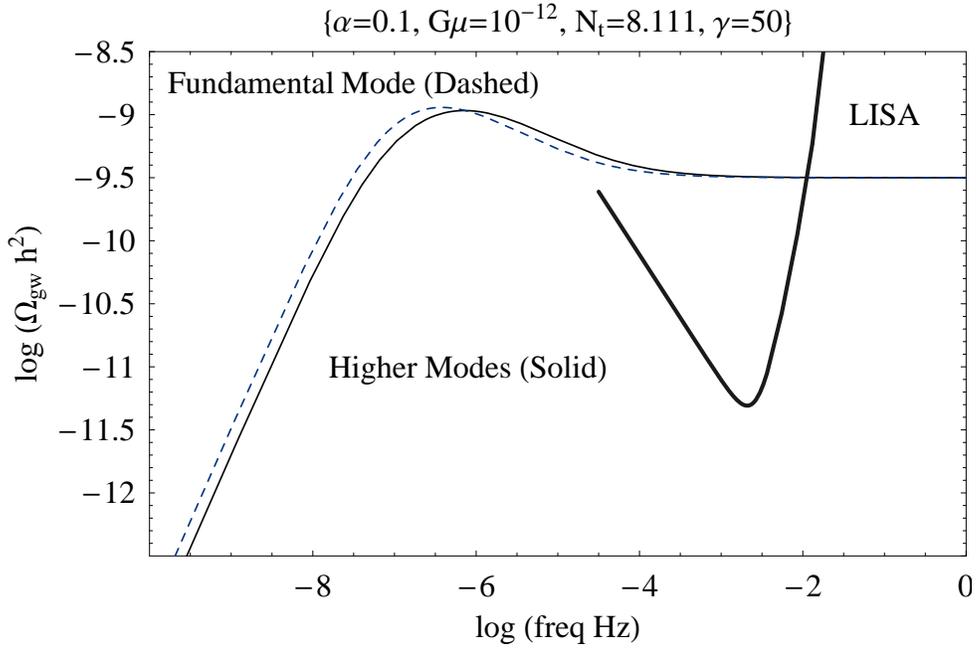}
       \caption{\label{fig7}The gravitational wave energy density per log frequency for $\gamma$=50, $G\mu=10^{-12}$, $\alpha=0.1$; plotted in one case (dashed) with only the fundamental mode, while for another case (solid) with the first six modes.  The dashed curve has all power radiated in the fundamental mode, while the solid curve combines the fundamental with the next five modes: $P_1=25.074,\ P_2=9.95,\ P_3=5.795,\ P_4=3.95,\ P_5=2.93,\ P_6=2.30$.  Note the slight decrease in the peak and its shift to a higher frequency.  The LISA sensitivity is added as a reference.}
    \end{figure}

Calculations are done which include higher mode behavior of the loops; this sends some of the gravitational wave energy into higher frequencies for each loop.  Adding higher modes tends to smooth out, to a larger degree, the curves and slightly lower the peaks on the spectra.  When applied to the power per volume per log frequency as in Fig.~\ref{fig2} we see a less steep high frequency tail at each time and a slight decrease in the amplitude of the peak.  From Fig.~\ref{fig7} a comparison of the fundamental mode versus the fundamental mode with higher modes included shows: a) the peak decreases, b) the peaked portion widens slightly, and c) the red noise amplitude remains relatively unchanged.  

Fig.~\ref{fig7} demonstrates a calculation in which $\sim50\%$ of the power is in the fundamental mode and the power coefficients go as $n^{-4/3}$ in the higher modes.  For the fundamental mode $P_1=25.074$, and the next five modes are $P_2=9.95$, $P_3=5.795$, $P_4=3.95$, $P_5=2.93$, and $P_6=2.30$.  Their sum is $\sum P_n=\gamma=50$.  The amplitude of the peak shifts from $1.202\times10^{-9}$ to $1.122\times10^{-9}$, a 6.7\% decrease.  The frequency of the peak shifts more substantially, from $3.162\times10^{-7}\:Hz$ to $7.080\times10^{-7}\:Hz$, an increase of 124\%.  This increase in frequency of the peak does not affect the limits of detection by LISA because the amplitude shrinks very little.  Note that the high frequency red noise region is essentially unchanged, and for $G\mu=10^{-12}$ this is the region in which LISA is sensitive. 

Plots of spectra for other $G \mu$ with high frequency modes show changes similar to those in Fig.~\ref{fig7}.  The peak amplitude drops slightly and the frequency of the peak is increased.  Overall, the limits from Table~\ref{table1} remain unchanged by the introduction of 50\% of the power in the higher frequency modes.  

Cusps and kinks in the loops are responsible for high frequency ``bursts'' of gravitational wave energy~\cite{da00,da01,da05,sie}.  They are two different manifestations of string behavior: cusps are catastrophes that occur approximately once every oscillation, emitting a directed burst of gravitational radiation, while kinks are small wrinkles which propagate on the loops, emitting a higher frequency of directed gravitational energy.  The light strings make it less likely to detect these random bursts; in particular they are more difficult to pick out from their own confusion background~\cite{ho}.  Since cusps, kinks, and higher modes add an $f^{-1/3}$ tail to the power spectrum of each loop, it is reasonable to expect that they will have a similar effect on $\Omega_{gw}$.  The general behavior of high frequency contributions is represented in Fig.~\ref{fig7}.

                               \section{Millisecond Pulsar Limits}

    \subsection{Limits on Background}

Millisecond pulsars have been used as sources to indirectly measure a gravitational radiation background for a number of years \cite{det,sti,kaspi,lommen,thors,mch}.  Limits on the strain, and correspondingly the energy density spectrum $\Omega_{gw}$, have been estimated recently at three frequencies: $1/20\:yr^{-1}$, $1/8\:yr^{-1}$, and $1\:yr^{-1}$~\cite{jenet}.  The limits on the strain vary with the frequency dependence assumed for the source of gravitational radiation at the  corresponding frequencies:
\begin{equation}
\label{eqs:Aheqn}
h_c(f)=A \left(\frac{f}{yr^{-1}}\right)^{\beta},
\end{equation}
where $\beta$ is the power of the frequency dependence of the source and is denoted as $\alpha$ in F.A. Jenet, et al.,\cite{jenet}.  A table of values is provided in that source for various cosmological sources of stochastic gravitational radiation, with $\beta=-7/6$ used for cosmic strings.  From~\cite{jenet}, the cosmic string limits are:
\begin{subequations}
\label{eqs:limits}
\begin{eqnarray}
\Omega_{gw}(1\ yr^{-1}) h^2    &\leq& 9.6\times10^{-9},\\
\Omega_{gw}(1/8\ yr^{-1}) h^2  &\leq& 1.9\times10^{-8},\\
\Omega_{gw}(1/20\ yr^{-1}) h^2 &\leq& 2.6\times10^{-8}.
\end{eqnarray}
\end{subequations}

Results of our calculations show that for heavier strings at the measured frequencies $\beta$ is approximately -1, but for lighter strings it varies from this value.  The variation of $\beta$ is summarized in Table~\ref{table2}, which has been calculated at $f=1/20\:yr^{-1}$ and $f=1\:yr^{-1}$.  This change in $\beta$ will result in limits differing slightly only at low frequencies from those in Eqs.~\ref{eqs:limits} \cite{jenet}.

In Figures~\ref{fig1} and \ref{fig4} the limits for millisecond pulsars \cite{jenet} are shown, with allowed values of $\Omega_{gw}$ indicated by the direction of the arrow.  For large strings $\alpha=0.1$, the maximum allowed $G \mu$ is about $10^{-9}$, Fig.~\ref{fig1}.  Decreasing $\alpha$ decreases the amplitude of the background allowing for heavier strings within the pulsar measurement limits.  A more complete list is displayed in Table~\ref{table1}.

  \subsection{Dependence of Characteristic Strain on Frequency}

From the curves of Fig.~\ref{fig1} one can find the frequency dependence of the characteristic strain as given by $h_c(f)=A(f/yr^{-1})^{\beta}$.  Again, $\beta$ is the power law dependence of $h_c$ on frequency, and is generally written as $\alpha$ with a numerical value of -7/6 \cite{jenet,mag,mag00}.  In this paper $\beta$ is used to prevent confusion with the length of loops.  If $\Omega_{gw}$ is known, then using Eqn.~\ref{eqs:charstrain} we find:
\begin{equation}
\Omega_{gw}(f)=\frac{2 \pi^2 A^2}{3 H_o^2} \frac{f^{2(1+\beta)}}{(3.17\times10^{-8}\:Hz)^{2 \beta}}.
\end{equation}
With this equality $\beta$ is found by computing $d\ln(\Omega_{gw})/d\ln f$ at $f=1\:yr^{-1}=3.17\times10^{-8}Hz$ and setting this equal to $2(1+\beta)$.  The results are listed in Table~\ref{table2}.

\begin{table}
 \caption{\label{table2}$\beta$ for given $G\mu$ at $f=1/20\:yr^{-1}$ and $f=1\ yr^{-1}$, $\alpha=0.1$, $\gamma=50$, where $h_c\propto f^\beta$.}

\begin{tabular}{|c| c| c|}
\hline
     & $\beta$ & $\beta$ \\
 $G\mu$ & $f=1/20\:yr^{-1}$ & $f=1\:yr^{-1}$ \\
\hline
$10^{-6}$& -1.00306 & -1.00031 \\

$10^{-8}$& -1.05314 & -1.00894 \\

$10^{-9}$& -1.05985 & -1.03853 \\

$10^{-10}$& -0.931749 & -1.06593 \\

$10^{-11}$& -0.735778 & -0.987882 \\

$10^{-12}$& -0.682874 & -0.778679 \\

$10^{-13}$& -0.676666 & -0.688228 \\

$10^{-14}$& -0.676034 & -0.676054 \\

$10^{-15}$& -0.675971 & -0.674794 \\

$10^{-16}$& -0.675965 & -0.674668 \\
\hline
 \end{tabular}
 
 \end{table}

From Fig.~\ref{fig1} the change in slope with $G\mu$ is evident at the measured frequency $f_o=1\ yr^{-1}$.  For large values of $G\mu$, $f_o$ is in the flat red noise section of the spectrum so the slope of $\Omega_{gw}$ goes to zero, and $\beta$ approaches -1.  As the string tension is reduced, $\beta$ is measured first at the high frequency tail so the slope is negative, and thus $\beta$ is more negative.  Once over the peak on the high frequency side, the slope becomes positive, thus $\beta$ must be greater than -1.  The characteristic strain power law dependence $\beta$ then approaches a value of -0.67.  

This dependence on $G\mu$ changes the value of the constant $A$ in Eqn.~\ref{eqs:Aheqn} and thus the limits on $\Omega_{gw}$ \cite{jenet}.  Higher mode dependence only modestly changes the values of $\beta$ since it approaches the limits -1 and -2/3 for high and low frequencies respectively.  The frequency at which it takes on these values depends upon $G \mu$ as seen in Fig.~\ref{fig1} 

                                  \section{Future LISA Sensitivity}

   \subsection{Physical Principles}

The pending launch of LISA will give  new opportunities to observe gravitational waves in general, and a stochastic background in particular~\cite{ci,allen96,cornish,lisa}.  There are a number of theoretical sources for this background, including cosmic strings \cite{mag,kosow,ferrari99,postov,ho4}.  LISA sensitivity has been modeled and methods for improving sensitivity by calibrating noise and integrating over a broad band have been proposed \cite{shane,hoben,corn}.  The cosmic string background density spectrum calculated here is compared to the potential for detection by LISA. 

Detectors such as LISA measure the strain on spacetime, so a relation between the background spectrum and strain is needed.  The rms strain is denoted by $h_{rms}(f)$ or $\bar{h}(f)$, and for an isotropic, unpolarized, stationary background~\cite{jenet,hoben,mag00,allen96}.  This causes problems since typically $h_{rms}$ is the average value of $h_{ij}(x)$ over several wavelengths divided by 2.  It is better to use the spectral density $S_h(f)$, see Appendix D for details of these relationships.  
\begin{equation}
\Omega_{gw}(f)=\frac{4 \pi^2}{3 H_o^2} f^3 S_h (f).
\end{equation}
Another measure is the characteristic strain spectrum $h_c(f)^2=2f S_h(f)$, when substituted gives:
\begin{equation}
\label{eqs:charstrain}
\Omega_{gw}(f)=\frac{2 \pi^2}{3 H_o^2} f^2 h_c^2(f).
\end{equation}
In the literature, $S_h(f)$ is often used and referred to as the spectral density \cite{mag00}.  These relations are important as they relate the energy density spectrum $\Omega_{gw}(f)$, with the strain that LISA will detect.  This limit on the detectable strain is shown as a curve on our plots and is taken from~\cite{shane}.  

Here the LISA sensitivity curve is calculated with the following parameters: SNR=1.0, arm length=5$\times10^9$ m, optics diameter=$0.3$ m, laser wavelength=$1064$ nm, laser power=$1.0$ W, optical train efficiency=0.3, acceleration noise=$3\times10^{-15}$ m/(s$^2\sqrt{Hz}$), position noise=$2\times10^{-11}$ m/$\sqrt{Hz}$.

Extracting the background signal by combining signals over a broad band and an extended time can increase the sensitivity of the detector.  One method known as the ``Sagnac technique'' is shown on plots of the background \cite{hoben,corn}, for example see Fig.~\ref{fig1}.

   \subsection{Limits for LISA Sensitivity}

For both Figures~\ref{fig1} and \ref{fig4} there are limits on the detectability of the background due to strings of various mass densities by LISA from~\cite{shane}.  From Fig.~\ref{fig1} with $\alpha=0.1$ the minimum value of the string tension measurable is $\sim10^{-16}$.  In Fig. 7.4, which plots a smaller $\alpha=10^{-5}$, the minimum is $G\mu\sim10^{-12}$.  These values and others are listed in Table~\ref{table1}.  Sagnac observables increase our available spectrum and change the detectable string tensions as indicated \cite{hoben}. 

Shown in Table~\ref{table1} are the limits on string tensions for string sizes $\alpha$ of 0.1 to $10^{-6}$, along with the corresponding minimum tension that LISA can detect.  Galactic and extra-Galactic binaries confusion limit is taken into account for this table.  For $\alpha<10^{-2}$ the maximum string tension will be greater than $10^{-8}$, although these values are excluded by WMAP and SDSS~\cite{wy}.  It is clear from the table and figures that decreasing $\alpha$ increases the maximum string tension allowed as well as increasing the minimum string tension detectable by LISA.  

 \begin{table}
 \caption{\label{table1}Millisecond Pulsar Limits and LISA sensitivity for $\gamma=50$.  WMAP and SDSS \cite{wy} have excluded $G\mu\geq10^{-7}$.}
 
 \begin{tabular}{|c |c |c|}
  \hline
            & Millisecond Pulsar  &  Minimum LISA \\
  $\alpha$  & Limits, $G\mu<$   &   Sensitivity, $G\mu\geq$\\   
\hline
0.1        & $10^{-9}$ & $10^{-16}$ \\

$10^{-2}$  & $10^{-8}$  & $10^{-15}$\\

$10^{-3}$  & -          & $10^{-14}$\\

$10^{-4}$  & -          & $10^{-13}$\\

$10^{-5}$  & -          & $10^{-12}$\\

$10^{-6}$  & -          & $10^{-11}$\\
\hline
 \end{tabular}
 
 \end{table}

   \subsection{Confusion Noise from Binary Systems}

Galactic white dwarf binaries add a background ``confusion'' that shrouds portions of LISA's spectrum.  This background has been studied and predictions given for its spectrum~\cite{nel}.  There are also extra-Galactic binary sources which similarly create confusion noise~\cite{farm}.  These confusion limits are shown on plots of the gravitational radiation spectra, and they affect the minimum values of string tension that are detectable.

                                   \section{Conclusions}

Broadly speaking, the results here confirm with greater precision and reliability the estimates given in \cite{ho}.  Our survey of spectra provides a set of observational targets for current and future projects able to detect stochastic gravitational wave backgrounds.  For light strings, the string tension significantly affects the  spectrum of gravitational radiation from cosmic string loops. In addition to a reduction in the overall radiation density,   the most significant change is  the shift of the peak of the spectra to higher frequencies for light strings. Near the current pulsar limits, the peak happens to lie close to the frequency range probed by the pulsars, so spectra as computed here are needed as an input prior for establishing the value of the limits themselves.  For much lighter strings,  the peak is at too high a frequency for strings  to be detectable by pulsar timing as the spectrum falls below other confusion backgrounds.  Lighter strings will be detectable by LISA.   For the lightest detectable strings the peak happens to lie in the LISA band and the detailed spectrum must again be included in mounting observational tests.  The spectrum from strings is quite distinctive and not similar to any other known source.  

A high probability of forming loops with stable, non-self-intersecting orbits, as suggested by recent simulations, leads to larger string loops at formation which in turn  give an increased output of gravitational wave energy and improved possibilities for detection for very light strings.  Recent simulations have shown stable radiating loops form at a significant fraction of the horizon, of order $\alpha=0.1$;  for loops of this size our calculations show that  the maximum string tension allowed by current millisecond pulsar measurements is $G\mu<10^{-9}$, and  the minimum value detectable by LISA above estimated  confusion noise is $G\mu\approx10^{-16}$.  In field theories,  the string tension  is typically related to the scale by $G\mu\propto\Lambda_s^2/m_{pl}^2$, so  the maximum detectable scale in Planck masses currently allowed by millisecond pulsars is $\Lambda<10^{-4.5}$, or around the Grand Unification scale; with LISA, the limit will be about  
 $\Lambda_s \sim 10^{-8}$ or $10^{11}\:GeV$.   These results suggest that gravitational wave backgrounds are already a uniquely deep probe into the phenomenology of Grand  Unification and string theory cosmology, and will become more powerful in the future.  The most important  step in establishing these arguments for an unambiguous limit on fundamental theories will be a reliable quantitative calculation of $\alpha$.


 \chapter{Conclusion}

The results of this study indicate that cosmic strings may have a strong gravitational wave signal from loops in the millisecond frequency range, corresponding to the heart of the detection band for the \textit{Laser Interferometer Space Anetenna} (LISA).  This signal comes from both individual cosmic string loops within the Milky Way and a stochastic background from past and present loops throughout the universe.  For light strings the galactic loops produce a large average signal that acts as a ``confusion noise''.

Observations from WMAP, \textit{Sloan Digital Sky Survey} (SDSS), and millisecond pulsars have precluded heavier strings, so this study has extended the range of previous investigations and undertaken some new ones.  We have extended the range over previous investigations into the stochastic background of gravitaitonal waves, and have done a new investigation into the detectability of the gravitaitonal radiation from the normal mode oscillations of cosmic string loops.

From our calculations of the stochastic background it is found that for large loops string tensions down to $G\mu \approx 10^{-18}$ may be detectable, see Fig.~\ref{fig1}.  For these large loops the upper limit is also set at $G\mu < 10^{-9}$ by analysis of millisecond pulsars.  This further constrains string formation scenarios below the Planck scale.  In field theory the energy scale of the phase transition is given by $\Lambda_s^2=G\mu m_{pl}^2$ where $m_{pl}=1.22 \times 10^{19}$ GeV is the Planck mass.  From our large loops we have and upper bound of $\Lambda_s < 10^{-4.5} m_{pl}$ and LISA has a sensitivity down to $\Lambda_s \approx 10^{-8} m_{pl}$.  

For the loops within the galaxy, our calculations indicate that light and large strings may have a significant signal in bulk, see Fig~\ref{fig63}.  The signal from this confusion noise from large loops is detectable by LISA down to a string tension of $G\mu=10^{-19}$, which is lower than previous esimates.  The energy scale detectable is down to $\Lambda_s = 10^{-9.5} m_{pl}$ or $10^{9.5}$ GeV.

Thus cosmic strings can probe fundamental theories of physics from energies of $10^{14.5}$ GeV to $10^{9.5}$ GeV ($10^{-4.5} m_{pl}$ to $10^{-9.5} m_{pl}$).  Theories with energy scales in this range include:  
    \begin{itemize}
\item{GUT scale: $10^{16}$ GeV=$10^{-3} m_{pl}$}
\item{Peccei-Quinn scale: $10^8$ GeV = $10^{-11} m_{pl}$}
\item{B-L symmetry breaking scale: some theories $10^5$ to $10^{10}$ GeV  ($10^{-14}$ to $10^{-9} m_{pl}$)}
\item{Superstring brane cosmologies:  $10^{-6}< G\mu < 10^{-11}$ }
        \end{itemize} 
        
Thus the gravitational wave signal from cosmic string loops is a unique probe of field theory and string theory.  Further refinement in the value of the loop size $\alpha$ is required to eliminate uncertainties in these results.

%
%
%

%
\appendix

   \chapter{General Relativity}
   
 In this appendix we write out some of the basic results from general relativity.  A more complete description can be found in any general relativity text~\cite{schutz,wein,hartle,carroll}

First the Einstein-Hilbert action:
\begin{equation}
S_{E}= \frac{1}{16 \pi G}   \int \sqrt{-g}\; R\; d^4x
\end{equation}
where $R$ is the Ricci scalar and $g$ is the trace of the metric.  We define the Christoffel symbol by,
       \begin{equation}
\Gamma_{\mu\nu}^{\lambda}=\frac{1}{2}g^{\lambda\alpha} \left( \partial_{\mu}g_{\alpha\nu}+ \partial_{\nu}g_{\alpha\mu}-\partial_{\alpha}g_{\mu\nu} \right)
       \end{equation}
and the Riemann tensor,
   \begin{eqnarray}
R^{\mu}_{\nu\rho\sigma}&=&\partial_{\rho}\Gamma_{\nu\sigma}^{\mu}- \partial_{\sigma}\Gamma_{\nu\rho}^{\mu}+ \Gamma_{\alpha\rho}^{\mu}\Gamma_{\nu\sigma}^{\alpha}-\Gamma_{\alpha\sigma}^{\mu} \Gamma_{\nu\rho}^{\alpha}.
   \end{eqnarray}
The Ricci tensor is $R_{\mu\nu}=R^{\alpha}_{\mu\alpha\nu}$, and the Ricci scalar is given by $R=g^{\mu\nu}R_{\mu\nu}$.

We define the matter action as:
\begin{equation}
S_M=\int \sqrt{-g}\;T^{\mu\nu}g_{\mu\nu} d^4x,
\end{equation}
where $T^{\mu\nu}$ is the stress-energy (or momentum-energy) tensor.  The stress-energy tensor measures the energy, energy flux, momentum density, and stress at a point in spacetime.  The components are given by: $T_{00}$= energy density, $T_{0j}$=energy flux, $T_{j0}$= momentum density, $T_{ii}$= pressure, and $T_{ij}$= shear stress.  In matrix form:
\begin{equation}
T_{\mu\nu}= \left( \begin{array}{c|ccc}
\rho           &                         &  \text{energy flux}       &    \\
\hline
               &  P_x                    &        & \text{shear stress}   \\
\vec{p}        &                         &  P_y   &    \\
\text{density} & \text{shear stress}     &        & P_z
\end{array} \right).
\end{equation}

Thus the total gravitational action is the sum of the matter and Einstein actions, $S_g=S_E+S_M$.  Varying this action with respect to the space-time metric, $g_{\mu\nu}$ and setting it to zero, $\delta S=0$ we find:
\begin{equation}
  \label{eq:ein}
R_{\mu\nu}-\frac{1}{2}g_{\mu\nu}R=8\pi G T_{\mu\nu},
\end{equation}
which are the Einstein equations.
 
If we take the trace of Eq.~\ref{eq:ein} we get $R=-8\pi G T$.  Reinserting into Eq.~\ref{eq:ein} gives us,
\begin{equation}
\label{eq:ein2}
 R_{\mu\nu}=8 \pi G \left( T_{\mu\nu}-\frac{1}{2}g_{\mu\nu}T \right).
 \end{equation}
 
Eqs.~\ref{eq:ein} and ~\ref{eq:ein2} are the same, but Eq.~\ref{eq:ein2} is more useful if the stress-energy tensor is known, as in the analysis of cosmic expansion.


    \chapter{Cosmic Expansion}
    
In this appendix we review some basic results pertaining to the expansion of the universe, perhaps most importantly the dependence of the scale factor, $a(t)$, with cosmic time.  A more complete discussion can be found in any good text on the subject of cosmology~\cite{dodson,kolb,peacock,much,narl,borner,liddle,peebles}.
 
\section{Homogeneous and Isotropic Universe}    
The space-time interval is defined by
 \begin{equation}
ds^2=g_{\mu\nu}dx^{\mu}dx^{\nu},
 \end{equation}
where $g_{\mu\nu}$ is the metric of the space-time.

For a homogeneous and isotropic universe the Friedmann-Roberston-Walker (FRW) is a suitable metric which defines the spacetime interval, and is given by:
\begin{equation}
 ds^2= -dt^2+a(t)^2 \left( \frac{dr^2}{1-kr}-r^2 d\Omega^2 \right),
 \end{equation}
 where $t$ is the cosmic time, $r$ is a radial coordinate distance, $k$ is the curvature, and $a(t)$ is the scale factor.
 
 The coordinate distance vector $\vec{r}(t)$ is given by,
 \begin{equation}
     \vec{r}(t)=a(t)\vec{\xi},
 \end{equation}
 where $\vec{\xi}$ is a comoving coordinate that remains constant with the expansion.  Thus $a(t)$ defines the relative size and rate of expansion of the universe, and can be solved using the Einstein equations if the constituents of the universe are known, and thus the stress-energy tensor is known.
 
The Hubble law states that the velocity of a galaxy is proportional to its distance from the observer:
\begin{eqnarray}
\vec{v}&=&\frac{\dot{a}}{a}\; \vec{r},\\
       &=& H \vec{r},
\end{eqnarray}
where $H$ is the Hubble parameter and varies with cosmic time.  The Hubble law is not valid for very large redshifts, where cosmic expansion changes due to a transition in the dominant constituent of mass/energy.

Now we define the constituents of the universe and determine their effect on the expansion, and therefore calculate the scale factor, $a(t)$.  Einstein's equations,
\begin{equation}
R_{\mu\nu}=8 \pi G \left( T_{\mu\nu}-\frac{1}{2}g_{\mu\nu}T \right).
 \end{equation}
 are used with the FRW metric.
 
 Lets start with an isotropic ideal fluid, with the stress-energy given by,
 \begin{equation}
 T_{\mu\nu}=(P+\rho)u_\mu u_\nu + P g_{\mu\nu},
 \end{equation}
 where the $u_\mu$ are the four velocity of the fluid with respect to the observer.  If our reference frame is at rest with respect to the fluid, that is $u_\mu=(1,0,0,0)$ and locally flat we can write:
 \begin{equation}
 T_{\mu\nu}=(P+\rho)\delta_{0\mu}\delta_{0\nu} +P \eta_{\mu\nu}
 \end{equation}
 where $\eta_{\mu\nu}=(-1,1,1,1)$ is the Minkowski metric.  Taking the trace to use in Eq.~\ref{eq:ein2},
 \begin{eqnarray}
 tr(T_{\mu\nu})&=&T=T^\mu_{\;\mu}=\eta^{\mu\nu}T_{\mu\nu},\\
               &=& -\rho+3P.
 \end{eqnarray}

    \section{The Friedmann Equations}
    
 After a bit of calculating one finds the Friedmann equations:
 \begin{eqnarray}
      H^2 &=& \left(\frac{\dot{a}(t) }{a(t)}\right)^2=\frac{8 \pi G }{3}\rho-\frac{k}{a(t)^2},\\
      \frac{\ddot{a}}{a} &=& -\frac{4\pi G}{3}(3P+\rho),
  \end{eqnarray}
where $\rho$ is the sum of all mass/energy densities.  These equations describe the dynamics of an isotropic and homogeneous universe, with the first equation usually referred to as THE Friedmann equation.  

This first equation is often rewritten in terms of the ``criticial'' density, the density of the universe which results in a curvature of zero ($k$=0).  This is given by,
\begin{equation}
   \rho_c=\frac{3 H_0^2}{8 \pi G},
 \end{equation}
Where $H_0$ is the current value of the Hubble parameter.  We further define the densities in terms of the critical density,
\begin{eqnarray}
\Omega_j=\frac{\rho_j}{\rho_c},
\end{eqnarray}
where the $j$ represent the various components of the cosmic soup.  This leads to a rewrite of the Friedmann equation,
   \begin{eqnarray}
   H^2=H_o^2 \sum_j{\Omega_j}-\frac{k}{a^2}.
   \end{eqnarray}

Observations from WMAP have placed $\Omega_{\text{net}\,0}=1.02\pm.02$ and thus $k \approx 0$.  Regardless, the density of the universe is startlingly close to unity, a fact which still requires much explanation to be understood.
 
    \section{Cosmic Constituents}
    
 Now we introduce a small bestiary of the cosmos.  These components generally have densities that vary with temperature, and thus with the expansion.  In general we can write down the equation of state of each component in terms of the energy density,
 \begin{equation}
 P=w\; \rho
 \end{equation}
 where $w$ is a scalar.  The reader should be careful to discern the difference between the momentum and pressure.  As is the convention the same variable $p$ is used for both.  For this chapter we will use the upper case for the pressure. 
 
Conservation of the stress-energy tensor gives a relationship between the scale factor and density, which in the curved FRW spacetime must be written in covariant form:
\begin{eqnarray}
 \label{eq:conservation}
   D_{\mu}T^{\mu\nu}  &=& 0,\\
  \partial_{\nu} T^{\mu\nu}+\Gamma^{\mu}_{\nu \lambda}T^{\lambda\nu} + \Gamma^{\nu}_{\nu \lambda}T^{\mu\lambda} &=& 0,
    \end{eqnarray}
   
where $\Gamma^{\beta}_{\alpha \lambda}$ are the Christoffel symbols.  This can also be written in the form:
\begin{equation}
\frac{1}{\sqrt{-g}} \partial_{\mu}\sqrt{-g}\; T^{\mu\nu}=0,
\end{equation}
where $g$ is the determinant of the spacetime metric.

  Using the $\mu=0$ component of the stress-energy tensor leads to,
 \begin{equation}
    \label{eq:scaledynamics}
 \dot{\rho}+3(\rho+P)\frac{\dot{a}}{a}=0.
 \end{equation}
This equation gives the time dependance of the density of each constituent, and also describes the dependance of the densities on the scale factor (or volume).  

The constituents of the universe include radiation (utra-relativistic), matter (non-relativistic), vacuum or ``dark'' energy, neutrinos, gravitational waves, and cosmic strings of the infinite and loop variety.  It has been found via observation that gravitational waves and cosmic strings form a miniscule fraction of the cosmic energy density, and thus aren't considered when calculating the scale factor.  Since neutrinos have mass, it is likely that some of the species are non-relativistic at the present time.  This is a minimal consideration given that their contribution is at least five orders of magnitude smaller than the dark energy and matter at the present.  See Eqs.~\ref{constituents} for values of the density parameters, $\Omega$ of the various components. 

To find the equations of state of the components we start with the energy density by integrating over all possible microstates assuming a non-interacting fluid:
\begin{eqnarray}
\rho &=&  \int_\textbf{p} \frac{g\;d^3p}{(2 \pi \hbar)^{3}}\; \frac{\epsilon}{ e^{\frac{\epsilon-\mu}{kT}}\pm1},\\
     &=& \frac{4\pi g}{(2 \pi \hbar)^2} \int_m^{\infty} \frac{\sqrt{\epsilon^2-m^2}\; \epsilon^2 \; d\epsilon}{e^{\frac{\epsilon-\mu}{kT}}\pm1},\\
\end{eqnarray}
where $\epsilon^2=p^2 + m^2$ is the energy of each particle, $g$ is the number of internal degrees of freedom, and the plus sign is for fermions and  minus for bosons.  For a non-relativistic gas the the energy density is approximately just given by the mass density times $c^2$.  For an ultra-relativistc gas $m=0$ and the results simplifies greatly.

To find the pressure we must sum over all particles in a hemisphere that collide with some area $\Delta A\; \hat{n}$.  These particles then impart mometum to the surface given by $2(\vec{p}\cdot \hat{n})$ for a total contribution to the pressure given by
\begin{eqnarray}
\Delta P &=& \frac{2 (\vec{p} \cdot \hat{n}) \Delta N}{\Delta t \Delta A},\\
         &=& \frac{p^2}{2 \pi \epsilon} \left( \frac{\sqrt{\epsilon^2-m^2}\; \epsilon^2 \; d\epsilon}{e^{\frac{\epsilon-\mu}{kT}}\pm1}\right) \int_{\Omega} \cos^2(\theta) \sin(\theta) d\theta d\phi,\\
          &=& \frac{p^2}{3 \epsilon} \left( \frac{\sqrt{\epsilon^2-m^2}\; \epsilon^2 \; d\epsilon}{e^{\frac{\epsilon-\mu}{kT}}\pm1}\right).
\end{eqnarray}
The expression in parenthesis is the number density of particles with energy $\epsilon$ to $\epsilon + d\epsilon$.

To find the total pressure we integrate over the energy states,
\begin{eqnarray}
P &=& \int_m^{\infty} \frac{p^2}{3 \epsilon}\; \frac{\sqrt{\epsilon^2-m^2}\; \epsilon^2 }{e^{\frac{\epsilon-\mu}{kT}}\pm1}\; d\epsilon,\\
   &=& \frac{\rho}{3}-\frac{m^2 g}{6 \pi^2} \int_m^{\infty}  \frac{\sqrt{\epsilon^2-m^2}}{e^{\frac{\epsilon-\mu}{kT}}\pm1}\; d\epsilon.
\end{eqnarray}
Immediately we see that massless (ultra-relativistic) particles have an equation of state given by $P=\rho/3 $ regardless of the particle species or chemical potential.

For the equations of state for the constituents of the universe we have:
\begin{eqnarray}
  \text{Non-relativistic matter},\; w=0,\;              & &\rho\propto a^{-3},\\
  \text{Ultra-relativistic  matter},\; w=\frac{1}{3},\; & & \rho \propto a^{-4}\\
  \text{Vacuum energy (Dark energy)},\; w=-1,\;         & & \rho=\text{constant}.
 \end{eqnarray}
The results for the scale factor were found from Eq.~\ref{eq:scaledynamics}.  These relations indicate a universe composed of ``dark enegy'' will eventually have the expansion dominated by it.  Both relativistic and non-relativistic matter density shrink as the universe expands, which corresponds to the observation that the very early universe was dominated by radiation, which was replaced by non-relativistic matter, finally to be overtaken today by vacuum energy.  

A description of calculation of the scale factor for this work is found in Appendix D.

 \section{Entropy of the Universe}
 
  We also define the entropy by,
\begin{equation}
S=s\: a(t)^3,
\end{equation}
where $s$ is the entropy density.  From the standard cosmological model homogeneity implies that entropy is conserved as there are no temperature differences to drive heat flow, thus $dS/dt = 0$.




   
 

    \chapter{Signal Detection}
    
 In this appendix we characterize the signal on a detector and the mathematical fomalism used in this process.  From an individual detector we can characterize the output by $s(t)=h(t)+n(t)$ where $h(t)= F_+ h_+ + F_{\times}h_{\times}$ is the gravitational wave signal in terms of its two polarizations, and $n(t)$ is intrinsic noise.  In order to extract the noise and to discern the signal in both space and time, we take the following steps.

           \section{Spectral Density and Characteristic Strain}

In the transverse traceless gauge we can define the wave in terms of its fourier components, in TT gauge by,
\begin{eqnarray}
   \label{eq:hij}
h_{ij}(t,\textbf{x})&=&\int_{-\infty}^{\infty} df \int_{\Omega} d\hat{k} \int_0^{2\pi} d\psi \; \big( \tilde{h}_+(f,\hat{k},\psi) e_{ij}^+(\hat{k},\psi)+{}\nonumber \\
    &&{}+ \tilde{h}_{\times}(f,\hat{k},\psi) e_{ij}^{\times}(\hat{k},\psi) \big) e^{-i2\pi f(t-\hat{k}\cdot \textbf{x})},
\end{eqnarray}
where, $\tilde{h}_A(-f,\hat{k},\psi) = \tilde{h}^{*}_A(f,\hat{k},\psi)$.  Also $\hat{k}$ is a unit vector pointed in the direction of propagation and $\psi$ is the angle of polarization of the wave.  The two polarization tensors have the orthogonality relation $e_{ij}^A(\hat{k})e^{ij}_{A'}(\hat{k})=2\delta^{AA'}$, no sum over $A$.  Now we define the \textit{spectral density} $S_h(f)$ from the ensemble average of the fourier amplitudes,
\begin{eqnarray}
\left\langle \tilde{h}^*_A(f,\hat{k},\psi) \tilde{h}_{A'}(f',\hat{k}',\psi') \right\rangle=\delta_{AA'} \delta(f-f')\nonumber\\ 
\times \frac{\delta^2(\hat{k},\hat{k}')}{4\pi} \frac{\delta(\psi-\psi')}{2\pi} \frac{1}{2} S_h(f),
\end{eqnarray}
where $\delta^2(\hat{k},\hat{k}')=\delta(\phi-\phi') \delta(\cos (\theta)-\cos(\theta'))$. 

Now we write down the average of the real waves we get,
\begin{eqnarray}
\left\langle   h_{ij}(t,\textbf{x})h^{ij}(t,\textbf{x})\right\rangle &=&2\; h_{rms}^2,\\
     &=& 2\int_{-\infty}^{\infty} df S_h(f),\\
     &=& 4 \int_{f=0}^{f=\infty} d(\ln f)\:f \:S_h(f).
\end{eqnarray}

The characteristic strain is defined by,
\begin{equation}
 \label{strain}
 \left\langle   h_{ij}(t,\textbf{x})h^{ij}(t,\textbf{x}) \right\rangle=2 \int_{f=0}^{f=\infty} d(\ln f)\:h_c^2(f).
\end{equation}
Therefore, 
\begin{eqnarray}
h_c^2(f)=2 f \:S_h(f).
\end{eqnarray}
 From above we see that $S_h(f)$ gives the rms value of $h$ within a range of $f$ if the strain is confined to a small region of $f$:
 \begin{equation}
 h^2 \approx \Delta f \; S_h(f),
 \end{equation}   
 or more broadly for the Fourier transform,
 \begin{equation}
 \left| \tilde{h}(f) \right|^2 \approx T \; S_h(f).
 \end{equation}
 where T is the total integration time of observation.  
 
 For LISA the Fourier transform is defined as~\cite{shane1},
 \begin{equation}
   \tilde{h}(f)=\frac{1}{\sqrt{T}} \int_{-T/2}^{T/2} dt\; h(t) e^{i 2 \pi f t},
  \end{equation}
 which results in the equality,
 \begin{equation}
  S_h(f)=  \left| \tilde{h}(f) \right|^2.
 \end{equation}
 In this relationship, the spectral density is often defined by the new variable, $S_h(f)=h_f^2$.
    
    \section{Noise and Strain Sensitivity}

The fourier components of the noise satisfy,
\begin{equation}
\left\langle \tilde{n}_i^*(f) \tilde{n}_j(f') \right\rangle=\frac{1}{2}\delta(f-f')\delta_{ij}S^{(i)}_n(f),
\end{equation}
where $S_n(f)$ is the spectral noise density, dimensions Hz$^{-1}$.  So we get,
\begin{equation}
\left\langle n^2(t) \right\rangle=\int_0^{\infty}df S_n(f). 
\end{equation}
 We can also define the strain sensitivity as $\sqrt{S_n(f)}$.

        \section{Single Detector Response}

The response from detetor is given by $S(t)=h(t)+n(t)$ where $h$ is the signal from the waves and $n$ is the noise.  
\begin{equation}
h(t)=D^{ij}h_{ij}(t),
\end{equation}
where $D^{ij}$ is the \textit{detector tensor}.  Examples include an interferometer given by $D^{ij}=\frac{1}{2}(\hat{m}_i\hat{m}_j-\hat{n}_i\hat{n}_j)$, and a bar detector $D_{ij}=\hat{\ell}_i\hat{\ell}_j$.  Note that $m$ and $n$ are unit vectors along the axes of the arms, not necessarily orthogonal.  From eqn.~\ref{eq:hij} we get,
\begin{eqnarray}
h(t)&=&\sum_{A=+,\times}\int_{-\infty}^{\infty}df \int_{\Omega} d\hat{k} \int_0^{2\pi} d\psi \nonumber\\  
      && {}\times \tilde{h}_A(f,\hat{k},\psi) D^{ij} e_{ij}^A(\hat{k},\psi) e^{-i2\pi f(t-\hat{k}\cdot \vec{x})}.
\end{eqnarray}
We can also define the \textit{detector pattern functions} as,
\begin{equation}
F_A(\hat{k},\psi)=D^{ij} e_{ij}^A(\hat{k},\psi),
\end{equation}
 therefore,
\begin{eqnarray}
h(t)&=&\sum_{A=+,\times}\int_{-\infty}^{\infty} df \int_{\Omega} d\hat{k} \int_0^{2\pi} d\psi \nonumber\\     
     &&{}\times \tilde{h}_A(f,\hat{k},\psi) F_A(\hat{k},\psi) e^{-i2\pi f(t-\hat{k}\cdot \vec{x})},
\end{eqnarray}
and its Fourier transform,
\begin{eqnarray}
\tilde{h}(f)&=&\sum_{A=+,\times} \int_{\Omega} d\hat{k} \int_0^{2\pi} d\psi \nonumber\\
   &&{}\times \tilde{h}_A(f,\hat{k},\psi) F_A(\hat{k},\psi) e^{i2\pi f\hat{k}\cdot \vec{x}}.
\end{eqnarray}

For a stochastic background we find that $<h(t)>=0$ so we want to look at $<h^2(t)>$,
\begin{eqnarray}
\left\langle h^2(t) \right\rangle &=& F \int_{0}^{\infty} df S_h(f),\\
F& \equiv & \int \frac{d\hat{k}}{4\pi} \int \frac{ d\psi}{2\pi} \sum_{A=+,\times} F^A(\hat{k},\psi)F^A(\hat{k},\psi),
\end{eqnarray}
where $F$ is the \textit{pattern function} of the detector and $S_h(f)$ is the spectral density.  These equations are ultimately independent of $\psi$ for a stochastic background, so when calculating backgrounds they are omitted in future formulae.

\section{Signal to Noise}

A gravitational wave of Fourier transform $\tilde{h}(f)$ incident on a dector has a signal to noise ratio of,
\begin{equation}
(SNR)^2=4\int_0^{\infty} df \frac{|\tilde{h}(f)|^2}{S_n(f)}.
\end{equation}
This is an upper limit assuming optimal filtering.

    \subsection{Signal to Noise of a Single Detector for a Periodic Source}
    
A period source delivers a signal at the detector given by,
\begin{equation}
h(t)= F_+(\hat{k}, \psi) h_+(t)  +  F_{\times}(\hat{k}, \psi) h_{\times}(t),
\end{equation}
where $h_A(t)$ are periodic functions of time with some frequency $f_o$.  The two $h$'s may differ up to a phase $\phi$.

From the equation above we find,
\begin{eqnarray}
(SNR)^2 &=& \left|F_+(\hat{k}, \psi) h_{+,o}  +  F_{\times}(\hat{k}, \psi) h_{\times,o}e^{i \phi}\right|^2 \int_0^{\infty} df \frac{\delta(f-f_o) \delta(0)}{S_n(f)},\\
      &=& \left|F_+(\hat{k}, \psi) h_{+,o}  +  F_{\times}(\hat{k}, \psi) h_{\times,o}e^{i \phi}\right|^2  \frac{T}{S_n(f_o)},
\end{eqnarray}
where $T$ is the total observation time.

    \subsection{Signal to Noise of a Single Detector for the Stochastic Background}
  
To compute the signal to noise ratio we need both $\left\langle h^2(t)\right\rangle$ and $\left\langle n^2(t)\right\rangle$.  From equations above we note that:
\begin{eqnarray}
\left\langle s^2(t)\right\rangle &=& \left\langle n^2(t)\right\rangle + \left\langle h^2(t)\right\rangle,\\
                              &=& \int_0^{\infty} df(S_n(f)+F \; S_h(f)).
\end{eqnarray}

To calculate the signal to noise the signals must be put into bins corresponding to the observation time $T$ of frequency range $\Delta f=1/T$.  We then sum over the bins such that $\int S_n df = \sum S_n \Delta f$, and look in each freqency bin to find the signal to noise (SNR):
\begin{eqnarray}
  SNR &=& \frac{\left\langle h^2(t)\right\rangle}{\left\langle n^2(t)\right\rangle},\\
      &=& \frac{F\; S_h(f) \Delta f}{S_n(f) \Delta f},\\
      &=& \frac{F\; S_h(f)}{S_n(f)}.
\end{eqnarray}
The final result is independent of integration time.  Thus for a single detector searching for a stochastic background signal, it will either stand out immediately or cannot be detected over the noise.

The minimum detectable value of $S_h(f)$ for a given $SNR$ and $S_n(f)$ is given by,
\begin{equation}
  (S_h(f))_{min}=\frac{S_n(f)}{F} SNR.
\end{equation}
The corresponding minimim detectable spectral density $\Omega_{gw}$ is then found to be,
\begin{equation}
(\Omega_{gw}(f))_{min}=\frac{4 \pi^2 f^3}{3 H_o^2} \frac{S_n(f)}{F} SNR.
\end{equation}

    \section{Two-detector Correlations}

Now we want to look into multiple detectors, and calculate the correlations between these detectors.  For the $i^{th}$ detector we get the signal,
\begin{equation}
h_{(i)}(t)=\sum_{A=+,\times}\int_{-\infty}^{\infty} df \int_{\Omega} d\hat{k}\; \tilde{h}_A(f,\hat{k}) F_{(i)}^A(\hat{k}) e^{-i2\pi f(t-\hat{k}\cdot \vec{x}_i)},
\end{equation}
so we can write the fourier components of $h(t)$,
\begin{equation}
\tilde{h}_{(i)}(f)=\sum_{A=+,\times} \int_{\Omega} d\hat{k}\; \tilde{h}_A(f,\hat{k}) F_{(i)}^A(\hat{k}) e^{i2\pi f\hat{k}\cdot \vec{x}_i},
\end{equation}
So correlating two amplitudes from two different detectors will be given by,
\begin{equation}
S_{12}=\int_{-T/2}^{T/2} dt \int dt' s_1(t) s_2(t') Q(t,t').
\end{equation}
$Q$ will tend to fall off quickly for large $t-t'$, and the limit for large $T$,
\begin{equation}
S_{12}=\int_{-\infty}^{\infty} df \tilde{s}^*_1(f) \tilde{s}_2(f) \tilde{Q}(f).
\end{equation}
We want to find the ensemble average $<S_{12}>$, and the contribution from GW's is given by,
\begin{eqnarray}
\left\langle h_{(12)} \right\rangle&=&\int_{-\infty}^{\infty} df \left\langle \tilde{h}^*_{(1)} \tilde{h}_{(2)}     \right\rangle \tilde{Q}(f),\\
&=&\int_{-\infty}^{\infty} df \int d\hat{k} d\hat{k}'\: e^{2\pi i f\;(\hat{k}\cdot \vec{x}_1 -\hat{k}'\cdot \vec{x}_2)}\sum_{AA'}\times  \nonumber\\
&&  F_1^A(\hat{k}) F_2^A(\hat{k}')\left\langle \tilde{h}^*_A(f,\hat{k}) \tilde{h}_{A'}(f,\hat{k}') \right\rangle \tilde{Q}(f),\nonumber\\
  &=&T\int_{-\infty}^{\infty} df \frac{1}{2} S_h(f) \Gamma(f) \tilde{Q}(f), 
\end{eqnarray}
where $\delta(f-f)=\int_{-T/2}^{T/2} dt=T$ and
\begin{eqnarray}
\Gamma(f)=\int \frac{ d\hat{k}}{4\pi} \left(\sum_A F^A_1(\hat{k}) F^A_2(\hat{k}) \right) e^{2 \pi i f \hat{k}\cdot \Delta\vec{x}}.
\end{eqnarray}

   \subsection{Filtering}

We define the variation of $S_{12}$ as,
\begin{equation}
N \equiv S_{12}-\left\langle S_{12} \right\rangle,
\end{equation}
which leaves $<N>=0$.
\begin{eqnarray}
\left\langle  N^2 \right\rangle &=& \left\langle S^2_{12}\right\rangle-\left\langle S_{12}\right\rangle^2, \\
&=& \int_{-\infty}^{\infty} df df'\tilde{Q}(f) \tilde{Q}^*(f') \Big[ \left\langle  \tilde{S}_1^*(f) \tilde{S}_2(f) \tilde{S}_1(f') \tilde{S}_2^*(f')    \right\rangle  \nonumber\\
  && -\left\langle \tilde{S}^*_1(f) \tilde{S}_2(f')\right\rangle    \left\langle \tilde{S}_2^*(f) \tilde{S}_1(f')    \right\rangle \Big].
\end{eqnarray}

We are primarily interested in the regime where $n(t)>>h(t)$, and the noise in the two detectors is uncorrelated.  Then $S_i \approx n_i$, and we also assume a gaussian distribution,
\begin{equation}
\left\langle  N^2 \right\rangle = \frac{T}{4} \int_{-\infty}^{\infty} df \left|\tilde{Q}(f) \right|^2 S^2_n(f),
\end{equation}
where,
\begin{equation}
S_n(f)=\sqrt{ S^{(1)}_n(f) S^{(2)}_n(f) }.
\end{equation}

For the signal to noise ratio we define,
\begin{equation}
SNR=\left( \frac{\left\langle S_{12} \right\rangle}{\sqrt{\left\langle N^2 \right\rangle}} \right)^{1/2}.
\end{equation}  Allen defines it without the square root.

We are looking for a filter function $\tilde{Q}(f)$ that maximizes the SNR.  We can define an inner product using the conventions above as,
\begin{equation}
(A,B)=\int_{\infty}^{-\infty} df A^*(f) B(f) S^2_n(f).
\end{equation}
Which gives us,
\begin{equation}
\left\langle N^2 \right\rangle=\frac{T}{4}(\tilde{Q},\tilde{Q}),    
\end{equation}
and
\begin{equation}
\left\langle h_{(12)} \right\rangle = \frac{T}{2} \left(\tilde{Q},\frac{ \Gamma S_h }{S_n^2}  \right)
\end{equation}

Then we must maximize,
\begin{equation}
(SNR)^4=\frac{\left\langle h_{(12)} \right\rangle^2}{\left\langle N^2 \right\rangle}   = T \left(\tilde{Q},\frac{ \Gamma S_h }{S_n^2}  \right)^2 \frac{1}{( \tilde{Q},\tilde{Q})}.
\end{equation}
The solution is given by,
\begin{equation}
\tilde{Q}(f)=k\frac{\Gamma(f)S_h(f)}{S_n^2(f)},
\end{equation}
with $k$ an arbitrary constant.

       
               \chapter{Numerical Calculations for the Stochastic Background}

   \section{General}

For computations the relevant differential and integral equations are solved numerically.  Here is a list of the equations that are used, most of which have been discussed in the chapters above.  In addition there are several more that are expanded upon in this appendix.  Below note that $n(t,t')$ is the number density with time $t$ at an observation time $t'$, while $n(f',t')$ is the number density as a function of frequency and time $t'$.  The time $t$ is always earlier than the time $t'$, which is the time of observation of the quantity in question.
\begin{eqnarray}
\dot{a}&=&\sqrt{\frac{8 \pi G}{3} \rho a^2},\\
\rho&=&\rho_{rad}+\rho_{\nu}+\rho_{matter}+\rho_\Lambda 
              \nonumber\\
     & &{}+\rho_\infty+\rho_L+\rho_{gw},\\
L(t_c,t)&=&\alpha c H(t_c)^{-1}-G \mu \frac{t-t_c}{c},\\
f(t,t')&=&\frac{2c}{L(t,t')},\\
n(t,t')&=&\frac{N_t}{\alpha} \left(\frac{H(t)}{c}\right)^3 \left(\frac{a(t)}{a(t')}\right)^3,\\
\rho_{gw}(f)&=&\gamma G \mu^2 c \int_{t_0}^{t_{univ}} \frac{a(t')^4}{a(t_{univ})^4} 
            \nonumber\\
              & &\times\:n(f\frac{a(t')}{a(t_{univ})},t') dt'.
\end{eqnarray}
Recent observations allow us to neglect loops $\rho_L$, infinite strings $\rho_{\infty}$, and gravitational waves $\rho_{gw}$ in the Friedmann equation, so these terms are removed \cite{pdg}. 

     \section{Scale Factor}

Calculations of the scale factor are done using the standard cosmological model with a homogeneous isotropic spacetime defined by the Friedmann-Robertson-Walker metric, solved using Einstein's equations~\cite{ba,dodson,kolb}.  As an addition the effects of a cosmological constant on the cosmic string gravitational wave background are computed.  The energy densities used to describe the dynamics of the scale factor are radiation, neutrinos, matter, and a cosmological constant.  The equation of motion for the scale factor is determined by the Friedmann equation.
\begin{equation}
\label{friedmann}
\left(\frac{\dot{a}}{a}\right)^2=\frac{8\pi G}{3}(\rho_{rel}+\rho_m)+\frac{\Lambda}{3},
\end{equation}
where the $\rho$'s are functions of the scale factor. We then integrate the above equation to find $a(t)$.

The relativistic and matter densities follow the relations:
\begin{subequations}
\begin{eqnarray}
\rho_{rel}=\frac{(\rho_{rel})_o a_o^4}{a(t)^4},\\
\rho_{m}=\frac{(\rho_{m})_o a_o^3}{a(t)^3}.
\end{eqnarray}
\end{subequations}
The gravitational radiation density also goes as $a(t)^{-4}$. 

In order to compute the scale factor the Friedmann equation is transformed into an integral equation and solved numerically.   The main source of error is the relativistic treatment of neutrinos for all time, but this will incur a very small error in more recent times.  Given that neutrinos have mass, it is possible several species will be non-relativistic at the present time.  This is a minimal consideration as the matter and dark energy contributions are larger by at least five orders of magnitude when the neutrinos are non-relativistic.  The following parameters are taken from the Particle Data Group, 2006~\cite{pdg}:
\begin{subequations}
   \label{constituents}
\begin{eqnarray}
  \Omega_{\Lambda}&=&0.76,\\
\Omega_{matter}&=&0.24,\\
  \Omega_{rad}&=&4.6\times10^{-5},\\
 \Omega_{\nu}&=&3.15\times10^{-5},\\
           h&=&0.73,\\
      \rho_c&=&\frac{3 H_o^2}{8 \pi G},
\end{eqnarray}
\end{subequations}
where $\Omega_j=\rho_j /\rho_c$, and the constant $h$ is given by $H_o=100h\ km\ s^{-1}Mpc^{-1}$.

    \section{Gravitational Radiation from Loops}

At a given time $t'$, there is an earlier time $t$ before which all of the loops have decayed, which we denote $t_e$.  To find this value at any time $t'$, Eqn.~\ref{eqs:length} is set to zero giving $L(t,t')=0$, and  $t=t_e$ is calculated numerically.  Note that $t_e$ is a function of $t'$.  For very early times $t_e(t')$ can be lower than the initial time in our sum over cosmic history $t_0$.  In this case $t_0$ was used instead of the value for $t_e(t')$. 

For many of the calculations it is assumed each loop radiates at only one frequency, $f=2c/L$.  Then a numerical list of values is computed for $t=t_e$ to $t=t'$, $\{L(t,t'),n(t,t')\}$, and an interpolation function is created from the list.   This is the number density as a function of length and time $t'$, $n(L,t')$.  Next a list of frequency and number density is created, $\{2c/L,n(L,t')\}$, with lengths running from $L(t_e,t')$ to $L(t',t')$.  An interpolation function is made from this list, and this results in the number density as a function of time $t'$, $n(f',t')$.  But the frequencies must be redshifted to the current time $t_{univ}$ so the function is $n(fa(t')/a(t_{univ}),t')$.

Then we must integrate over all $t'$ from the earliest time of loop formation $t_0$ to the present age of the universe $t_{univ}$.  This integral,
\begin{equation}
\rho_{gw}(f)=\gamma G \mu^2 c \int_{t_0}^{t_{univ}} \frac{a(t')^4}{a(t_{univ})^4}\ n(f\frac{a(t')}{a(t_{univ})},t') dt',
\end{equation}
is transformed into a sum,
\begin{equation}
\rho_{gw}(f)=\gamma G \mu^2 c \sum_{t'=t_0}^{t_{univ}} \frac{a(t')^4}{a(t_{univ})^4}\ n(f\frac{a(t')}{a(t_{univ})},t') \Delta t'.
\end{equation}
Suitable $\Delta t'$'s are chosen to ensure both computability and to minimize errors.  In general we solve all equations in logarithmic bins so $\Delta t'$ will vary in accordance.  Using small values of $\Delta t$ to facilitate computation speed will result in relatively small error; choosing $\Delta t=0.1$ in log scale gives a 10\% smaller value for $\Omega_{gw}$ compared to $\Delta t=0.001$ or $\Delta t=0.01$ in log scale.  Values smaller than $\Delta t=0.01$ give little or no change to the background.  After the sum the result is the gravitational wave density as a function of frequency at the present time.

To include higher modes the power coefficients $P_j$ must be included in the calculation.  The subscript $j$ is used instead of $n$ here to prevent confusion with the number density, $n(f',t')$.  At each time $t'$ the number density, and thus the time rate of change of the energy density, of loops is calculated exactly as above, but more frequencies are included.  Each number density at frequency $f_j$ is summed and weighted with $P_j$, and this number density is then integrated over cosmic history,
\begin{eqnarray}
  \rho_{gw}(f)&=&G\mu^2 c \int_{t_0}^{t_{univ}}dt' \frac{a(t')^4}{a(t_{univ})^4}
                 \nonumber\\
                 & &\times\left\{\sum_{j=1}^{\infty} P_j\: n(f_j\frac{a(t')}{a(t_{univ})},t')\right\} .
\end{eqnarray}
The overall integral is then transformed into a sum as before.  The sum of the power coefficients is $\sum P_j=\gamma$, where the $P_j\propto j^{-4/3}$.

The final result is to calculate $\Omega_{gw}(f)$ at the present time.  This is done by finding the current energy density of gravitational waves as a function of frequency, and then taking the derivative with respect to f:
\begin{equation}
\Omega_{gw}(f)=\frac{f}{\rho_c}\frac{d\rho_{gw}}{df}.
\end{equation}
This is then plotted as $\log(\Omega_{gw}(f)\ h^2)$ vs. $\log(f/Hz)$, as seen in Figures 7.1, $7.3-7.7$.  This quantity is, again, the density of stochastic  gravitational wave energy per log of frequency in units of the critical density and is related to the strain on space by the gravity waves.

\raggedbottom\sloppy

\end{document}